\documentclass[twocolumn]{aastex701}

\usepackage{amsmath}
\usepackage{amsfonts}
\usepackage{graphicx}
\usepackage{rotating}
\usepackage{natbib}
\usepackage{txfonts}
\usepackage{color}
\usepackage{wrapfig}
\usepackage{amssymb}
\usepackage{color}
\usepackage{colortbl}
\usepackage{mhchem}

\shorttitle{MAGOS: Hot Cores in the LMC} 
\shortauthors{T. Shimonishi et al.} 

\begin{document}

\title{MAGellanic Outflow and chemistry Survey (MAGOS): Hot cores in the LMC}

\author[orcid=0000-0002-0095-3624]{Takashi Shimonishi}
\affiliation{Institute of Science and Technology, Niigata University, Ikarashi-ninocho 8050, Nishi-ku, Niigata 950-2181, Japan}
\email{shimonishi@env.sc.niigata-u.ac.jp}  

\correspondingauthor{Takashi Shimonishi} 
\email{shimonishi@env.sc.niigata-u.ac.jp}

\author[orcid=0000-0002-6907-0926]{Kei E. I. Tanaka}
\affiliation{Department of Earth and Planetary Sciences, Institute of Science Tokyo, Meguro, Tokyo, 152-8551, Japan}
\email{kei.tanaka@eps.sci.isct.ac.jp}

\author[orcid=0000-0001-7511-0034]{Yichen Zhang}
\affiliation{Department of Astronomy, School of Physics and Astronomy, Shanghai Jiao Tong University, 800 Dongchuan Road, Shanghai 200240, China}
\affiliation{State Key Laboratory of Dark Matter Physics, School of Physics and Astronomy, Shanghai Jiao Tong University, Shanghai 200240, China}
\affiliation{Key Laboratory for Particle Astrophysics and Cosmology (MOE) / Shanghai Key Laboratory for Particle Physics and Cosmology, Shanghai 200240, China}
\email{yczhang.astro@gmail.com}

\author[orcid=0000-0002-2026-8157]{Kenji Furuya}
\affiliation{RIKEN Pioneering Research Institute, 2-1 Hirosawa, Wako-shi, Saitama 351-0198, Japan}
\email{kenji.furuya@riken.jp}

\author[0000-0002-8691-4588]{Yu Cheng}
\affil{National Astronomical Observatory of Japan, 2-21-1 Osawa, Mitaka, Tokyo, 181-8588, Japan}
\email{ycheng.astro@gmail.com}

\author[orcid=0000-0001-5817-6250]{Asako Sato}
\affiliation{Institut de Ciències de l’Espai (ICE-CSIC), Campus UAB, Can Magrans S/N, E-08193 Cerdanyola del Vallès, Catalonia, Spain}
\email{asako@ice.csic.es}

\begin{abstract}  
The Large Magellanic Cloud (LMC) provides a key laboratory for exploring the diversity of star formation and interstellar chemistry under subsolar metallicity conditions. 
We present the results of a hot core survey toward 30 massive protostellar objects in the LMC using the Atacama Large Millimeter/submillimeter Array (ALMA) at 350~GHz. 
Continuum imaging reveals 36 compact sources in total, among which line analyses identify 9 hot cores and 1 hot-core candidate, including two newly identified sources. 
We detect \ce{CO}, \ce{HCO+}, \ce{H^{13}CO+}, \ce{HC^{15}N}, \ce{HC3N}, \ce{SiO}, \ce{SO}, \ce{SO+}, \ce{NS}, \ce{SO2}, \ce{^{34}SO2}, \ce{^{33}SO2}, \ce{CH3OH}, \ce{^{13}CH3OH}, \ce{HCOOH}, \ce{HCOOCH3}, \ce{CH3OCH3}, \ce{C2H5OH}, \ce{H2CCO} (tentative), and hydrogen recombination lines from hot cores. 
\ce{CH3OCH3}, a complex organic molecule larger than \ce{CH3OH}, is detected for the first time in a hot core outside the LMC bar region. 
All hot cores show stronger emission in the high-excitation SO line compared to non-hot-core sources, suggesting that its strong detection will be useful for identifying hot-core candidates in the LMC. 
Chemical analysis reveals a spread of more than two orders of magnitude in \ce{CH3OH} abundances, with some sources deficient in COMs. 
In contrast, \ce{SO2} is detected in all hot cores, and its abundance shows a good correlation with rotational temperature. 
The hot cores without \ce{CH3OH} detections are all located outside the LMC bar region and are characterized by either high luminosity or active star formation in their surroundings. 
A combination of locally low metallicity, active star formation in the vicinity, and high protostellar luminosity may jointly trigger the COM-poor hot core chemistry observed in the LMC.
\end{abstract} 

\keywords{astrochemistry --- ISM: molecules --- stars: protostars --- Magellanic Clouds --- radio lines: ISM}

\section{Introduction} \label{sec_intro} 
Massive stars have played an important role throughout cosmic history \citep[see e.g.,][and references therein]{ZY07,Tan14,Mot18,Beu25}. 
The first stars born immediately after the Big Bang provided the first heavy elements to the Universe. 
In the present-day Galaxy, massive stars are the primary sources of ultraviolet (UV) radiation, turbulent energy, and heavy elements, which govern the dynamical and chemical evolution of the interstellar medium (ISM). 
Massive stars also play a significant role in shaping the molecular complexity of the ISM, as various interstellar molecules are frequently detected along sightlines toward embedded massive protostars \citep[e.g.,][]{Her09}. 
Understanding the metallicity dependence of the physical and chemical processes in massive star formation is particularly important because the galactic metallicity increases with cosmic evolution. 

Owing to its high resolution and sensitivity, the Atacama Large Millimeter/submillimeter Array (ALMA) has enabled observations of nearby low-metallicity galaxies, the Large and Small Magellanic Clouds (LMC/SMC), at molecular cloud core scales of $\sim$0.1~pc.
The LMC is an ideal laboratory for low-metallicity star formation studies, owing to its proximity \citep[49.97 $\pm$ 1.11 kpc,][]{Pie13}, moderately face-on geometry \citep[$\sim$35$^{\circ}$,][]{vdM01}, and sub-solar metallicity environment \citep[$Z$ $\sim$1/2--1/3 of the solar metallicity $Z_\odot$; e.g., ][]{Rus92, Duf82, Wes90, And02, Rol02}. 
Recent ALMA observations have indeed detected protostellar activities, such as outflows, in the LMC and SMC \citep[e.g.,][]{Fuk15,ST16,Tok22b,ST23}. 

Hot molecular cores are compact ($\lesssim$0.1~pc), dense ($\gtrsim$10$^6$~cm$^{-3}$), and hot ($\gtrsim$100~K) protostellar sources, which appear in the early evolutionary stage of star formation \citep[e.g.,][and references therein]{Kur00,Naz25}. 
The chemistry at this stage is thought to be driven by the warm-up and sublimation of ice mantles of dust grains formed during the earlier, colder stages, as well as by subsequent high-temperature gas-phase reactions \citep[e.g.,][]{Cha92, NM04, Gar08b}. 
A wide variety of molecular species, including complex organic molecules (COMs), are often detected in solar-metallicity hot cores owing to their chemically-rich nature \citep[e.g.,][]{Her09}. 
A comprehensive understanding of the chemical compositions of hot cores across different metallicities is essential for elucidating how natal environments influence the chemical evolution of protostellar sources. 

With ALMA, hot cores have now been detected in low-metallicity environments, including the LMC \citep{ST16,ST20,Sew18,Sew22a,Gol24,Bro25}, the SMC \citep{ST23}, and the extreme outer Galaxy \citep{ST21,Ike25}. 
Among LMC hot cores, the abundance of CH$_3$OH, one of the simplest COMs and a classical hot core tracer, shows a large scatter: in some sources, it is roughly consistent with the metallicity-scaled abundance of their Galactic counterparts, while in others, it is significantly depleted beyond what would be expected from metallicity differences alone. 
The reason behind the chemical diversity of organic molecules in the LMC remains unclear; however, since CH$_3$OH is predominantly formed through surface reactions on cold dust grains \citep[e.g.,][]{Wat07}, astrochemical simulations suggest that variations in chemical evolution during the ice formation stage may influence the abundances of organic molecules \citep{Ach15,Ach16,Ach18,Pau18,ST20}. 
In contrast to organic molecules, sulfur-bearing species, SO and SO$_2$, are commonly detected in LMC/SMC hot cores with similar abundances \citep[][]{ST23}. 
This suggests that these inorganic molecules may serve as useful hot core tracers in the LMC, replacing CH$_3$OH, which is commonly used as a hot core tracer in solar-metallicity environments. 
However, the number of hot core samples observed in low-metallicity environments, including the LMC, remains limited, highlighting the need for systematic survey observations.

The ``MAGellanic clouds Outflow and chemistry Survey (MAGOS)'' is a comprehensive project aimed at exploring various aspects of star formation within the LMC and SMC with ALMA (PI: K. Tanaka; Project code 2019.1.01770.S). 
Using MAGOS data, \citet{ST23} have reported the detection of two new hot cores in the SMC. 
A series of studies have also used MAGOS data to identify and classify filamentary structures associated with protostars in the LMC \citep{Tok23} and the SMC \citep{Tok25}. 
This paper presents the results of the hot core survey for $\sim$30 LMC massive protostellar objects. 

This paper is structured as follows. 
Section~\ref{sec_tarobsred} describes the details of target selection, observations, and data reduction. 
Section~\ref{sec_res} presents the spectra and line intensity distributions of LMC hot cores obtained from the observations. 
Section~\ref{sec_ana} outlines the analysis of column densities based on the detected molecular lines and continuum emission. 
Section~\ref{sec_disc} discusses the molecular species that serve as hot core tracers in the LMC, compares the associated physical properties and molecular compositions, and examines the possible factors contributing to the chemical diversity of organic molecules in the LMC. 
Finally, Section~\ref{sec_sum} provides a summary of this work.

\section{Observations and Data Reduction} \label{sec_tarobsred} 
\subsection{Target selection} \label{sec_tar} 
The target sources were selected from the catalog of young stellar objects (YSOs) presented by \citet{Sea09}, which includes 277 sources. 
From this catalog, 30 infrared sources were chosen as the pointing centers for the present ALMA observations. 

To begin the selection process, we calculated the luminosities of the sources in the catalog. 
The sources in the \citet{Sea09} catalog include near-infrared and mid-infrared photometric data, which are based on the 2MASS and Spitzer SAGE catalogs \citep{Skr06,Mei06}. 
Far-infrared data were obtained by performing positional cross-matching between the \citet{Sea09} sources and the Herschel HERITAGE catalog \citep{Mei10}, resulting in spectral energy distributions (SEDs) covering the 1--500~$\mu$m range. 
The bolometric luminosities ($L_{\mathrm{bol}}$) were derived by directly integrating the fluxes at the available wavelengths: 1.2, 1.7, 2.2, 3.6, 4.5, 5.8, 8.0, 24, 100, 160, 250, 350, and 500~$\mu$m. 
The luminosities of the target YSOs are primarily dominated by far-infrared fluxes, for which the angular resolutions of Herschel/PACS at 100~$\mu$m and 160~$\mu$m are approximately 7$\arcsec$ and 11$\arcsec$, respectively \citep{Mei13}.

Next, we divided the sources into three luminosity bins: $10^{4}$–$10^{4.5}$~$L_\odot$ (119 sources), $10^{4.5}$–$10^{5}$~$L_\odot$ (57 sources), and $10^{5}$–$10^{5.5}$~$L_\odot$ (17 sources). 
Sources with luminosities below $10^{4}$~$L_\odot$ (84 sources) were excluded from the selection. 
Within each luminosity bin, we arbitrarily selected 10 sources, resulting in a total of 30 targets. 
Sources were chosen based on the presence of molecular outflows, hot-core signatures, or infrared spectral features indicative of early evolutionary stages, e.g., silicate/ice absorption, or polycyclic aromatic hydrocarbon (PAH) emission. 
Therefore, the final sample is not a purely random selection. 
We denote the 10 sources in the $10^{4}$–$10^{4.5}$ bin as Ll01 to Ll10, those in the $10^{4.5}$–$10^{5}$ bin as Lm01 to Lm10, and those in the $10^{5}$–$10^{5.5}$ bin as Lh01 to Lh10.

\begin{figure}[tp!]
\begin{center}
\includegraphics[width=8.5cm]{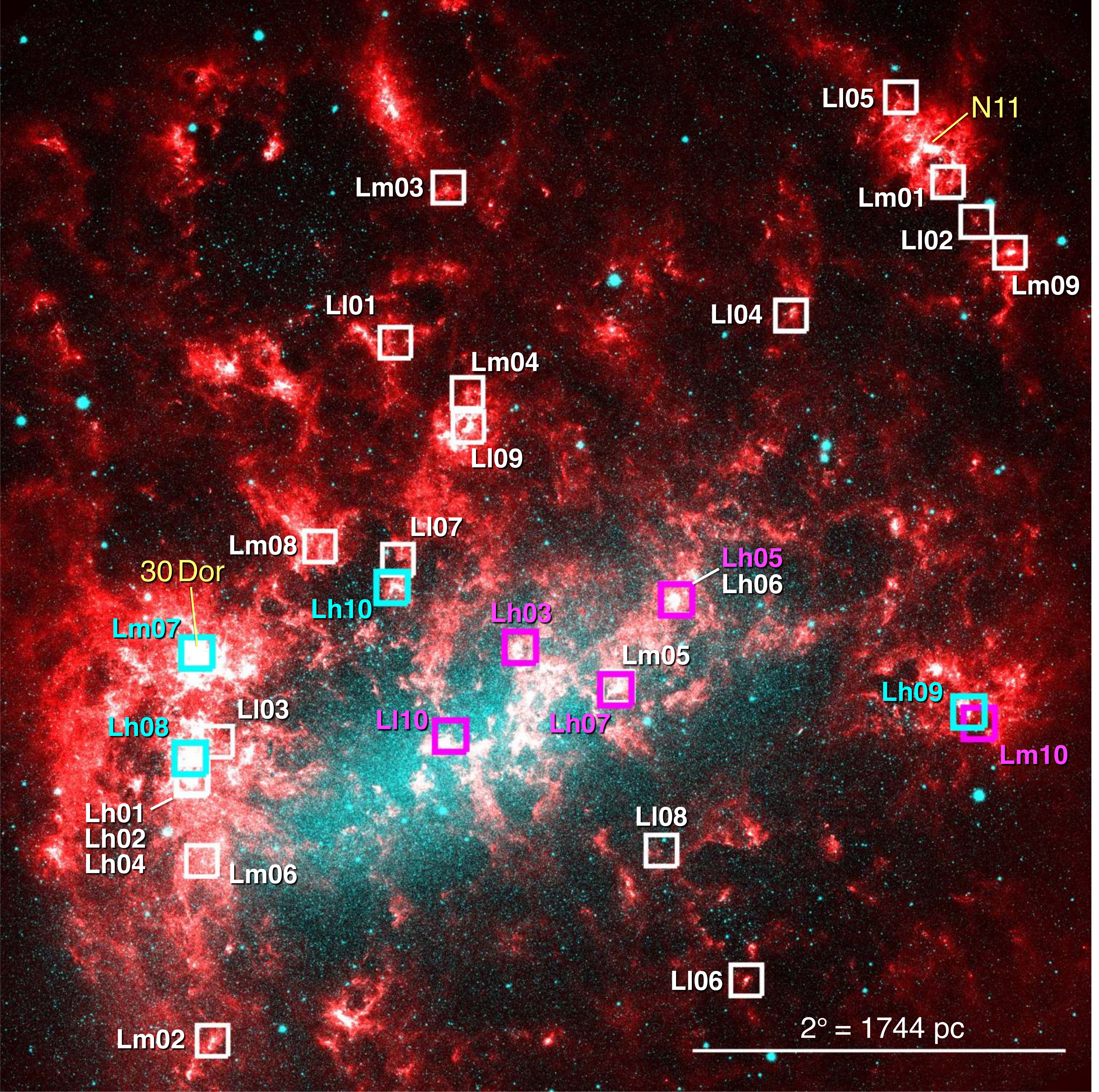}
\caption{
Spatial distribution of the observed massive protostars in the LMC. 
Hot cores with CH$_3$OH detection are indicated by magenta squares, those without CH$_3$OH detection are by cyan squares (including one hot core candidate), and the remaining sources are shown in white (see Section~\ref{sec_disc_distribution} for details). 
The background is a two-color image: light blue represents Spitzer/IRAC 3.6~$\mu$m data and red represents Herschel/PACS 160~$\mu$m data \citep{Mei06,Mei10}. 
The 3.6~$\mu$m emission mainly traces the stellar distribution, while the 160~$\mu$m emission traces the distribution of the ISM. 
The location of the 30 Dor and N11 regions are labeled in yellow. 
North is up, and east is to the left.
}
\label{distribution}
\end{center}
\end{figure}

The spatial distribution of the selected targets across the LMC is shown in Figure~\ref{distribution}. 
The targets are distributed widely across the LMC. 
Detailed information on the selected sources is summarized in Table~\ref{sourcelist}. 

Some of the infrared sources were resolved into multiple continuum sources in the present ALMA observations. 
In such cases, the luminosity was divided among the resolved components in proportion to their 850~$\mu$m continuum fluxes obtained. 
These resolved sources are labeled with suffixes, such as Lh05-1 and Lh05-2. 
In total, 36 continuum sources were identified in this survey.

\begin{deluxetable*}{lccccccl}
\tablecaption{Source List \label{sourcelist}}
\tabletypesize{\footnotesize} 
\tablewidth{0pt}
\tablehead{
\colhead{Source} & \colhead{R.A.} & \colhead{Decl.} & \colhead{$V_{\rm sys}$\tablenotemark{a}} & \colhead{Luminosity\tablenotemark{b}} & \colhead{HC\tablenotemark{c}} & \colhead{IR\tablenotemark{d}} & \colhead{Other Name/Reference}  \\
\colhead{} & \colhead{(ICRS)} & \colhead{(ICRS)} & \colhead{(km s$^{-1}$)} & \colhead{($10^4\,L_\odot$)} & \colhead{} & \colhead{} & \colhead{} 
}
\startdata
Lh10 & 05:26:46.61 & -68:48:47.03 & 250.8 & 23.0 & $\checkmark$ & S &  ST11 (1)   \\
Lh09 & 04:51:53.31 & -69:23:28.54 & 235.8 & 20.8 & $\checkmark$ & SE &  H72.96-69.39 (2)  \\
Lh08 & 05:39:38.93 & -69:39:10.80 & 238.1 & 17.3 & $\checkmark$ & PE &  N160A-mm (3)  \\
Lh07 & 05:13:25.16 & -69:22:45.44 & 239.6 & 15.5 & $\checkmark$ & P &  N113~A1 (4)  \\
Lh06 & 05:09:50.54 & -68:53:5.32 & 237.7 & 15.3 &  & P &  N105-1A  (5) \\
Lh05-1 & 05:09:51.99 & -68:53:28.40 & 242.9 & 7.2 & $\checkmark$ & PE & N105-2A (5) \\
Lh05-2 & 05:09:52.57 & -68:53:28.15 & 243.6 & 7.5 & $\checkmark$ & PE & N105-2B (5) \\
Lh05-3\tablenotemark{$\dag$} & 05:09:52.22 & -68:53:22.55 & 243.0 & -- & & -- & N105-2C (5) \\
Lh04 & 05:39:41.97 & -69:46:12.03 & 237.5 & 12.7 &  & PE &  N159W-S YSO-S (6)  \\
Lh03 & 05:19:12.31 & -69:09:7.32 & 265.6 & 11.6 & $\checkmark$ & S &  ST16 (7)  \\
Lh02 & 05:39:37.59 & -69:45:26.18 & 237.7 & 11.3 &  & PE & N159W-N~YSO-N (6)  \\
Lh01-1 & 05:39:36.71 & -69:45:38.00 & 237.2 & 6.0 &  & PE & N159W-N~MMS2d  (8) \\
Lh01-2 & 05:39:37.09 & -69:45:36.66 & 235.8 & 4.3 &  & PE & N159W-N~MMS2a  (8)  \\
Lm10 & 04:51:11.35 & -69:26:46.57 & 241.0 & 8.3 & $\checkmark$ & S &   \\
Lm09 & 04:52:9.26 & -66:55:21.69 & 272.6 & 8.0 &  & PE &  \\
Lm08 & 05:30:54.29 & -68:34:28.20 & 256.8 & 7.8 &  & S & ST5 (9) \\
Lm07 & 05:38:39.71 & -69:05:38.13 & 244.9 & 6.2 & ?  & PE &  S5-A (10) \\
Lm06 & 05:39:31.25 & -70:12:16.94 & 223.5 & 4.6 &  & P & ST1 (9) \\
Lm05-1 & 05:13:18.90 & -69:21:51.73 & 235.4 & 2.5 &  & PE &  \\
Lm05-2 & 05:13:18.93 & -69:21:47.44 & 237.5 & 1.7 &  & PE &  \\
Lm04 & 05:22:2.69 & -67:47:2.90 & 282.4 & 3.7 &  & P &  \\
Lm03 & 05:22:49.81 & -66:40:55.78 & 296.4 & 3.5 &  & PE &  \\
Lm02 & 05:39:55.70 & -71:10:1.27 & 230.8 & 3.5 &  & PE &  \\
Lm01 & 04:55:50.65 & -66:34:34.70 & 275.9 & 3.2 &  & S & ST13  (11) \\
Ll10 & 05:23:33.44 & -69:37:12.27 & 250.1 & 2.8 & $\checkmark$ & S & 14A (12)  \\
Ll09 & 05:22:1.92 & -67:57:57.86 & 283.6 & 2.8 &  & PE &  \\
Ll08 & 05:10:24.17 & -70:14:6.53 & 236.3 & 2.8 &  & S &  ST17 (13)  \\
Ll07-1 & 05:26:21.53 & -68:39:59.29 & 267.2 & 0.9 &  & P &  \\
Ll07-2 & 05:26:22.02 & -68:39:57.16 & 266.2 & 1.0 &  & P &  \\
Ll06 & 05:04:38.21 & -70:54:44.17 & 231.6 & 1.9 &  & PE &  \\
Ll05 & 04:58:42.49 & -66:08:35.55 & 275.3 & 1.5 &  & S &  \\
Ll04-1 & 05:03:56.15 & -67:20:39.50 & 270.9 & 1.4 &  & PE &  \\
Ll04-2\tablenotemark{$\dag$} & 05:03:55.86 & -67:20:44.97 & 267.9 & -- &  & -- &  Field 12 (12)  \\
Ll03 & 05:37:54.86 & -69:34:35.71 & 265.2 & 1.1 &  & S &  \\
Ll02 & 04:54:8.18 & -66:46:30.48 & 274.5 & 1.1 &  & PE &  \\
Ll01 & 05:26:1.24 & -67:30:11.98 & 287.1 & 1.1 &  & SE &  \\
\enddata
\textbf{Notes.} 
\tablenotetext{a}{Source velocities measured using the HCO$^+$(4-3) line for a 1.0 pc diameter region around each continuum peak. }
\tablenotetext{b}{If multiple submillimeter sources are associated with a single infrared source, their luminosities were estimated based on the ratio of their continuum fluxes. }
\tablenotetext{c}{``$\checkmark$'' and ``?'' indicate that the source is associated with a hot core or a hot core candidate, respectively. }
\tablenotetext{d}{Infrared spectral group classified in \citet{Sea09}. See Section~\ref{sec_disc_IRtype} for details. }
\tablenotetext{$\dag$}{These sources are spatially well separated from the infrared source, and their luminosities and infrared spectral types are unknown. }
\textbf{References: }
(1) \citet{ST16b}; 
(2) \citet{Nay19}; 
(3) \citet{Bro25}; 
(4) \citet{Sew18}; 
(5) \citet{Sew22a}; 
(6) \citet{Fuk15}; 
(7) \citet{ST20}; 
(8) \citet{Tok22a}; 
(9) \citet{ST}; 
(10) \citet{vanGel20}; 
(11) \citet{thesis}; 
(12) \citet{Gol24}; 
(13) \citet{ST16}
\end{deluxetable*}

\subsection{Observations} \label{sec_obs} 
The MAGOS observations were conducted with ALMA in October 2019 during Cycle 7. 
The sky frequency of 343.90--344.83, 344.89--347.70, 356.20--359.68 GHz, were covered by five spectral windows with a velocity resolution of 0.41--0.43 km s$^{-1}$. 
The total number of 12 m antennas used was 41--46, where the minimum--maximum baseline lengths were 14--783 m. 
The typical precipitable water vapor was 0.5--1.1 mm during the observations. 
The on-source integration time per target was 7.6 minutes and the total observation time for all LMC targets was 6.7 hours.

\subsection{Data reduction} \label{sec_red} 
Raw data was processed with the \textit{Common Astronomy Software Applications} (CASA) package \citep[version 6.2.1,][]{casa2022}. 
We performed the imaging process with the pipeline-calibrated visibility data provided by the observatory. 
The CASA task \texttt{tclean} was used for imaging and the masking was done using the auto-multithresh algorithm. 
The original synthesized beam size is 0$\farcs$30--0$\farcs$31 $\times$ 0$\farcs$37--0$\farcs$39 with the Briggs weighting and a robustness parameter of 0.5. 
For the spectral analysis in this work, all line images were restored to have a common circular beam size of 0$\farcs$40, which corresponds to 0.1~pc at the distance of the LMC. 
Continuum images have a typical beam size of 0$\farcs$30 $\times$ 0$\farcs$37. 
The maximum recoverable scale is about 4$\arcsec$. 
The images were corrected for the primary beam pattern using the \texttt{impbcor} task in CASA. 
The aggregate continuum image was constructed by selecting line-free channels. 
A typical rms noise level of continuum images is 0.1-0.2 mJy/beam. 
The continuum emission was subtracted from the spectral data using the \texttt{uvcontsub} task in CASA.

\section{Results} \label{sec_res} 
\subsection{Continuum and SO emission maps} \label{sec_img} 
To characterize the spatial distributions of submillimeter emission toward the observed protostellar objects, we first examine their continuum and SO(8$_8$–7$_7$) emission maps, which are shown in Appendix~\ref{sec_app_image_all}. 
We identified 36 continuum point sources in total, whose detailed information is summarized in Table \ref{sourcelist}. 
CO(3–2) and HCO$^{+}$(4–3) lines are detected in all sources, and HCO$^+$ images are presented in \citet{Tok23}.

\subsection{Spectra} \label{sec_spc} 
To further investigate the physical and chemical properties of the detected sources, we extracted spectra and continuum fluxes from the 0$\farcs$41 (0.1~pc) diameter circular region centered at each continuum peak. 
For Lh09, the extraction was performed for an elliptical region covering the continuum center (0$\farcs$62 $\times$ 0$\farcs$41, position angle = 65$\arcdeg$), since its high-excitation SO$_2$ emission is elongated. 
The rms noise levels of the spectral data are typically 0.24--0.30 K for sources located at the center of the observed field. 
We identified spectral lines with the aid of the Cologne Database for Molecular Spectroscopy\footnote{https://www.astro.uni-koeln.de/cdms} \citep[CDMS,][]{Mul01,Mul05} and the molecular database of the Jet Propulsion Laboratory\footnote{http://spec.jpl.nasa.gov} \citep[JPL,][]{Pic98}.

\subsection{Hot core identification} \label{sec_hcid} 
Hot cores are identified based on the presence of compact high-temperature  gas associated with continuum sources. 
In this work, gas temperatures are estimated by excitation analyses of SO$_2$ and/or CH$_3$OH lines (see Section~\ref{sec_ana}), and sources with rotational temperatures exceeding 100~K are classified as hot cores. 
Nine hot cores (Lh10, Lh09, Lh08, Lh07, Lh05-1, Lh05-2, Lh03, Lm10, and Ll10) and one hot-core candidate (Lm07) are identified in this survey (see Section~\ref{sec_disc_indiv} for their details). 

The spectra of the identified hot cores are shown in Appendix~\ref{sec_app_spec}, and the detected molecules are summarized in Table~\ref{tab_MolSum}. 
For the detected lines, we estimate the peak brightness temperatures, full widths at half maximums (FWHM), central velocities, and integrated intensities by fitting Gaussian profiles. 
The measured line intensities for all continuum sources are summarized in Appendix~\ref{sec_app_spec} for the selected transitions. 

\begin{figure*}[tp!]
\begin{center}
\includegraphics[width=17.5cm]{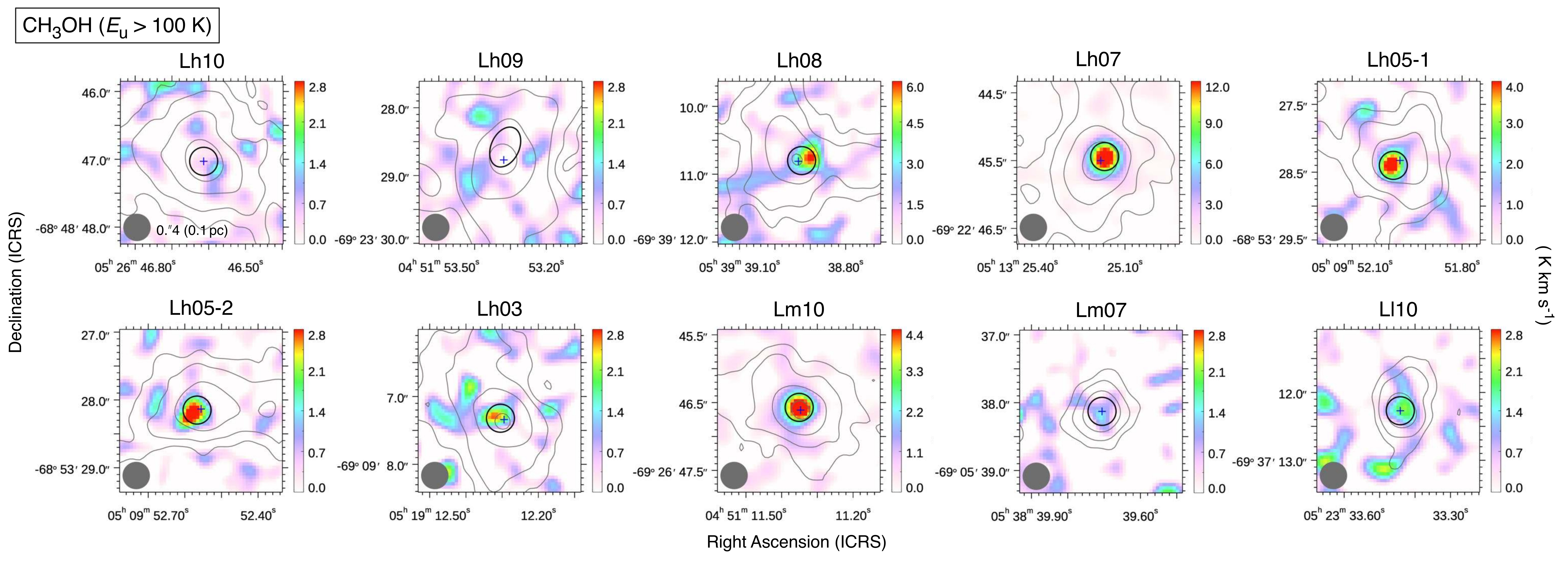}
\caption{
Integrated intensity distributions of the CH$_3$OH emission. 
High-excitation ($E_{u}$ $>$100 K) lines are stacked to reduce the noise level. 
Gray contours represent the 850~$\mu$m continuum distribution and the contour levels are 5$\sigma$, 10$\sigma$, 25$\sigma$, 100$\sigma$, 500$\sigma$ (only for Lh09) of the rms noise (0.1-0.2 mJy/beam). 
The spectra discussed in the text are extracted from the region indicated by the black open circle/ellipse. 
The blue cross represents the peak positions of continuum emission. 
The synthesized beam size of 0$\farcs$40 (0.1~pc) is shown by the gray filled circle in each panel. 
}
\label{img1}
\end{center}

\begin{center}
\includegraphics[width=17.5cm]{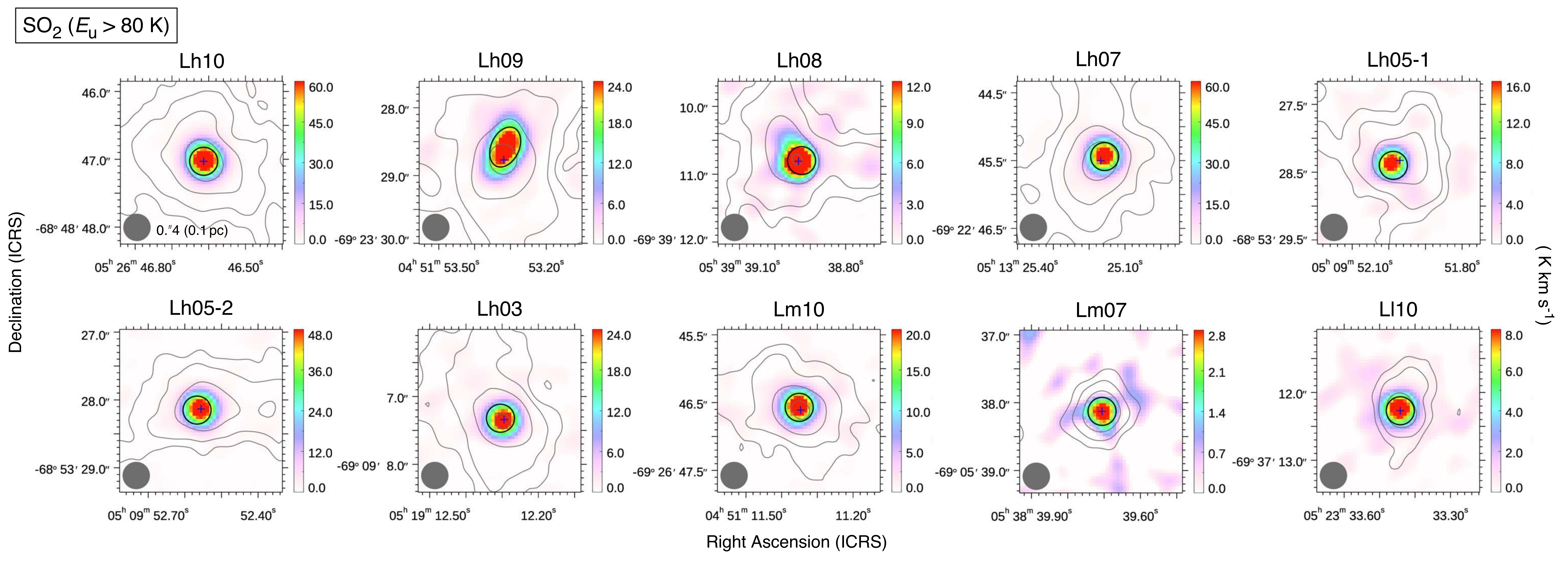}
\caption{
The same as in Figure \ref{img1}, but for the high-excitation ($E_{u}$ $>$80 K) SO$_2$ emission. 
}
\label{img2}
\end{center}

\begin{center}
\includegraphics[width=17.5cm]{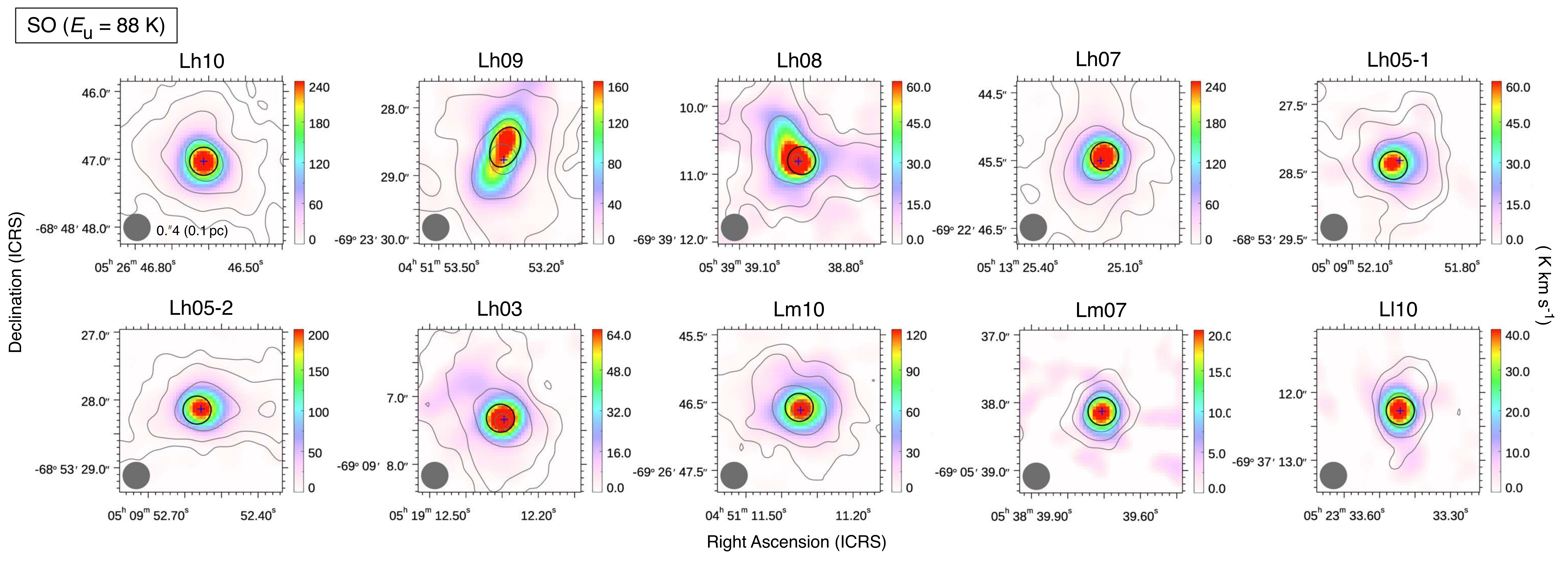}
\caption{
The same as in Figure \ref{img1}, but for the SO (8$_{8}$--7$_{7}$) line. 
}
\label{img3}
\end{center}
\end{figure*}

Figures \ref{img1}--\ref{img3} show integrated intensity distributions of high-excitation CH$_3$OH, SO$_2$, and SO lines for the hot cores. 
The images are constructed by integrating spectral data in the velocity range where the emission is detected. 
For CH$_3$OH and SO$_2$, high-excitation lines are stacked to reduce the noise level. 
The intensity peaks of these molecular emission coincide with those of the continuum emission as shown in the figures. 
In all hot cores, the spatial offsets between the peaks of the high-temperature molecular gas emission and the continuum sources are smaller than half the beam size, suggesting that the present observations are tracing hot-core regions in the vicinity of embedded protostars.

\begin{deluxetable*}{ l c c c c c c c c c c c c c c c c}
\tablecaption{Summary of detected species \label{tab_MolSum}}
\tabletypesize{\footnotesize} 
\tablehead{
\colhead{Source}  & \colhead{\ce{CO}}  &  \colhead{\ce{HCO+}} & \colhead{\ce{H^{13}CO+}} &  \colhead{\ce{HC^{15}N}}   &  \colhead{\ce{SiO}}  &  \colhead{\ce{SO}}  & \colhead{\ce{SO2}}  &  \colhead{\ce{^{34}SO2}}  &  \colhead{\ce{^{33}SO2}}  &  \colhead{\ce{SO+}}   &  \colhead{\ce{NS}} &  
\colhead{\ce{CH3OH}}  &  \colhead{\ce{HCOOH}}  &  \colhead{\ce{CH3OCH3}}   &  \colhead{\ce{HC3N}}  & \colhead{RRL\tablenotemark{a}}  
}
\startdata 
Lh10 & $\checkmark$ & $\checkmark$ & $\checkmark$ & ? & $\checkmark$ & $\checkmark$ & $\checkmark$ & $\checkmark$ & $\checkmark$ & $\checkmark$ & -- & -- & -- & -- & -- & $\checkmark$ \\
Lh09 & $\checkmark$ & $\checkmark$ & $\checkmark$ & -- & $\checkmark$ & $\checkmark$ & $\checkmark$ & $\checkmark$ & $\checkmark$ & $\checkmark$ & -- & -- & -- & -- & -- & $\checkmark$ \\
Lh08 & $\checkmark$ & $\checkmark$ & $\checkmark$ & $\checkmark$ & $\checkmark$ & $\checkmark$ & $\checkmark$ & $\checkmark$ & -- & -- & -- & ? & -- & -- & -- & ? \\
Lh07\tablenotemark{$\dag$} & $\checkmark$ & $\checkmark$ & $\checkmark$ & $\checkmark$ & $\checkmark$ & $\checkmark$ & $\checkmark$ & $\checkmark$ & $\checkmark$ & $\checkmark$ & $\checkmark$ & $\checkmark$ & $\checkmark$ & $\checkmark$ & $\checkmark$ & -- \\
Lh05-1 & $\checkmark$ & $\checkmark$ & $\checkmark$ & $\checkmark$ & $\checkmark$ & $\checkmark$ & $\checkmark$ & $\checkmark$ & $\checkmark$ & -- & -- & $\checkmark$ & $\checkmark$ & -- & -- & -- \\
Lh05-2 & $\checkmark$ & $\checkmark$ & $\checkmark$ & $\checkmark$ & $\checkmark$ & $\checkmark$ & $\checkmark$ & $\checkmark$ & -- & $\checkmark$ & -- & $\checkmark$ & -- & -- & ? & -- \\
Lh03 & $\checkmark$ & $\checkmark$ & $\checkmark$ & NA & $\checkmark$ & $\checkmark$ & $\checkmark$ & $\checkmark$ & $\checkmark$ & $\checkmark$ & -- & $\checkmark$ & -- & -- & -- & -- \\
Lm10 & $\checkmark$ & $\checkmark$ & $\checkmark$ & $\checkmark$ & $\checkmark$ & $\checkmark$ & $\checkmark$ & $\checkmark$ & -- & $\checkmark$ & $\checkmark$ & $\checkmark$ & $\checkmark$ & $\checkmark$ & -- & -- \\
Lm07 & $\checkmark$ & $\checkmark$ & $\checkmark$ & $\checkmark$ & --                  & $\checkmark$ & $\checkmark$ & -- & -- & -- & -- & --                     & -- & -- & -- & -- \\
Ll10 & $\checkmark$ & $\checkmark$ & $\checkmark$ & $\checkmark$ & $\checkmark$ & $\checkmark$ & $\checkmark$ & -- & -- & -- & -- & $\checkmark$ & -- & -- & -- & -- \\
\enddata
\tablecomments{``$\checkmark$": detection, ``?'': tentative detection, ``--'': non-detection, "NA": data not available because the line falls outside the frequency coverage. }
\tablenotetext{a}{Radio Recombination Line (RRL), H37$\gamma$. }
\tablenotetext{$\dag$}{In Lh07, \ce{^{13}CH3OH}, \ce{HCOOCH3}, and \ce{C2H5OH} are also detected, and \ce{H2CCO} is tentatively detected. 
}
\end{deluxetable*}

\begin{figure*}[tbp!]
\begin{center}
\includegraphics[width=12.5cm]{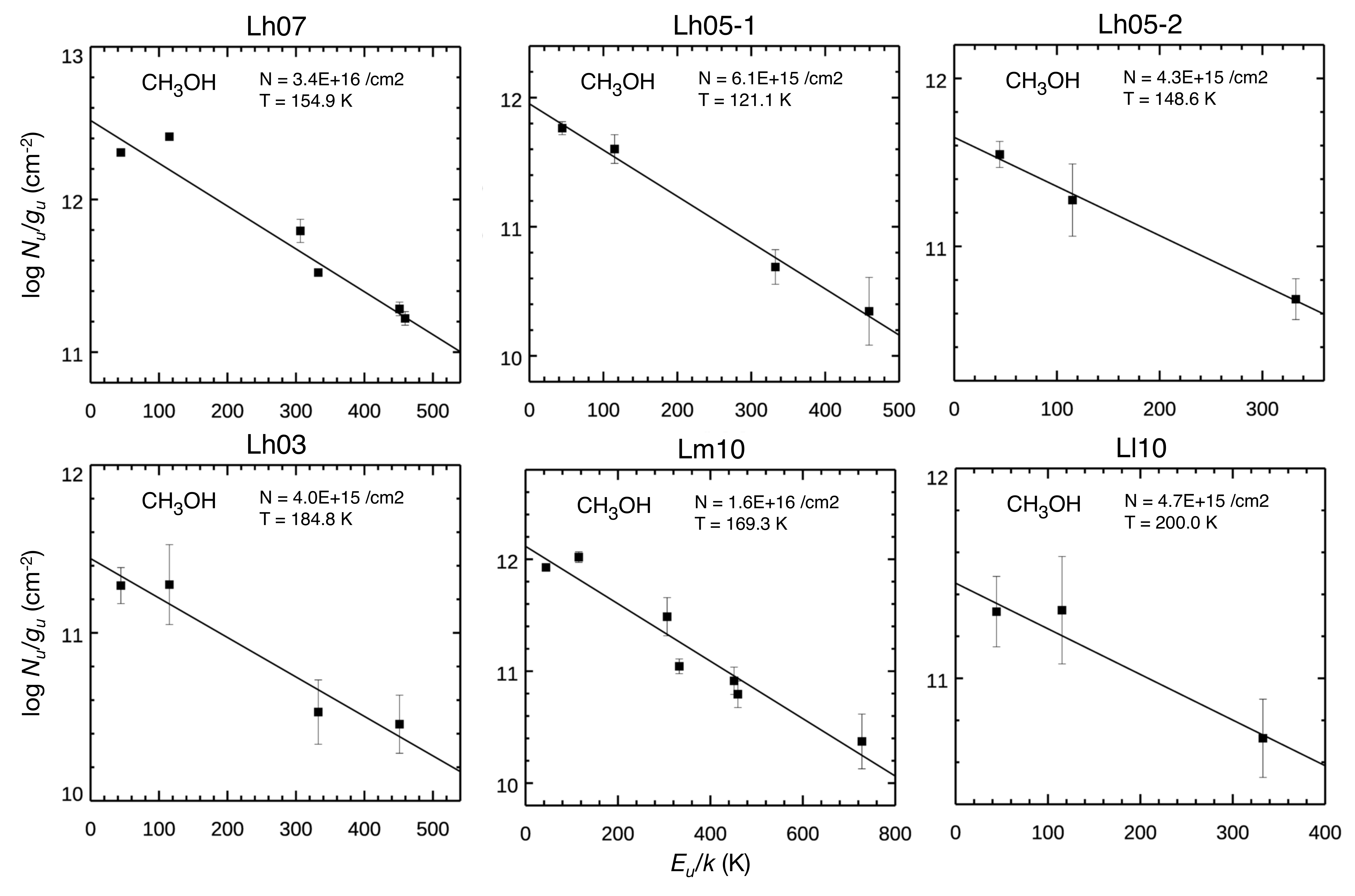}
\caption{
Results of rotation diagram analyses for CH$_3$OH. 
The solid lines represent the fitted straight line.  
Derived column densities and rotation temperatures are shown in each panel. 
}
\label{rd1}
\end{center}
\end{figure*}

\begin{figure*}[tbp!]
\begin{center}
\includegraphics[width=17.5cm]{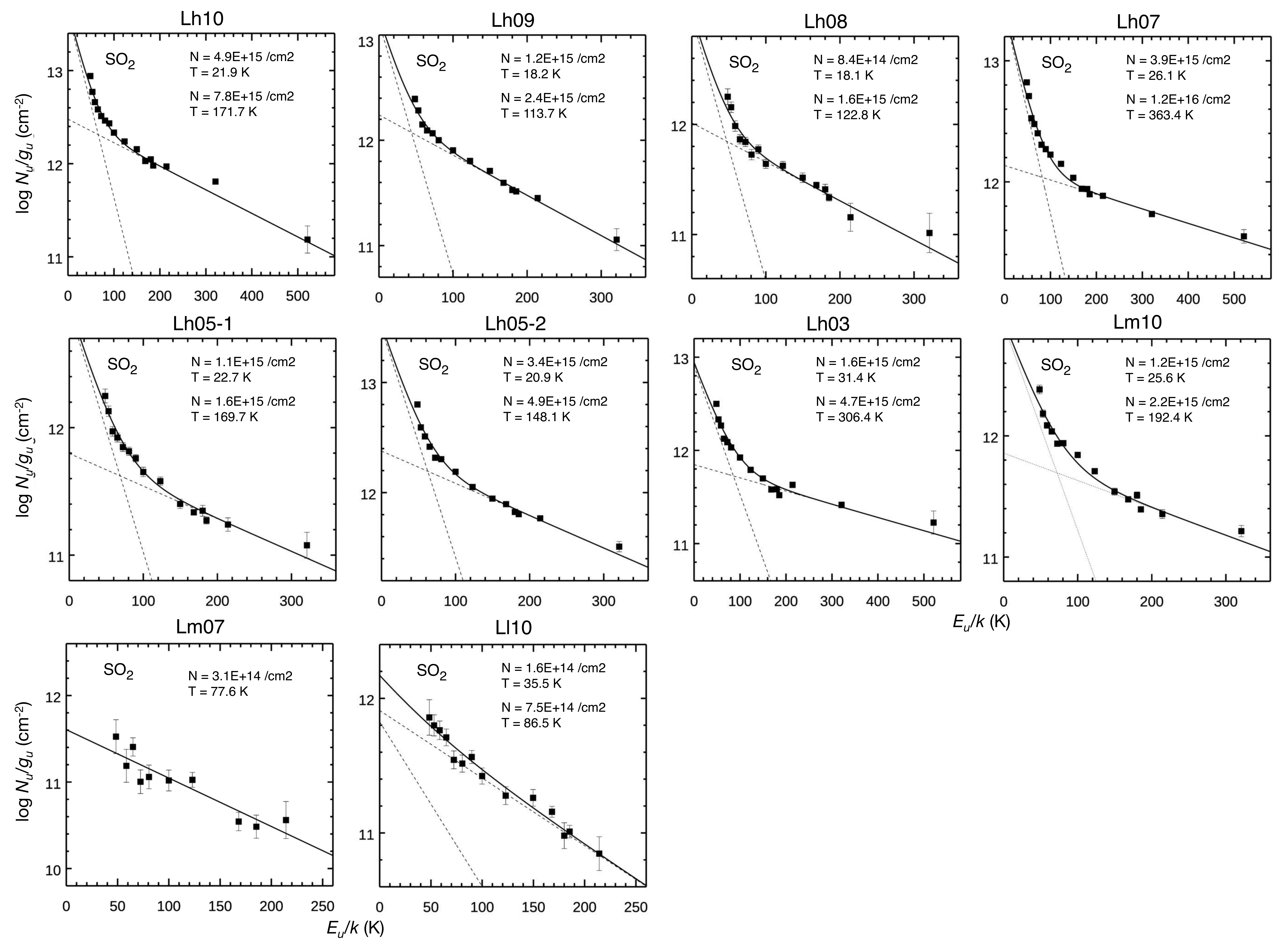}
\caption{
Results of rotation diagram analyses for SO$_2$. 
A straight-line fit is separated to low- and high-temperature components, which are shown by the dotted lines (see Section \ref{sec_rd} for details). 
The solid lines represent the sum of the two temperature components. 
Derived column densities and rotation temperatures are shown in each panel.  
}
\label{rd2}
\end{center}
\end{figure*}

\begin{figure*}[tbp!]
\begin{center}
\includegraphics[width=17.5cm]{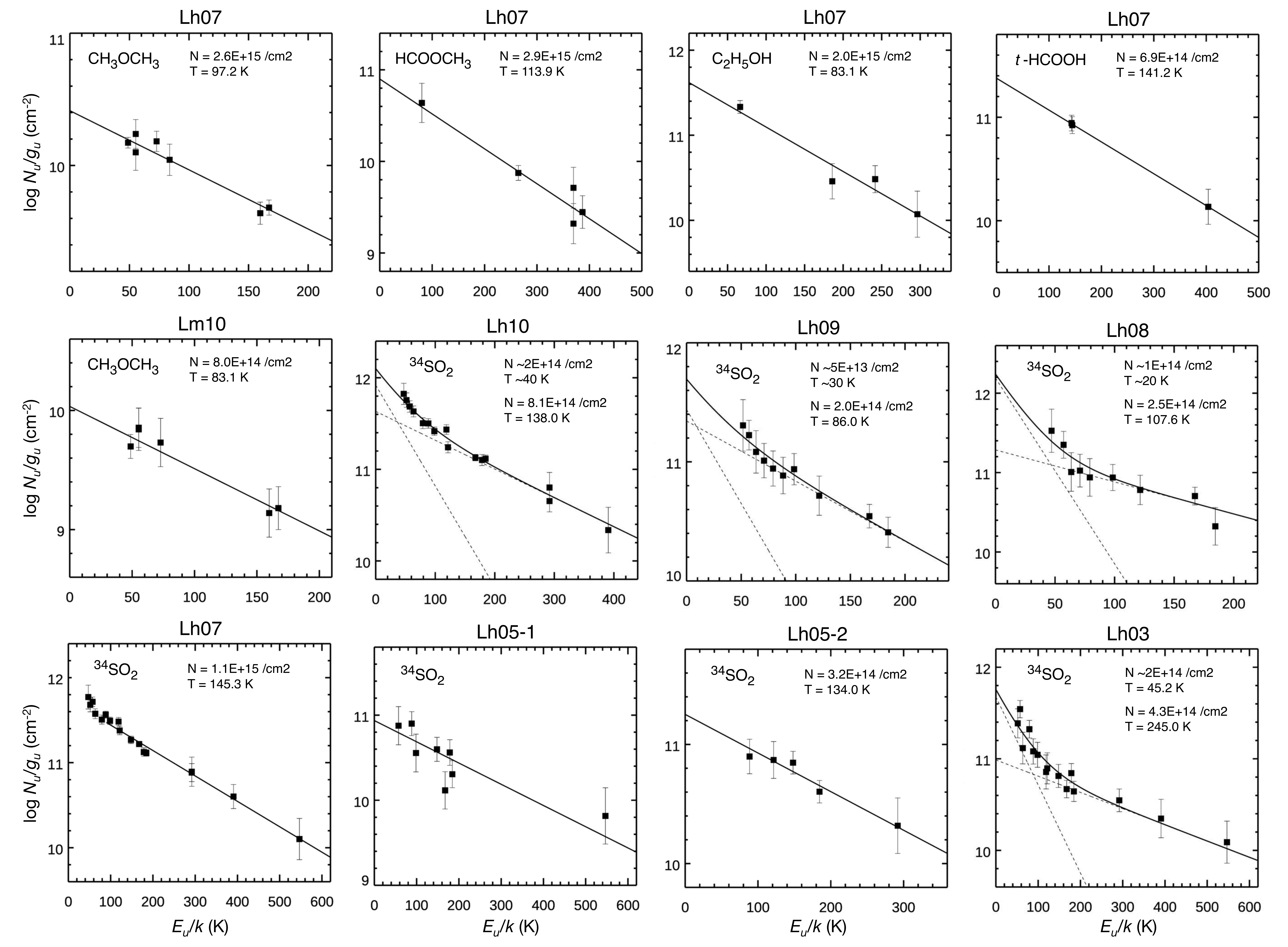}
\caption{
Results of rotation diagram analyses for other molecular species. 
The solid lines represent the fitted straight line. 
For $^{34}$SO$_2$ in Lh03, Lh08, Lh09, Lh10, a straight-line fit is separated to low- and high-temperature components, which are shown by the dotted lines. 
Derived column densities and rotation temperatures are shown in each panel. 
}
\label{rd3}
\end{center}
\end{figure*}

\section{Data Analysis} \label{sec_ana} 
\subsection{Rotation diagram analysis of \ce{CH3OH} and \ce{SO2}} \label{sec_rd}
Multiple high-excitation lines are detected for hot-core sources, and these lines are used for the excitation analysis. 
We use the rotation diagram method with the following standard formulae to estimate column densities and temperatures \citep[e.g., ][]{Sut95}: 
\begin{equation}
\log \left(\frac{ N_{u} }{ g_{u} } \right) = - \left(\frac {\log e}{T_{\mathrm{rot}}} \right) \left(\frac{E_{u}}{k} \right) + \log \left(\frac{N}{Q(T_{\mathrm{rot}})} \right),  \label{Eq_rd1}
\end{equation}
where 
\begin{equation}
\frac{ N_{u} }{ g_{u} } = \frac{ 3 k \int T_{\mathrm{b}} dV }{ 8 \pi^{3} \nu S \mu^{2} }, \label{Eq_rd2} \\ 
\end{equation}
and $N_{u}$ is a column density of molecules in the upper energy level, $g_{u}$ is the degeneracy of the upper level, $k$ is the Boltzmann constant, $\int T_{\mathrm{b}} dV$ is the integrated intensity estimated from the observations, $\nu$ is the transition frequency, $S$ is the line strength, $\mu$ is the dipole moment, $T_{\mathrm{rot}}$ is the rotational temperature, $E_{u}$ is the upper state energy, $N$ is the total column density, and $Q(T_{\mathrm{rot}})$ is the partition function at $T_{\mathrm{rot}}$. 
Eq.~\ref{Eq_rd1} is derived from Eq.~\ref{Eq3} assuming $T_{\mathrm{rot}}$ $>>$ $T_{\mathrm{bg}}$. 
We assume an optically thin condition and the local thermodynamic equilibrium (LTE). 
All the spectroscopic parameters required in the analysis are extracted from the CDMS or JPL database. 

The results of the analyses are shown in Figure \ref{rd1} for CH$_3$OH, Figure \ref{rd2} for SO$_2$, and Figure \ref{rd3} for other molecular species. 
For SO$_2$ and $^{34}$SO$_2$, a straight-line fit is separated into low- and high-temperature regimes ($E_{u}$ $\lesssim$80 K and $E_{u}$ $\gtrsim$90 K), because different temperature components are seen in the diagram. 
The derived rotational temperatures of CH$_3$OH or SO$_2$ exceed 100 K, suggesting the presence of high-temperature gas associated with each protostar.

\subsection{Column densities of other molecules} \label{sec_n}
Column densities of other molecules are derived with the assumed temperatures. 
The following equation is used based on the standard treatment of optically thin emission in the LTE \citep[e.g.,][]{Yam17}: 
\begin{equation}
N = \frac{3 k Q(T_{\mathrm{rot}}) \int T_{\mathrm{b}} dV}{8 \pi^{3} \nu S \mu^{2}} \Bigg{\{} 1 - \frac{\exp(h \nu / kT_{\mathrm{rot}})-1}{\exp(h \nu / kT_{\mathrm{bg}})-1} \Bigg{\}}^{-1} \exp\Bigg{(}\frac{E_u}{kT_{\mathrm{rot}}}\Bigg{)}  \label{Eq3}, 
\end{equation}
where 
$N$ is the total column density, 
$k$ is the Boltzmann constant, 
$T_{\mathrm{rot}}$ is the rotational temperature, 
$T_{\mathrm{bg}}$ is the cosmic microwave background temperature (2.73 K) , 
$Q(T_{\mathrm{rot}})$ is the partition function at $T_{\mathrm{rot}}$, 
$\int T_{\mathrm{b}} dV$ is the integrated intensity estimated from the observations, 
$\nu$ is the transition frequency, 
$S$ is the line strength, $\mu$ is the dipole moment, 
$h$ is the Planck constant, 
and 
$E_{u}$ is the upper state energy. 
All the spectroscopic parameters required in the analysis are extracted from the CDMS database, except for \ce{SO+}, \ce{HCOOCH3}, and \ce{CH3CHO}, which are based on the JPL database. 

The assumed rotational temperatures are summarized in the table in Appendix \ref{sec_app_n}. 
$T_{\mathrm{rot}}$ for SO and \ce{SO+} are assumed to be the same as the high-temperature component of \ce{SO2}, presuming that these sulfur-bearing species trace the same gas in a hot core. 
$T_{\mathrm{rot}}$ for the other complex species (i.e., CH$_3$CHO, HC$_3$N, and H$_2$CCO in Lh07; HCOOH in Lh05-1 and Lh05-2; HCOOH, HCOOCH$_3$, C$_2$H$_5$OH, CH$_3$CHO in Lm10) are assumed to be the same as that of CH$_3$OH. 

The rotational temperature ($T_{\mathrm{rot}}$) of \ce{SiO} is assumed to be equal to that of high-temperature SO$_2$ or CH$_3$OH, or their average when both are available, assuming that SiO originates from a hot core. 
A 50$\%$ uncertainty is applied to the assumed $T_{\mathrm{rot}}$ of SiO, and this is propagated into the column density calculation. 
\ce{HCO+} exhibits extended spatial distributions \citep[see][]{Tok23}, and is therefore unlikely to trace the high-temperature gas component of hot core regions. 
Therefore, the column density of H$^{13}$CO$^+$ was calculated by varying the rotational temperature between 30 K and 60 K, assuming that it traces a relatively cold and extended gas component.

\subsection{H$_2$ column density and abundances } \label{sec_h2} 
The column density of molecular hydrogen ($N_{\mathrm{H_2}}$) is estimated from the dust continuum data based on the standard treatment of optically thin emission. 
We use the following equation: 
\begin{equation}
N_{\mathrm{H_2}} = \frac{F_{\nu} / \Omega}{2 \kappa_{\nu} B_{\nu}(T_{d}) Z \overline{M}_w m_{\mathrm{H}}} \label{Eq_h2}, 
\end{equation}
where $F_{\nu}/\Omega$ is the continuum flux density per beam solid angle as estimated from the observations, $\kappa_{\nu}$ is the mass absorption coefficient of dust grains coated by thin ice mantles taken from \citet{Oss94} and we use 1.96 cm$^2$ g$^{-1}$ at 850 $\mu$m, $T_{d}$ is the dust temperature and $B_{\nu}(T_{d})$ is the Planck function, $Z$ is the dust-to-gas mass ratio, $\overline{M}_w$ is the mean atomic mass per hydrogen \citep[1.41, according to][]{Cox00}, and $m_{\mathrm{H}}$ is the hydrogen mass. 
The continuum fluxes are measured as beam-averaged values over the same regions as the molecular lines, which cover both continuum and molecular line peaks.
We use the dust-to-gas mass ratio of 0.0027, which is derived by dividing the Galactic value (0.008) by three, based on the metallicity difference of the LMC. 
The dust temperature is assumed to be the same as the $T_{\mathrm{rot}}$ of high-temperature SO$_2$ or CH$_3$OH, or their average if both are available, assuming that the dust emission mainly arises from a hot-core region. 
The actual dust temperature in a hot core is uncertain and may differ from the assumed value. 
Here, we adopt a 30$\%$ uncertainty for $N_{\mathrm{H_2}}$, which corresponds to the variation in $N_{\mathrm{H_2}}$ that would result from a factor of 1.5 deviation in the assumed dust temperature. 

For Lh09, its continuum emission may be dominated by free-free emission even at 850 $\mu$m, as pointed out in \citet{Nay19} based on mm and cm continuum flux of this region. 
With the above method, the $N_{\mathrm{H_2}}$ of Lh09 is estimated to be 2 $\times$ 10$^{24}$ cm$^{-2}$, but this should be considered as an upper limit, given a possible contribution from free-free emission. 
The $N_{\mathrm{H_2}}$ values of high-luminosity hot-core sources (Lh07, Lh08, Lh10) range from 4 $\times$ 10$^{23}$ to 8 $\times$ 10$^{23}$ cm$^{-2}$. 
Considering this range, we here assume $N_{\mathrm{H_2}}$ = 8 $\pm$ 4 $\times$ 10$^{23}$ cm$^{-2}$ for Lh09. 

The 850 $\mu$m continuum brightnesses for all sources are summarized in the table in Appendix \ref{sec_app_spec}. 
The derived H$_2$ column densities ($N_{\mathrm{H_2}}$) for hot-core sources are listed in the table presented in Appendix \ref{sec_app_n}. 
Fractional molecular abundances relative to H$_2$ are calculated using these $N_{\mathrm{H_2}}$ values and are also included in the table. 

The $N_{\mathrm{H_2}}$ values of the present hot-core sources range from 1 to 8 $\times$ 10$^{23}$ cm$^{-2}$, corresponding to gas number densities of $n_{\mathrm{H_2}}$ = 0.5--4 $\times$ 10$^6$ cm$^{-3}$, assuming a source diameter of 0.1 pc and a uniform spherical gas distribution around each protostar. 
Since the target sources are not spatially well resolved, these gas densities should be regarded as lower limits.

\begin{figure}[tbp!]
\begin{center}
\includegraphics[width=6.8cm]{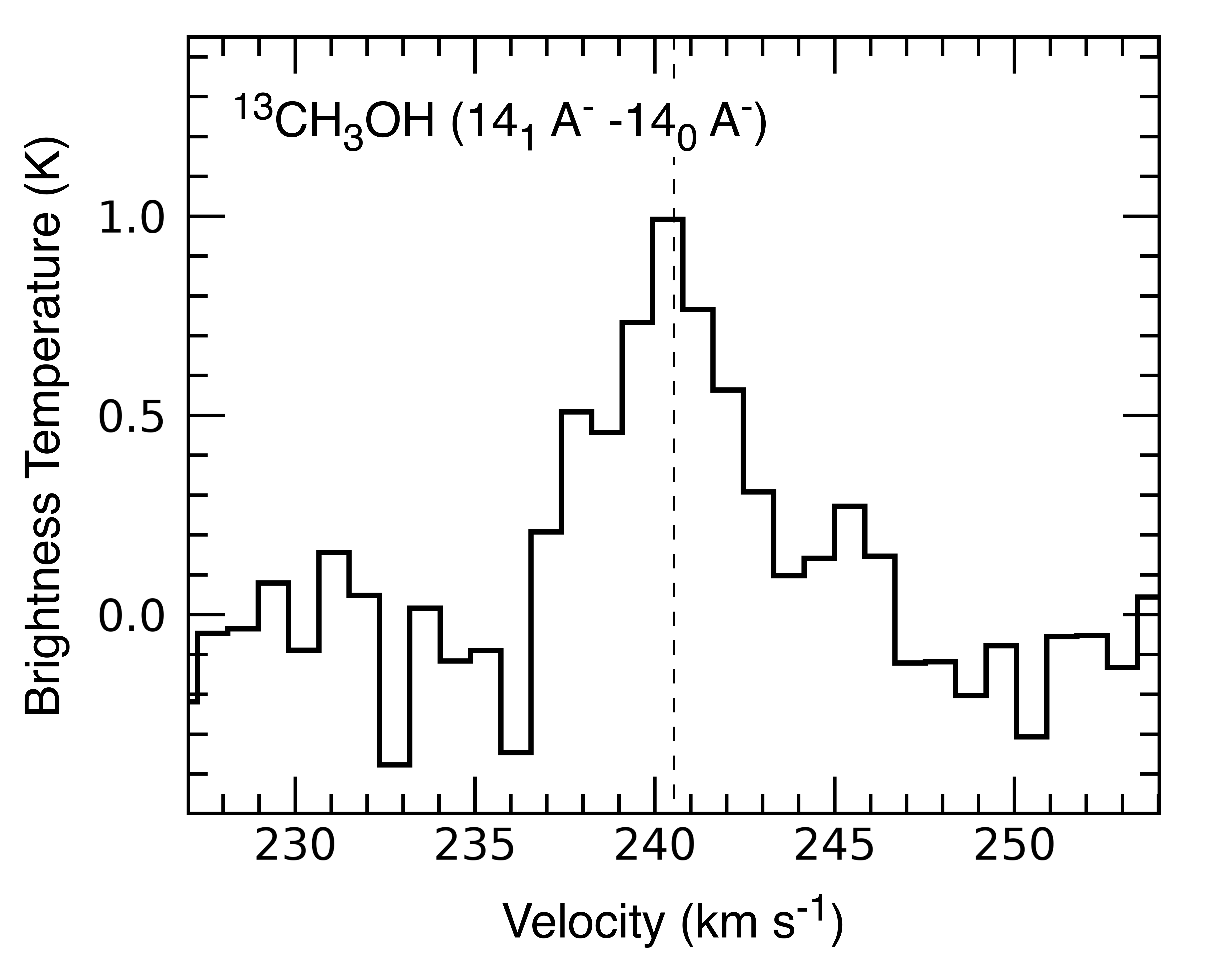}
\caption{
$^{13}$CH$_3$OH(14$_{1}$ A$^-$--14$_{0}$ A$^-$) line detected in Lh07. 
The vertical dotted line indicates the mean velocity of detected $^{12}$CH$_3$OH lines. 
The signal-to-noise ration of the line peak is 5.6~$\sigma$. 
}
\label{13CH3OH}
\end{center}
\end{figure}

\subsection{Optical thickness of CH$_3$OH and SO$_2$ lines} \label{sec_disc_tau} 
Isotopologues of CH$_3$OH and SO$_2$ are detected in several sources, and we therefore estimate the optical depth of the main species in this section. 
For $^{13}$CH$_3$OH, one clean line, $^{13}$CH$_3$OH(14$_{1}$ A$^-$--14$_{0}$ A$^-$, 347.1882830 GHz, $E_{u}$ = 254 K) is detected in Lh07, which is the most COM-rich hot core (Figure \ref{13CH3OH}). 
The $^{13}$CH$_3$OH(7$_{2}$ E--6$_{1}$ E, 357.6579540 GHz) line is also detected; however, it is blended with SO$_2$(9$_{4,6}$--9$_{3,7}$, 357.6718206 GHz) and $^{33}$SO$_2$(20$_{0,20}$--19$_{1,19}$, 357.6588389 GHz), and thus is not used in the following discussion. 
Using the rotational temperature derived from $^{12}$CH$_3$OH, the column density of $^{13}$CH$_3$OH is calculated, yielding a $^{12}$CH$_3$OH/$^{13}$CH$_3$OH column density ratio of $(3.4 \times 10^{16}) / (1.0 \times 10^{15}) = 34$. 
This value is slightly lower than, the $^{12}$C/$^{13}$C ratio of HCN previously reported for the LMC \citep[$49 \pm 5$,][]{Wan09}. 
This trend is in line with carbon isotope fractionation in dense gas, which tends to enrich \ce{CH3OH} in $^{13}$C and to deplete HCN in $^{13}$C \citep[e.g.,][]{ichimura24}. 
Even if we assume that the intrinsic $^{12}$C/$^{13}$C ratio of CH$_3$OH is 49, the optical depth of the $^{12}$CH$_3$OH emission is estimated to be $\tau \sim 0.7$ \citep[cf.][]{Gol99}.
Even in the most COM-rich source, $^{12}$CH$_3$OH emission appears to be only moderately thick. 
In other sources, the CH$_3$OH column densities are more than a factor of two lower than in Lh07, and $^{13}$CH$_3$OH lines are not detected, suggesting that the $^{12}$CH$_3$OH emission is optically thin in these sources. 

$^{34}$SO$_2$ is detected in seven hot cores, and analysis using rotational diagrams has been carried out. 
The derived $^{32}$SO$_2$/$^{34}$SO$_2$ column density ratios are 11 (Lh03), 7 (Lh05-1), 15 (Lh05-2), 11 (Lh07), 6 (Lh08), 11 (Lh09), and 9 (Lh10), considering only the high-temperature components.  These values are about 1.0 to 2.5 times lower than the $^{32}$S/$^{34}$S ratio reported for the LMC \citep[$\sim$15,][]{Wan09,ST20,Gon23}. 
Based on these results, the optical depth of the $^{32}$SO$_2$ emission is estimated to be up to $\tau \sim2.2$. 

In the following discussion sections, the column densities of CH$_3$OH and SO$_2$ are, when available, based on $^{13}$CH$_3$OH and $^{34}$SO$_2$ data, respectively. We adopt $^{12}$C/$^{13}$C = 49 and $^{32}$S/$^{34}$S = 15, as reported in the references above. 
Uncertainties of 30~$\%$ and 10~$\%$ are assumed for the carbon and sulfur isotope ratios, respectively, and these are propagated into the column density calculations.

\subsection{Notes on Individual Hot Cores} \label{sec_disc_indiv}
In this survey, nine hot cores and one hot-core candidate are identified based on the presence of compact, high-temperature gas associated with continuum sources. 
One hot core (Lm10) and one hot-core candidate (Lm07) are newly identified in this survey. 
A brief description of each source is provided below. 

\subsubsection{Lh10}
This source was reported as a hot core in \citet{ST16b} and represents the first hot core identified outside the Milky Way. 
Only H$_2$CO has been detected as an organic molecule in the previous study; COMs such as CH$_3$OH have not been detected to date, whereas sulfur-bearing species like SO and SO$_2$ are abundant. 
This study confirms these results, and the derived molecular abundances are consistent with those reported by \citet{ST16b}.

\subsubsection{Lh09}
This source is located in a region reported by \citet{Nay19} as a super star cluster candidate, where active star formation has been observed. 
Unlike the other hot cores, the high-excitation SO$_2$ emission is elongated and slightly offset to the northwest from the continuum source.
No COMs are detected in this work. 
Recent JWST MIRI observations have obtained infrared spectra around Lh09, in which hydrogen recombination lines, H$_2$ lines, and fine-structure lines have been detected \cite{Nay24c}. 

\subsubsection{Lh08}
This source shows tentative signs of CH$_3$OH emission in the multiline stacked images (see Figure~\ref{img1}), however, high-$E_u$ CH$_3$OH lines are not significantly detected in the spectra. 
On the other hand, \citet{Bro25} reported it as a hot core and detected high-temperature CH$_3$OH gas, albeit with a low abundance. 
In the present study, we conservatively provide an upper limit based on the non-detection of the CH$_3$OH(16$_1$ A$^-$–15$_2$ A$^-$) line. 
This upper limit is comparable to the value reported by \citet{Bro25}. 
High-temperature SO$_2$ gas is clearly detected in both studies.

\subsubsection{Lh07}
This source has the highest CH$_3$OH abundance among the currently known hot cores in the LMC. 
It was reported as a hot core by \citet{Sew18}, and COMs larger than CH$_3$OH, namely CH$_3$OCH$_3$ and HCOOCH$_3$ were detected. 
In addition to these species, the present study also detects HCOOH, HC$_3$N, C$_2$H$_5$OH, and tentatively H$_2$CCO. 
The CH$_3$OH and SO$_2$ column densities derived in this work are approximately three times higher than those reported by \citet{Sew22a}. 
This difference is likely due to our use of isotopologues for deriving column densities or different spectral coverage.

\subsubsection{Lh05-1,2}
These two hot cores are located in the same region. 
They are separated by only about 3$\arcsec$ (0.7~pc) and have nearly identical systemic velocities, suggesting a common origin from the same molecular cloud. 
Compared to Lh05-1, Lh05-2 exhibits broader molecular line widths; for example, the FWHM of SO$_2$(19$_{1,19}$–18$_{0,18}$) is 3.9 km s$^{-1}$ for Lh05-1 and 11.6 km s$^{-1}$ for Lh05-2. 
Although their molecular abundances are similar, these results indicate significant differences in gas kinematics within the hot cores. 
Both sources were reported as hot cores by \citet{Sew22a}, and the molecular abundances obtained in this study are consistent with theirs.

\subsubsection{Lh03}
This source was previously analyzed in detail by \citet{ST20} based on ALMA Band 6 and 7 spectra. 
That study detected CH$_3$OH and CH$_3$CN but no larger COMs, which is consistent with the results of the present survey. 
Our CH$_3$OH abundance is approximately twice as high, likely due to the limited number of CH$_3$OH lines covered in our survey. 
\citet{ST16} conducted VLT/ISAAC observations targeting the 3.53 $\mu$m CH$_3$OH ice band, but no detection was made ($N$(CH$_3$OH)/$N$(H$_2$O) $<$6~$\%$). 
Our SO$_2$ column density is consistent with that reported by \citet{ST20}.

\subsubsection{Lm10}
This hot core is newly discovered in the present survey. 
It is the second most COM-rich source after Lh07 among the present samples, with detections of organic molecules such as CH$_3$OH, HCOOH, and CH$_3$OCH$_3$. 
This represents the first detection of COMs larger than CH$_3$OH in a hot core located outside the LMC bar region. 
It is situated near Lh09, a COM-poor hot core (Fig.~\ref{distribution}). 
Recent JWST MIRI observations have obtained spectra of an infrared source near Lm10, in which the CO$_2$ ice absorption as well as emission features from PAHs, H$_2$, and fine-structure lines are detected \citep{Nay24c}.

\subsubsection{Lm07}
This source exhibits compact SO$_2$ gas at $\sim$80 K but no detection of CH$_3$OH. 
Since gas exceeding 100~K was not detected and the line intensities are weaker than those of other hot cores, we classify this source as a hot core candidate. 
It is situated in a highly active star-forming region near 30 Doradus.

\subsubsection{Ll10}
This is the least luminous hot core among those observed in this survey. 
It was reported as a hot core (source 14A) by \citet{Gol24}, and our molecular abundances are consistent with their results.

\section{Discussion} \label{sec_disc} 
\subsection{Distribution of hot cores within the LMC} \label{sec_disc_distribution} 
The spatial distribution of the observed hot cores within the LMC is shown in Figure~\ref{distribution}. 
The bar region of the LMC, characterized by its high stellar density and relatively high metallicity, is shown with a light-blue background in the figure. 
Among the present hot core samples (including one candidate), five out of ten hot cores are located within the bar region. 
In the direction of N11, the second-largest H\,\textsc{ii} region in the LMC after 30 Dor and located at the northwestern edge of the LMC, four sources (Ll05, Lm01, Ll02, Lm09) were observed in this survey, but no hot cores were detected. 

All COM-poor hot cores (Lh10, Lh09, Lh08, Lm07) are located outside the bar region. 
As pointed out by \citet{Gol24}, these regions are considered to have relatively low metallicity, which may suggest that environmental differences influence the chemical evolution of hot cores. 
However, Lm10, the second most COM-rich hot core after Lh07, is located near the COM-poor source (Lh09), and this area lies outside the bar region. 
This implies that low local metallicity may be a necessary condition for triggering COM-poor hot core chemistry; however, it is not a sufficient condition, and additional factors besides low metallicity may also play a role. 
Further discussion on the trigger of COM-poor hot-core chemistry in the LMC is presented in Section~\ref{sec_disc_molab2}.

\begin{figure*}[tbp!]
\begin{center}
\includegraphics[width=17.5cm]{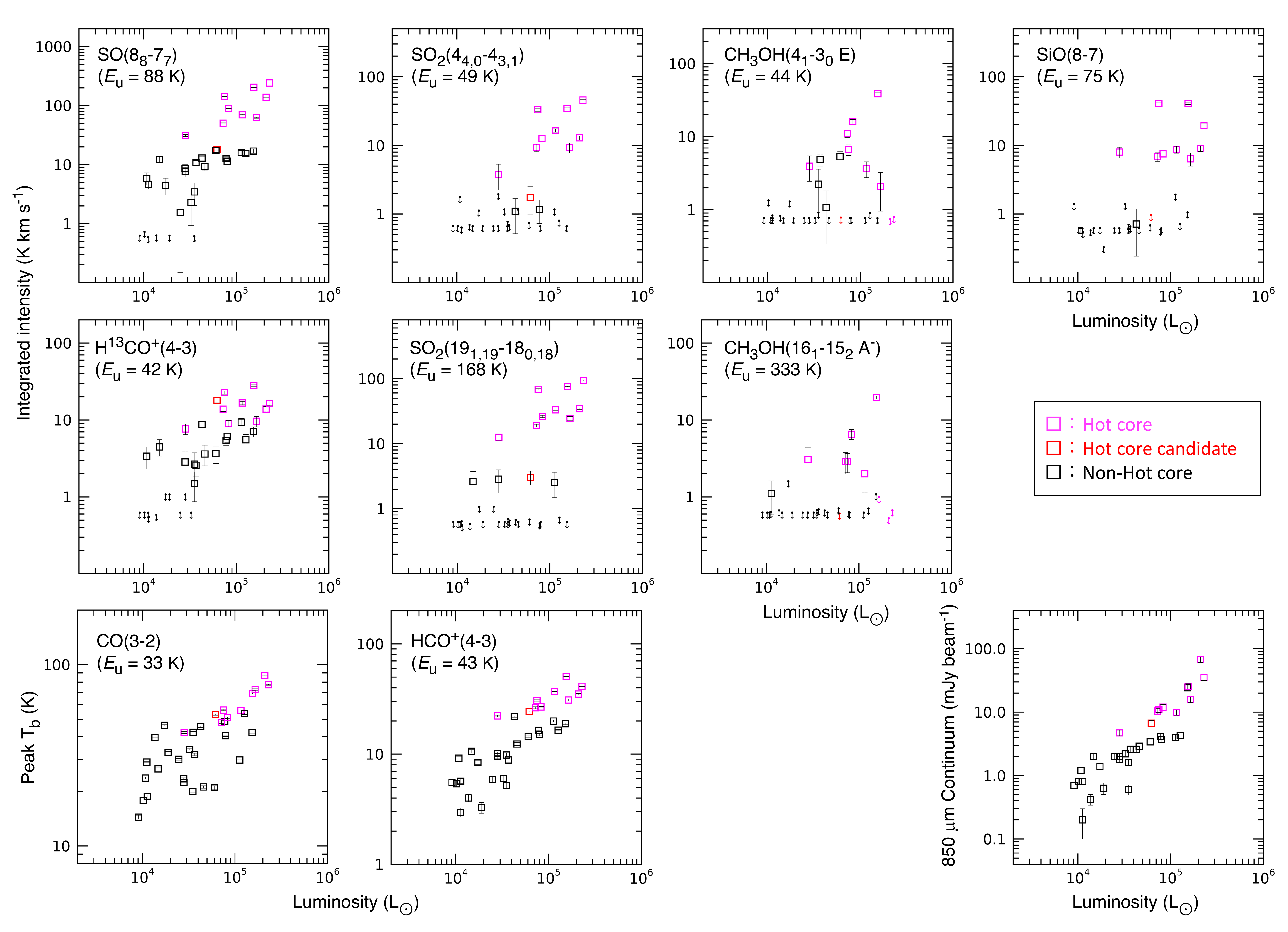}
\caption{
Line intensities vs. luminosities. 
The top two rows of panels show the integrated intensities, the bottom left panels show the peak intensities, and the bottom right panel shows the continuum brightness. 
Sources associated with a hot core and a hot core candidate are indicated in magenta and red, respectively. 
Upper limit points are indicated by dots with downward arrows. 
}
\label{vsIntensity}
\end{center}
\end{figure*}

\begin{figure*}[tbp!]
\begin{center}
\includegraphics[width=17.5cm]{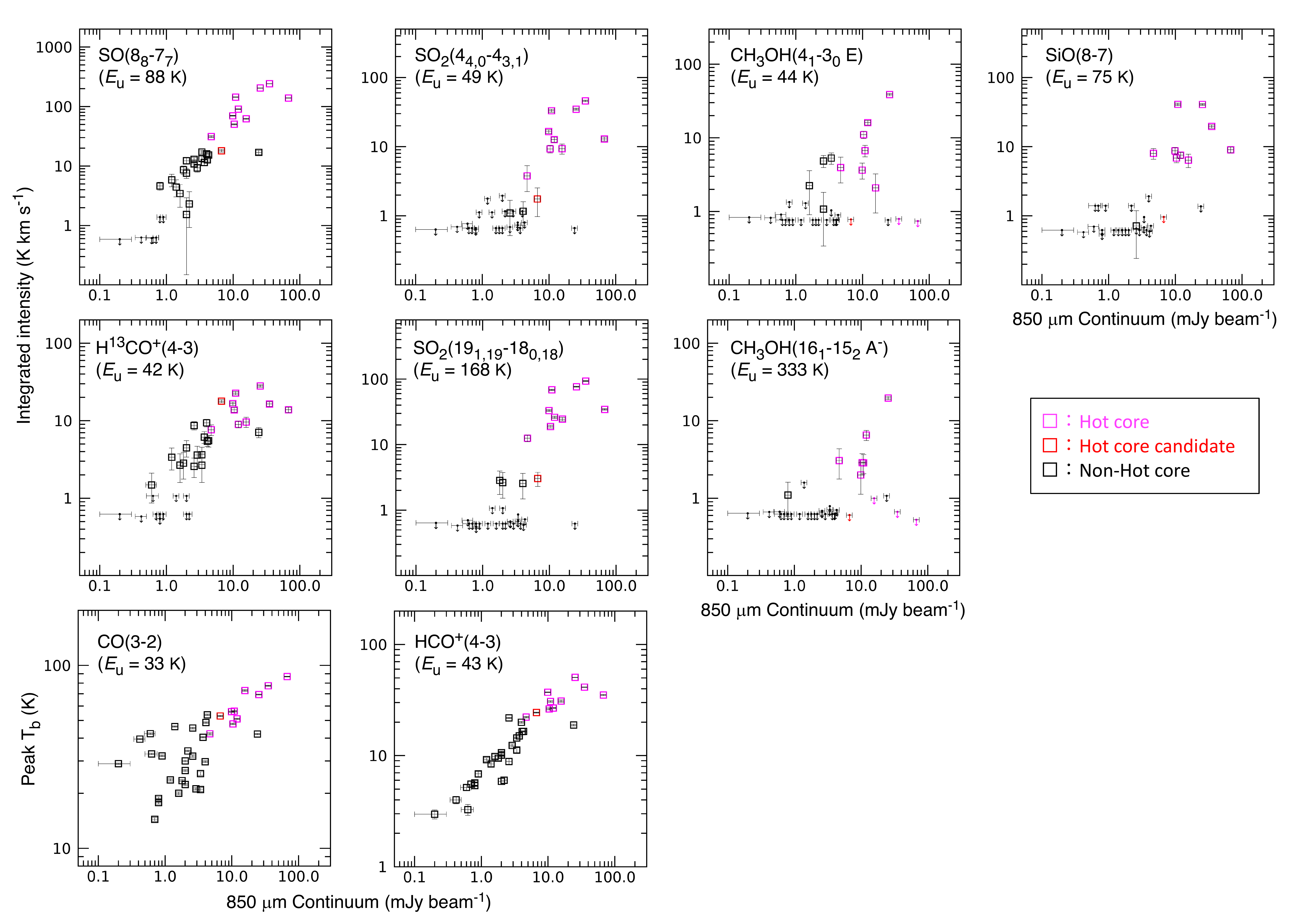}
\caption{
Same as in Figure~\ref{vsIntensity}, but the 850 $\mu$m continuum brightnesses are plotted on the horizontal axis. 
}
\label{vsIntensity_conti}
\end{center}
\end{figure*}

\subsection{Molecular tracer of LMC hot cores} \label{sec_disc_tracer} 
Identifying molecular tracers that can distinguish sources associated with hot cores provides valuable information for planning larger-sample surveys. 
Figure~\ref{vsIntensity} shows the relationships between integrated or peak line intensities and bolometric luminosities for all sources. 
Overall, sources associated with hot cores tend to exhibit stronger line intensities, and more luminous sources show stronger line emission. 

Of particular interest in these plots is the utility of the high-excitation SO line (8$_{8}$--7$_{7}$, $E_{u}$ = 88 K, $S\mu^{2}$ = 18.6 Debye$^2$) as a good tracer of LMC hot cores. 
As shown in the figure, all LMC hot cores show the strong high-excitation SO line, which clearly distinguish them from non-hot-core sources. 
All sources with SO integrated intensities greater than approximately 20 K km s$^{-1}$ are hot cores accompanied by high-excitation SO$_2$ and/or CH$_3$OH lines, whereas no such high-excitation lines are detected in sources below this SO threshold. 
The hot core candidate, Lm07, also shows a relatively high SO integrated intensity, placing it between hot-core and non-hot-core sources. 
Similar trends are also seen for the SO$_2$ ($E_{u}$ = 49 and 168 K) and SiO ($E_{u}$ = 75 K) lines, suggesting that strong emission in these lines, along with their spatial association with continuum sources, could serve as good evidence for the presence of hot cores. 
However, because these lines are generally weaker than the SO (8$_{8}$--7$_{7}$) line, the high-excitation and large-$S\mu^{2}$ SO line is likely more suitable for large-sample surveys. 

In the case of the high-excitation CH$_3$OH line ($E_{u}$ = 333 K), some hot cores show non-detections. 
This is likely due to the large source-to-source variation in the abundance of organic molecules in LMC hot cores (see Section~\ref{sec_disc_molab}). 
While CH$_3$OH is often used as a tracer of hot cores in nearby solar-metallicity star-forming regions, it may not serve as a universal tracer in LMC hot cores, which exhibit different chemical evolution. 
Regarding low-excitation CH$_3$OH, it is detected with similar intensities in some non-hot-core sources as in hot cores, indicating that it is not necessarily associated exclusively with hot cores. 
Low-excitation CH$_3$OH gas may be released via non-thermal desorption mechanisms, such as shock spattering, which likely contribute to its detection in non-hot-core sources. 

For the H$^{13}$CO$^+$ line ($E_{u}$ = 42 K), the separation between hot-core and non-hot-core sources is less clear compared to the SO line, making it difficult to identify hot cores based solely on the intensity of H$^{13}$CO$^+$. 
For the CO ($E_{u}$ = 33 K) and HCO$^+$ ($E_{u}$ = 43 K) lines, their peak intensities are shown in the figure, as these lines are considered to be optically thick. 
The HCO$^+$ peak intensities show a good correlation with luminosity (correlation coefficient = 0.84), likely reflecting the presence of warmer and denser gas surrounding more luminous protostars. 
Even in the CO and HCO$^+$ plots, sources associated with hot cores tend to show higher line intensities overall. 
However, the separation between hot-core and non-hot-core sources is not as prominent as with the high-excitation SO line. 
Particularly, at the lower-luminosity end, it is difficult to identify hot-core sources based solely on CO or HCO$^+$ peak intensities. 

The continuum emission show similar behavior to CO and HCO$^+$ (bottom right of Fig.~\ref{vsIntensity}). 
The continuum fluxes show a good correlation with luminosity (correlation coefficient = 0.89) and tend to be higher in hot core sources. 
However, since continuum flux alone does not provide sufficient information on the temperature of molecular gas, high-excitation molecular line tracers are more reliable indicators for hot core identification.

We note that luminosity information is missing for some sources, and that the infrared data used to derive luminosities have significantly lower spatial resolution compared to the ALMA data (see Section~\ref{sec_tar}). 
In contrast, ALMA submillimeter continuum traces compact components well. 
Therefore, we also plot the line intensities against the 850~$\mu$m continuum brightnesses in Figure~\ref{vsIntensity_conti}. 
As mentioned in the previous paragraph, luminosities and 850~$\mu$m continuum brightnesses show a good correlation; thus, similar trends are observed in this comparison as in the one with luminosity discussed above. 
For the high-excitation SO line, the integrated intensities show a good correlation with the 850~$\mu$m continuum, with a correlation coefficient of 0.87. 
This correlation is somewhat stronger than that seen in the comparison with luminosity (i.e., correlation coefficient = 0.74 in the SO panel in Figure~\ref{vsIntensity}). 
For the HCO$^+$ line, the correlation is even tighter, with a correlation coefficient of 0.93.
Other lines also exhibit trends similar to those observed in the comparison with luminosity.

\subsection{Molecular abundances vs. physical properties} \label{sec_disc_mol_vs_phys} 
To investigate the physical factors contributing to the chemical diversity of hot cores, this section compares the physical properties and chemical compositions of LMC hot cores. 

\subsubsection{vs. Rotational temperatures} \label{sec_disc_Trot} 
\begin{figure}[tbp!]
\begin{center}
\includegraphics[width=8.7cm]{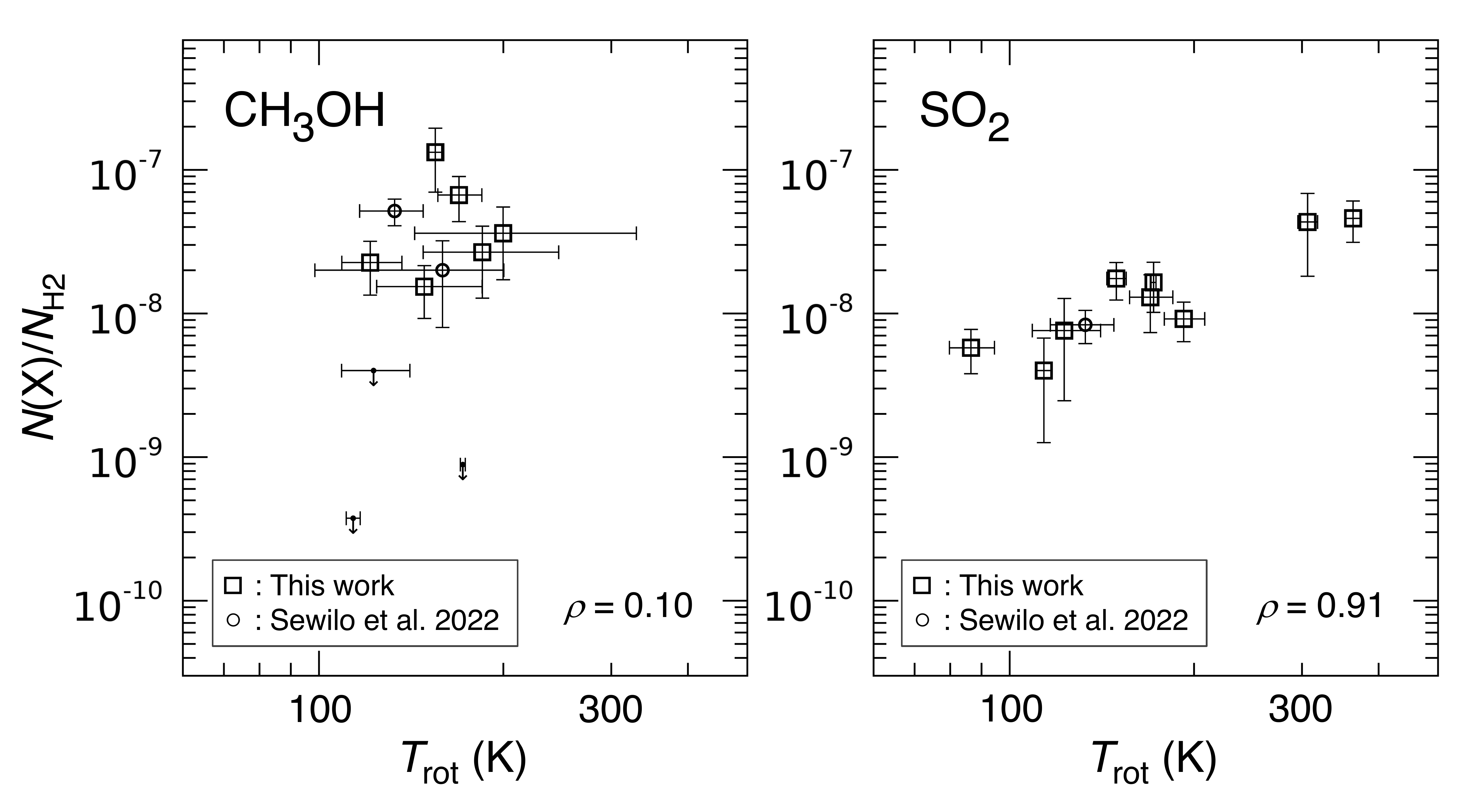}
\caption{
Molecular abundances vs. rotational temperatures for hot cores (open squares). 
In the CH$_3$OH panel, two hot core samples from \citet{Sew22a} (N105 3B and N113 B3) are shown as open circles. 
For sources in which CH$_3$OH was not detected, the rotational temperature of SO$_2$ is plotted instead of unknown CH$_3$OH temperatures. 
Upper limits are indicated by black dots with downward arrows. 
The correlation coefficient ($\rho$) is shown in the lower right corner of each panel. 
}
\label{vsTrot}
\end{center}
\end{figure}

Figure~\ref{vsTrot} shows the relationship between molecular abundances and rotational temperature for CH$_3$OH and SO$_2$ in LMC hot cores. 
A strong correlation is observed for SO$_2$ (correlation coefficient = 0.91), suggesting that temperature rise in hot core regions contribute to enhanced SO$_2$ abundances. 
This trend is likely related either to an additional supply of sulfur from dust grains in high-temperature gas or to enhanced SO$_2$ synthesis via high-temperature gas-phase chemistry \citep[e.g.,][]{NM04}. 

In contrast, CH$_3$OH shows no significant correlation with $T_\mathrm{rot}$. 
The correlation coefficient based on eight hot cores with high-temperature CH$_3$OH detections, including two literature samples from \citet{Sew22a}, is 0.10. 
Furthermore, there are hot cores in which CH$_3$OH is not detected despite the presence of high-temperature gas traced by SO$_2$. 

Sublimation of ice mantles is a key process that significantly influences the chemical composition of hot cores, particularly for species like CH$_3$OH, which are predominantly formed via surface reactions on dust grains during the cold prestellar stage \citep[e.g.,][]{McC23}. 
Binding energies on water ice surfaces have been reported to be $\sim$2000~K for SO \citep[e.g.,][]{Ngu24} and $\sim$5000~K for SO$_2$ and CH$_3$OH \citep[e.g.,][]{Per22,Das18}, respectively. 
These values correspond to sublimation temperatures of $\sim$40~K for SO and $\sim$100~K for SO$_2$ and CH$_3$OH, respectively. 
The rotational temperatures of SO$_2$ measured for hot cores without CH$_3$OH detections (Lh10, Lh09, and Lh08) suggest that these sources possess physical conditions suitable for ice sublimation. 
However, CH$_3$OH is deficient in the sublimated high-temperature gas in these sources. 
This implies that factors other than ice sublimation, such as the efficiency of organic molecule formation during the ice-mantle growth stage, or molecular destruction in the hot core phase, may significantly affect the CH$_3$OH abundance in LMC hot cores.

\subsubsection{vs. Line widths} \label{sec_disc_FWHM} 
\begin{figure}[tbp!]
\begin{center}
\includegraphics[width=8.7cm]{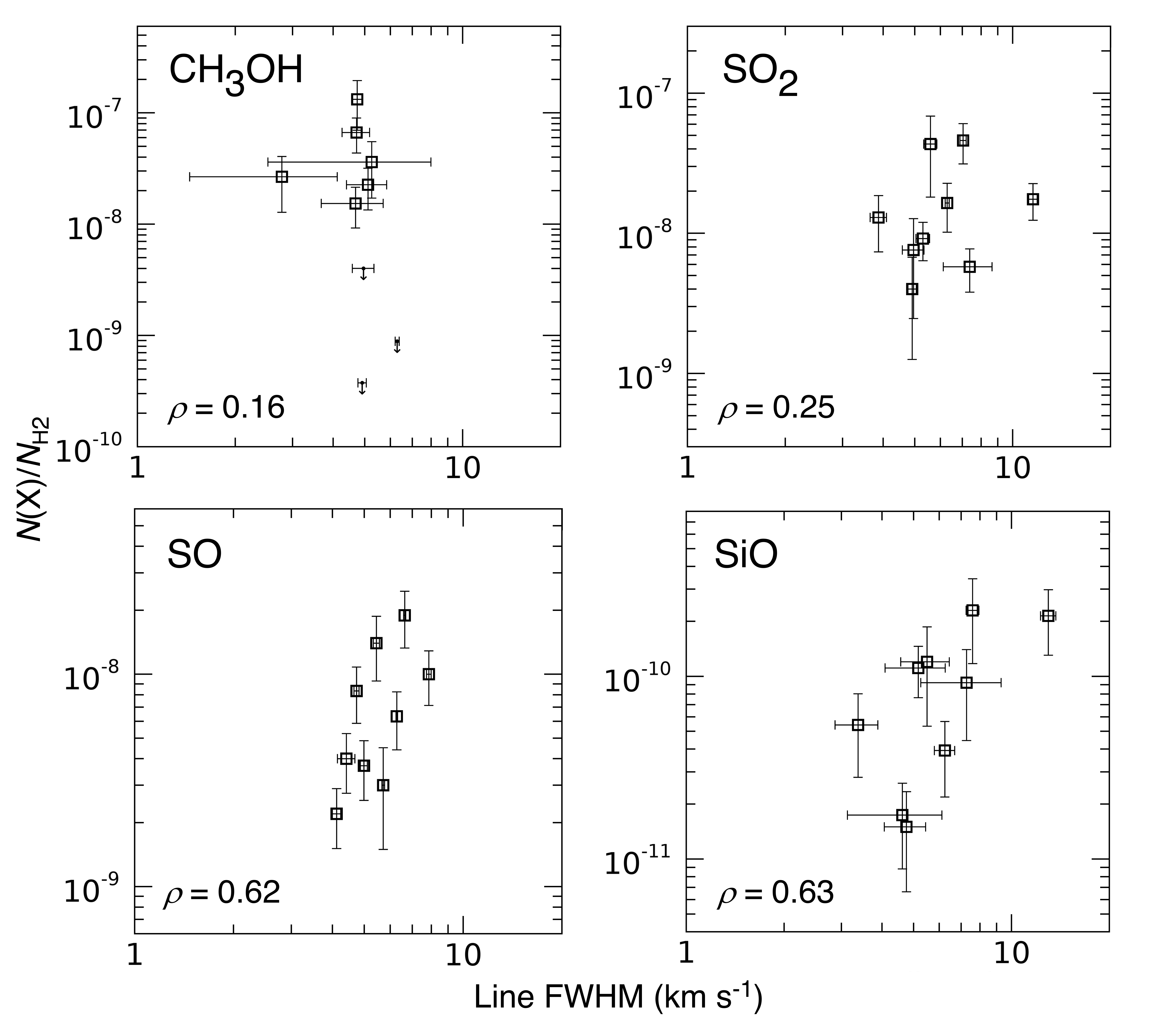}
\caption{
Molecular abundances vs. line FWHM widths for hot cores. 
Upper limit points are indicated by black dots with downward arrows. 
The line FWHMs are measured using the following transitions: CH$_3$OH(4$_{1}$ E--3$_{0}$ E), SO$_2$(19$_{1,19}$--18$_{0,18}$), SO(8$_{8}$--7$_{7}$), and SiO(8--7). 
In sources where CH$_3$OH was not detected, the FWHM of SO$_2$ is plotted instead to represent the line width. 
The correlation coefficient ($\rho$) is shown in the lower left corner of each panel. 
}
\label{vsFWHM}
\end{center}
\end{figure}

Figure~\ref{vsFWHM} shows the relationship between line FWHM and molecular abundances. 
No clear correlation is observed for CH$_3$OH and SO$_2$. 
For SO and SiO, a weak positive correlation is seen (correlation coefficient $\sim$0.6), but it cannot be regarded as a strong one. 
This suggests that there is no strong link between non-thermal gas motions in hot-core regions and their chemical composition. 

\begin{figure}[tbp!]
\begin{center}
\includegraphics[width=6.5cm]{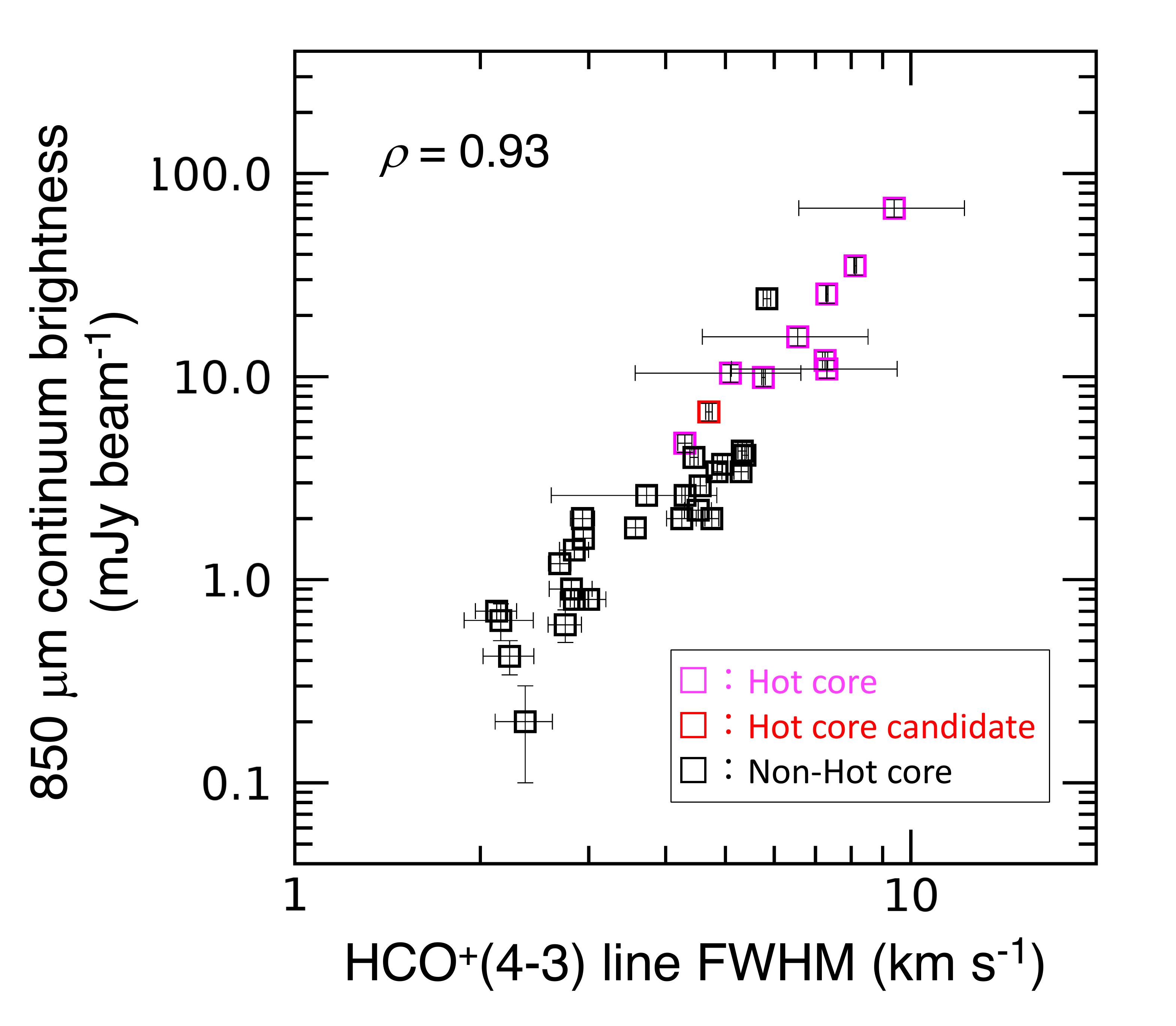}
\caption{
850 $\mu$m continuum brightness vs. FWHM of the HCO$^+$(4-3) line. 
Sources associated with hot cores and a hot-core candidate are shown in magenta and red, respectively, while other sources are shown in black. 
The correlation coefficient ($\rho$) is shown in the upper-left corner. 
}
\label{Cont_vs_FWHM}
\end{center}
\end{figure}

In shock regions associated with protostellar outflows, molecular abundances are expected to increase. 
However, with the present survey data obtained at a resolution of 0.1~pc, chemical and kinematic signatures attributable to shock regions would not be fully captured, or would be diluted. 
Figure~\ref{Cont_vs_FWHM} presents a comparison between the FWHM of the HCO$^+$(4-3) line and the continuum brightness. 
As shown in the figure, a strong correlation is found regardless of the presence or absence of hot cores (correlation coefficient = 0.93). 
This correlation may reflect that more massive cores induce stronger feedback, resulting in more pronounced non-thermal gas motions. 
At the present spatial resolution, the line widths of cores appear to be primarily governed by their mass. 
Future higher-resolution observations are expected to reveal the chemical and physical properties of molecular gas associated with shock regions around LMC hot cores (Chen et al., in prep.).

\subsubsection{vs. Infrared spectral characteristics} \label{sec_disc_IRtype} 
\begin{figure}[tbp!]
\begin{center}
\includegraphics[width=8.7cm]{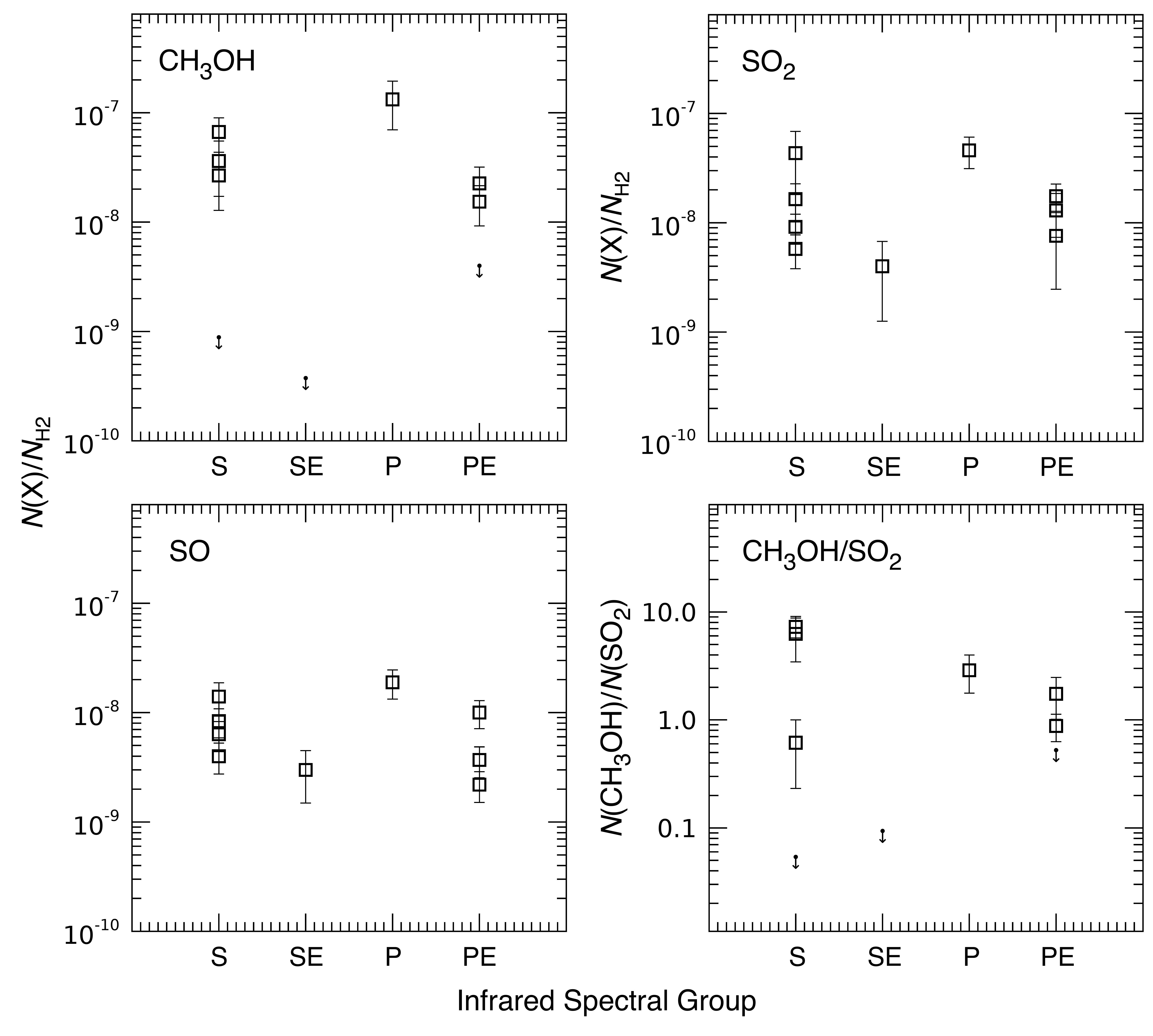}
\caption{
Molecular abundances vs. infrared spectral groups classified in \citet{Sea09} for the present hot cores. 
Upper limit points are indicated by black dots with downward arrows. 
The bottom right panel shows the abundance ratio of CH$_3$OH normalized by SO$_2$. 
See Section \ref{sec_disc_IRtype} for details. 
}
\label{vsIR}
\end{center}
\end{figure}

Figure~\ref{vsIR} shows a comparison between the molecular compositions of hot cores and their infrared spectral characteristics. 
\citet{Sea09} classified massive YSOs in the LMC based on 5--38~$\mu$m spectra obtained with Spitzer/IRS. 
In their classification, the sources were grouped into five categories, S, SE, P, PE, and E. 
Sources that did not fall into any of these categories were labeled as F, and those for which classification was not possible due to limited spectral coverage were labeled as U. 
The S group includes sources dominated by silicate and ice absorption features, while the SE group shows similar features along with fine-structure lines from ionized gas. 
The P group is characterized by prominent PAH emission features, and the PE group exhibits both PAH features and fine-structure lines. 
The E group consists of sources in which fine-structure lines from ionized gas dominate the spectrum. 
In the MAGOS project, we targeted sources belonging to the S, SE, P, and PE groups, and identified at least one hot core in each group. 

As shown in Figure~\ref{vsIR}, there is no clear correlation between the infrared spectral group and molecular composition. 
Sources with low CH$_3$OH abundances are not confined to any specific infrared spectral group. 
\citet{Sea09} proposed that YSOs evolve in the sequence S $\rightarrow$ SE $\rightarrow$ P $\rightarrow$ PE, corresponding to an increasing prominence of infrared emission features. 
Within the current sample, there is no apparent relationship between the molecular compositions of hot cores and the evolutionary stages inferred from their infrared spectral characteristics. 
To remove systematic uncertainties associated with $N_{\mathrm{H_2}}$ normalization, Figure~\ref{vsIR} also shows abundance ratios normalized by SO$_2$, but this comparison reveals a similar lack of correlation. 

It should be noted, however, that there is a significant difference in spatial scale between the regions from which molecular lines were extracted with ALMA and those observed in infrared spectroscopy with Spitzer. 
Given the slit width of Spitzer/IRS, the infrared spectra were extracted from regions approximately 4$\arcsec$ to 11$\arcsec$ in size, centered on the infrared point source. 
In contrast, the spectral extraction region in the present ALMA observations corresponds to 0$\farcs$41. 
Although the hot core sources are the dominant contributors to the submillimeter continuum flux, and possibly to the infrared flux, within the field of view, the Spitzer spectra are likely affected by non-negligible contamination from emission components outside the hot core regions. 
Such contamination can significantly affect the evaluation of the evolutionary stage of protostars based on absorption features in their infrared spectra.
Future high-spatial-resolution observations with JWST will be essential for obtaining infrared data that are directly comparable to the ALMA observations.

\subsubsection{vs. Star formation activities} \label{sec_disc_SF} 
\begin{figure}[tbp!]
\begin{center}
\includegraphics[width=8.7cm]{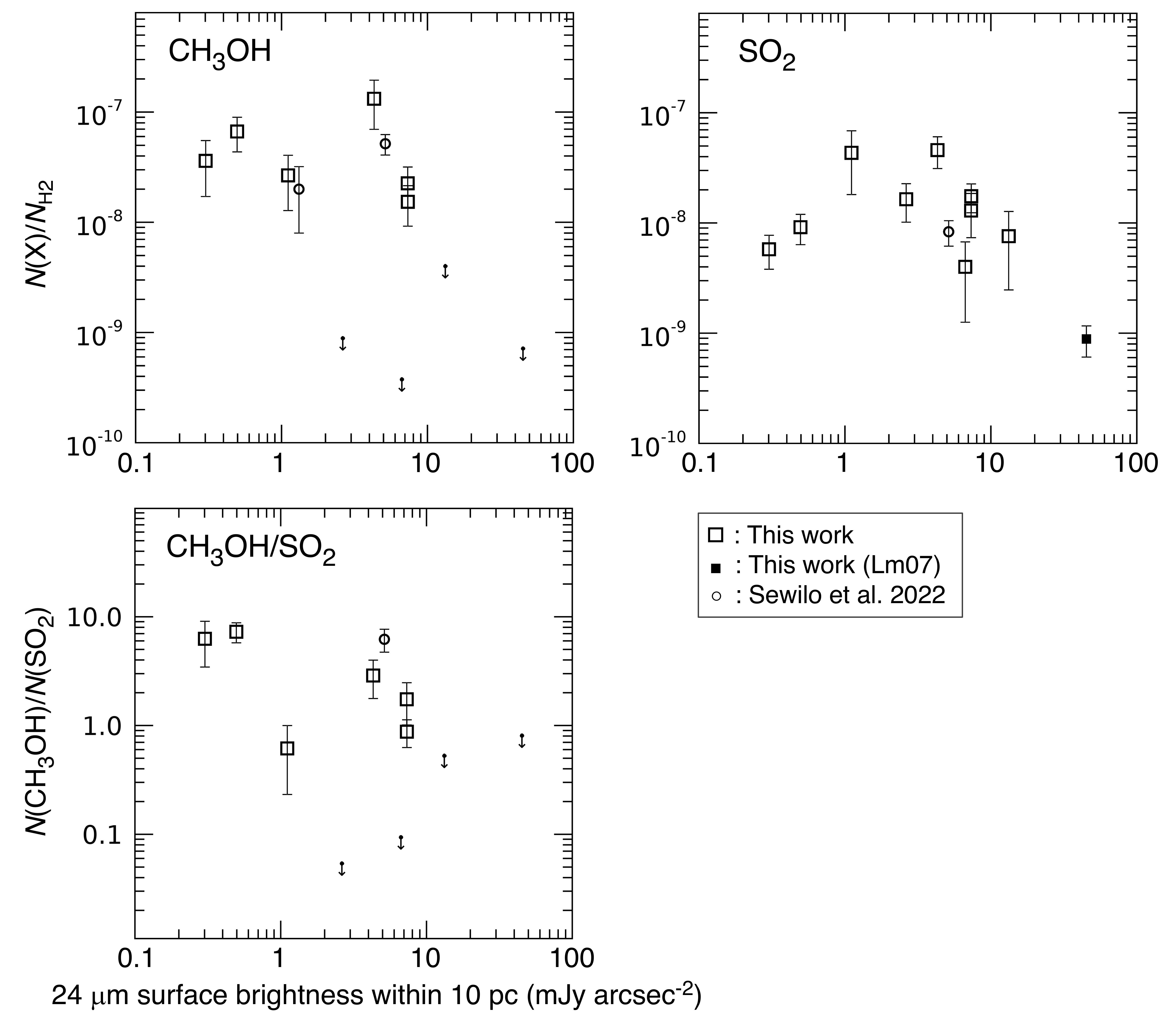}
\caption{
Molecular abundances vs. 24~$\mu$m brightness averaged over a 10~pc radius around each hot core. 
The hot cores observed in this work are shown as open squares, while the hot core candidate Lm07 is shown as a filled square. 
Hot core samples from \citet{Sew22b} are indicated by open circles. 
Upper limits are indicated by black dots with downward arrows. 
The bottom panel shows the abundance ratio of CH$_3$OH normalized by SO$_2$. 
See Section~\ref{sec_disc_SF} for details. 
}
\label{vsSF}
\end{center}
\end{figure}

Figure~\ref{vsSF} shows a comparison between the molecular compositions of hot cores and the degree of star formation activity in their surrounding environments. 
The 24~$\mu$m infrared emission primarily traces thermal emission from warm dust and is commonly used as a tracer of large-scale star formation rates \citep[e.g.,][]{Cal07,Ken09}. 
In this figure, we plot the 24~$\mu$m surface brightness averaged over a 10~pc radius around each hot core, using the 24$\mu$m emission map of the LMC obtained by Spitzer/MIPS \citep{Mei06}. 
The hot core candidate, Lm07, is also included in the plot, because it is located in the most active region near 30~Dor among the present hot core samples. 

Hot cores with no CH$_3$OH detection (Lh10, Lh09, Lh08, and Lm07) are all located in regions with relatively high 24~$\mu$m brightness ($\gtrsim$2 mJy arcsec$^{-2}$). 
A similar trend is seen in the plot of abundance ratios of CH$_3$OH normalized by SO$_2$. 
For SO$_2$, no apparent relationship is found between its abundance and the 24~$\mu$m brightness of the surrounding regions.

It cannot be ruled out that increased radiation fields associated with active star formation in the natal molecular cloud may contribute to triggering COM-poor hot core chemistry.
However, given that CH$_3$OH abundances vary by nearly two orders of magnitude among sources with similar 24~$\mu$m brightness, it is evident that the degree of surrounding star formation activity alone does not determine the presence or absence of organic molecules in hot cores.

\begin{figure}[tbp!]
\begin{center}
\includegraphics[width=8.5cm]{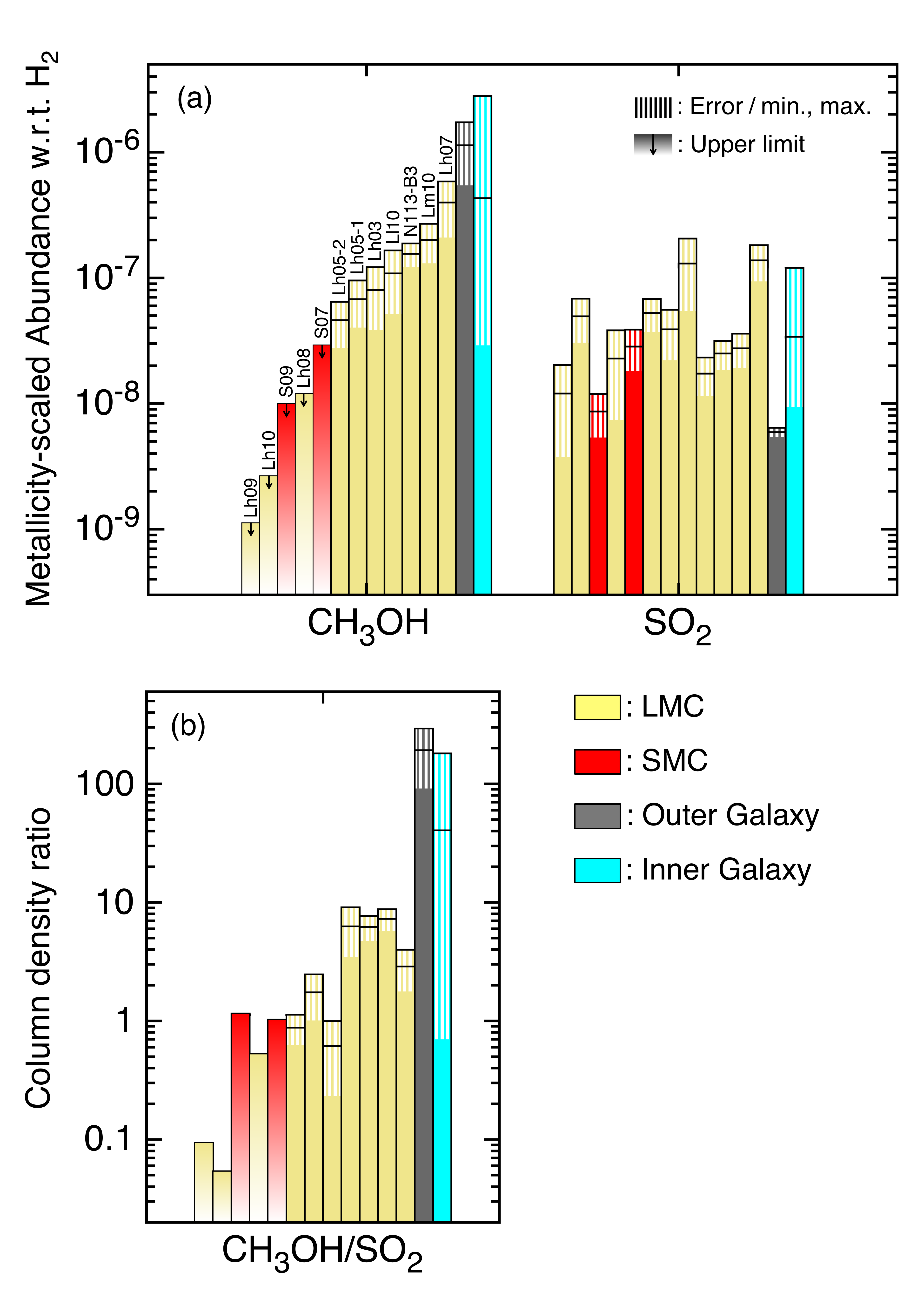}
\caption{
(a) Comparison of metallicity-scaled molecular abundances of hot cores in the LMC (yellow), SMC (red), outer Galaxy (gray), and inner Galaxy (cyan). 
Abundances in low-metallicity hot cores are scaled according to their metallicities to correct for differences in elemental abundances. 
We assume metallicities of 1/3, 1/5, and 1/4 $Z_\odot$ for the LMC, SMC, and the outer Galaxy, respectively. 
Only contributions from high-temperature gas components ($T_\mathrm{rot} >$100~K) are plotted. 
The area with thin vertical lines indicate the error bar for low-metallicity sources, while those of Galactic sources indicate the range between the maximum and minimum abundances among the compiled samples. 
Bars with color gradients and downward arrows represent upper limits. 
(b) Same as in panel (a), but showing the column density ratios of CH$_3$OH to SO$_2$. 
See Section~\ref{sec_disc_molab} for details on the plotted data and references. 
}
\label{abundance}
\end{center}
\end{figure}

\subsection{Molecular abundances variation within LMC hot cores} \label{sec_disc_molab} 
The large source-to-source variation of CH$_3$OH abundances reported in previous studies of the LMC is also confirmed in the present survey, which adopted a uniform spatial and spectral setup. 
Figure~\ref{abundance}(a) shows a comparison of metallicity-scaled abundances of CH$_3$OH and SO$_2$ in hot cores across environments with different metallicities. 

To focus on the high-temperature gas components associated with hot cores, this study derives $N_{\mathrm{H_2}}$ using dust temperatures estimated from CH$_3$OH and/or SO$_2$. 
For the SMC hot cores plotted in the figure, $N_{\mathrm{H_2}}$ has been recalculated using the same method as in this study, based on the data from \citet{ST23}. 
In the SMC, compact and warm gas components have only been detected in SO$_2$, while CH$_3$OH is detected in just one source, and only as an extended component at $\sim$40~K. 
Thus, for the SMC hot cores in this figure, we plot upper limits of CH$_3$OH estimated from the 7${_2}$ A$^+$--6${_2}$ A$^+$ transition ($E_u$ = 103~K). 
For the hot core in the extreme outer Galaxy, we plot values based on data obtained from the 0.1~pc region around the protostar reported by \citet{ST21}, with $N_{\mathrm{H_2}}$ recalculated using the same method as in this study. 
For comparison, we include Galactic hot core data: CH$_3$OH values are taken from 28 hot cores reported by \citet{Bae22}, while SO$_2$ values are taken from 10 hot cores reported by \citet{Sut95, Mac96, Hel97, Hat98b, Qin10, Zer12, Xu13}, and their median abundances are plotted in the figure. 

As shown in Figure~\ref{abundance}(a), CH$_3$OH abundances exhibit large source-to-source variations even after correcting for metallicity, suggesting that its abundance in hot cores cannot be explained by metallicity scaling alone. 
Such large variations are not seen in SO$_2$, another common hot core tracer. 
Among the LMC sources, the CH$_3$OH abundance varies by a factor of $>$350 between the most enriched and most depleted sources, while the variation for SO$_2$ is a factor of 12. 
The spread factor, defined as 10 to the power of the 1$\sigma$ standard deviation in log$_{10}$ space \citep{Naz22b}, is calculated to be $>$6.8 for CH$_3$OH and 2.2 for SO$_2$. 

Figure~\ref{abundance}(b) shows the abundance ratio of CH$_3$OH to SO$_2$. 
For comparison, Galactic hot cores with gas temperatures above 100~K from the massive protostar sample in \citet{vanGel22a} and \citet{San24} are also plotted. 
This comparison reveals a similar trend to that seen in the $N_{\mathrm{H_2}}$-normalized abundances. 
Among the LMC sources, the CH$_3$OH/SO$_2$ ratio varies by more than two orders of magnitude, with a spread factor greater than 5.4, comparable to that found for the $N_{\mathrm{H_2}}$-normalized abundances. 
This indicates that the large variation in CH$_3$OH abundances among LMC hot cores is not merely an artifact of the method used to derive $N_{\mathrm{H_2}}$.

\begin{figure*}[tpb!]
\begin{center}
\includegraphics[width=13cm]{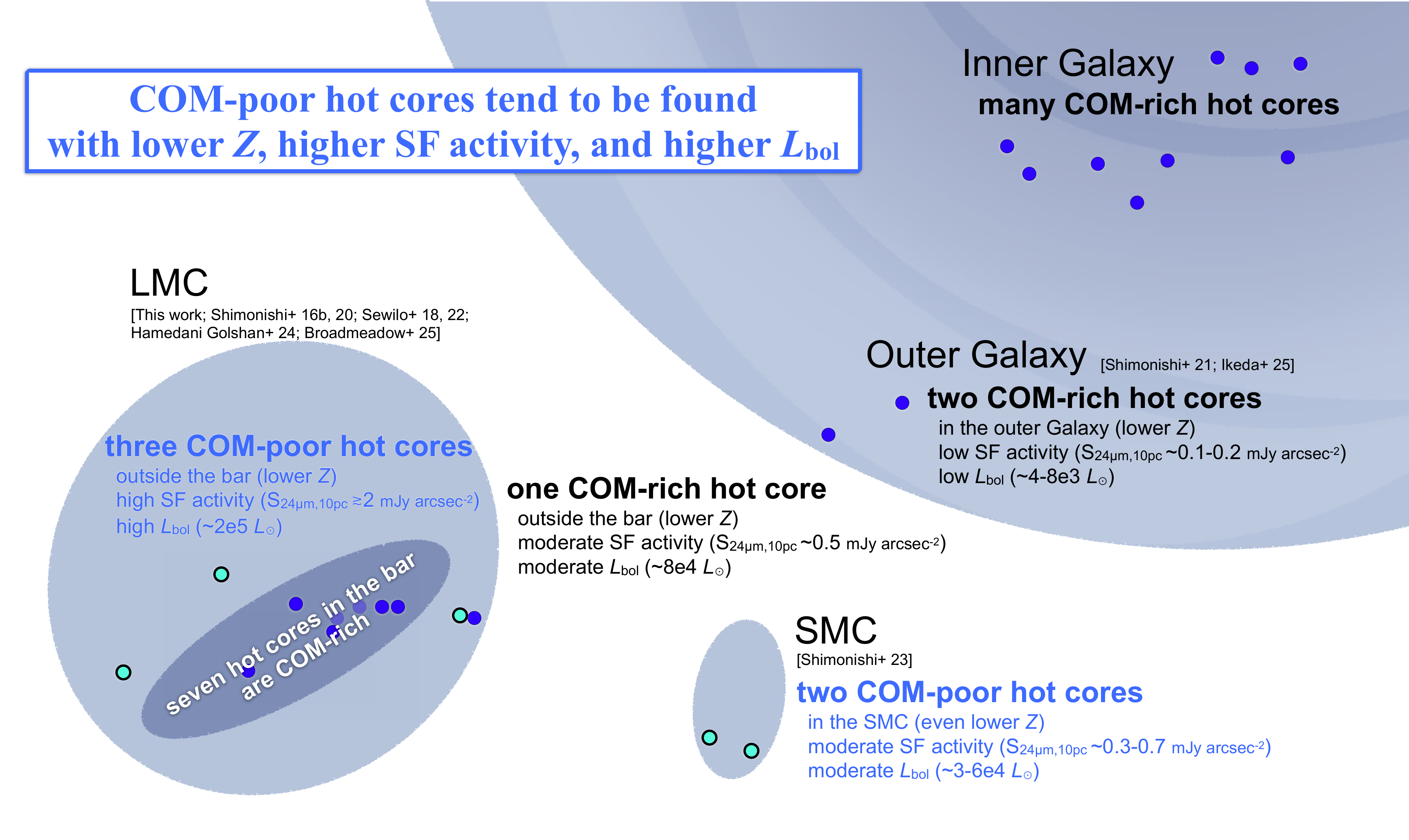}
\caption{
Schematic summary of COM-poor hot cores in low-metallicity environments. 
In the LMC, COM-poor hot cores are currently found only outside the bar region (i.e., lower metallicity, $Z$), where all of them have high bolometric luminosity ($L_{\mathrm{bol}}$) and are associated with high star-formation (SF) activity (S$_{\mathrm{24\mu m,10pc}}$ indicates 24~$\mu$m surface brightness averaged over a 10~pc radius around each hot core). 
In the even lower-metallicity SMC, both known hot cores are COM-poor despite having only moderate luminosity and star-formation activity. 
In contrast, the two known hot cores in the outer Galaxy are COM-rich despite the low metallicity, highlighting that low metallicity alone does not necessarily result in COM-poor chemistry. 
}
\label{conceptual}
\end{center}
\end{figure*}

\subsection{Trigger of COM-poor hot core chemistry} \label{sec_disc_molab2} 
The reason why the gas-phase abundance of CH$_3$OH exhibits such large variations among LMC hot cores, particularly the extreme COM-poor nature observed in some sources, which is not typically seen in solar-metallicity environments, remains under debate. 
Multiple factors may influence the abundance of COMs in hot cores, including observational factors such as optical depth or beam dilution effects, as well as intrinsic physical factors such as source size and structure, the degree of ice mantle sublimation, differences in evolutionary stage, variations in shock conditions, and local physical conditions. 

As discussed in Section~\ref{sec_disc_tau}, optical depth effects on CH$_3$OH are limited and unlikely to account for order-of-magnitude abundance variations. 
Shock conditions also appear to have a minor influence, as clear correlation is not observed between line width and line intensity, especially for CH$_3$OH and SO$_2$ (see Section~\ref{sec_disc_FWHM}). 
Differences in evolutionary stage may have an effect, but as discussed in Section~\ref{sec_disc_IRtype}, there is no clear correlation between the Spitzer spectral classifications and molecular abundances. 
This warrants further infrared studies with higher spatial resolution. 

The high-temperature components of most hot cores are not well spatially resolved in the present survey. 
The FWHM sizes of the continuum emission in hot cores derived from two-dimensional Gaussian fits range from 0$\farcs$4 to 0$\farcs$6, which is slightly larger than the typical continuum beam size (0$\farcs$30 $\times$ 0$\farcs$37). 
In contrast, the Gaussian FWHM sizes of the high-excitation SO$_2$ and CH$_3$OH emission, except for Lh09, are 0$\farcs$4--0$\farcs$5, which are roughly comparable to the 0$\farcs$40 beam size. 
One possible reason why the dust emission appears somewhat more extended is that the continuum contains contributions not only from the high-temperature dust in the hot core region but also from the surrounding, relatively low-temperature dust. However, it cannot be ruled out that the true emitting region of the high-temperature dust is intrinsically more extended than that of the high-temperature molecular gas. 
In this work, we derive beam-averaged column densities within the 0.1~pc (0$\farcs$41) diameter region under the assumption of a beam filling factor of unity for both the molecular gas and the continuum-based H$_2$ (see Section~\ref{sec_ana}). 
If the true spatial extent of the high-excitation molecular gas is significantly smaller than that of the high-temperature dust, the molecular abundances would be underestimated due to beam dilution. 
This point should also be kept in mind when comparing our results with those for interferometric observations of nearby star-forming regions, where the hot gas and dust emission are spatially well resolved. 

In nearby Galactic protostars, the dust opacity of the envelope and the presence of disks have been discussed as factors influencing COM intensity \citep[e.g.,][]{DeS20, Naz22a,Che25}. 
Such effects occur on spatial scales of a hundred to several thousand au. 
Given the spatial resolution of the present observations (0.1~pc), these dense substructures are unlikely to affect the observed CH$_3$OH intensity \citep[also see][for the effect of dust attenuation in massive protostars]{Naz23}. 
We note, however, that the current dataset does not allow us to assess how small-scale physical structures and the extent of hot gas emission are related to COM abundances. 
This issue will be addressed in a forthcoming paper based on high-resolution follow-up observations of the hot core sample presented here. 

A clear correlation is seen between SO$_2$ abundances and rotational temperature of hot cores, with warmer cores exhibiting higher SO$_2$ abundances (see Section~\ref{sec_disc_Trot}). 
This suggests that the source-to-source variation in SO$_2$ abundance seen in Figure~\ref{abundance}(a) is largely attributable to temperature differences. 
In contrast, CH$_3$OH shows no such trend with temperature, and some hot cores with high-temperature gas exhibit low CH$_3$OH abundances, implying that additional factors must be considered. 

Local environmental differences may influence COM abundances. 
Astrochemical simulations targeting the chemical evolution of hot cores in the LMC and SMC suggest that the dust temperature during the early ice formation stage can affect the gas-phase CH$_3$OH abundance in the subsequent hot core phase \citep{Ach18,ST20}. 
This is because solid CH$_3$OH formation is sensitive to dust temperature \citep[e.g.,][]{Cha12} and is suppressed on warmer dust surfaces \citep[also see][for the warm ice chemistry hypothesis based on ice observations of LMC protostars]{ST16}.
On the other hand, astrochemical simulations suggest that the solid SO$_2$ formation is less sensitive to the dust temperature compared to that of solid CH$_3$OH at least under the solar metallicity conditions \citep[see Fig.6 and 7 in][]{Fur24}. 
Variations in local metallicity and radiation field strength may introduce diversity in the physical conditions during the ice formation stage, leading to differences in CH$_3$OH abundance. 

The all COM-poor hot cores are located outside the LMC bar region, as discussed in Section~\ref{sec_disc_distribution}. 
However, the present observations also reveal a hot core, Lm10, which exhibit rich COM abundances, despite being located near a COM-poor hot core outside the bar. 
This would indicate that low metallicity is not a sufficient condition for triggering COM-poor hot core chemistry. 
As discussed in Section~\ref{sec_disc_SF}, all COM-poor hot cores are found in regions with active star formation. 
Intense interstellar radiation fields in such environments, in combination with low metallicity, may raise the dust temperature during the ice formation stage, potentially triggering COM-poor hot core chemistry. 

We also note that CH$_3$OH-non-detected hot cores are biased toward high-luminosity protostars. 
This suggests that UV-induced molecular destruction within hot cores may also contribute to the onset of COM-poor chemistry. 
If this is the case, the COM-poor hot core chemistry would originate in the protostellar stage rather than the prestellar stage. 

For reference, the hot core WB89-789 SMM1 in the extreme outer Galaxy, an environment with metallicity lower than that of the LMC, exhibits rich COM chemistry \citep{ST21}, and its CH$_3$OH abundance is consistent with metallicity scaling. 
Rich COM chemistry has also been suggested in another outer Galaxy hot core, Sh 2-283-1a SMM1 \citep{Ike25}. 
These sources differ from the LMC hot cores in that they have luminosities nearly an order of magnitude lower and are located in a region of relatively quiescent star formation (i.e., the outskirts of the Galaxy). 
The mid-infrared surface brightness within a 10 pc radius of these objects, as measured from the IRAS 25~$\mu$m data \citep{Neu84}, is about 0.1--0.2 mJy arcsec$^{-2}$. 
Such relatively quiescent conditions may facilitate efficient COM formation even in low-metallicity environments. 

Based on the above discussion, we speculate that a combination of multiple factors, including low metallicity, active star formation in the vicinity, and high protostellar luminosity, would jointly trigger the unique COM-poor hot core chemistry observed in some LMC sources. 
A schematic illustration summarizing the discussion is presented in Figure~\ref{conceptual}. 
Larger-sample hot-core surveys targeting low-metallicity systems, as well as more detailed investigations through astrochemical simulations incorporating a range of physical, structural, and environmental parameters, will be essential. 

Finally, to investigate the origin of organic molecules in the hot core stage and to assess environmental effects, detailed observations of solid CH$_3$OH toward low-metallicity hot cores are crucial. 
\citet{ST16} reported the detection of CH$_3$OH ice toward two massive embedded protostars (ST6 and ST10) in the LMC, based on ground-based infrared spectroscopy with the Very Large Telescope (VLT). 
More recently, \citet{Sew25}, using JWST, detected complex organic ices larger than CH$_3$OH toward one of these protostars, ST6. 
However, these protostars are not included in the present survey. 
Future comprehensive studies of both ice and gas toward low-metallicity hot cores will be essential for elucidating the formation of COMs and their connection to the hot core chemistry in such environments.

\section{Summary} \label{sec_sum} 
\begin{itemize}
  \item  We conducted a uniform hot-core survey toward 30 spectroscopically confirmed massive protostellar objects in the LMC, using ALMA at 350~GHz with a spatial resolution of 0.1~pc.

  \item  We detect 36 continuum sources in total. 
  Among them, nine are identified as hot cores and one as a hot-core candidate, based on the presence of compact ($\sim$0.1~pc) and high-temperature ($\gtrsim$100~K) \ce{SO2} and/or \ce{CH3OH} gas. 
  One new hot core (Lm10) is discovered, in which \ce{CH3OCH3} is detected, representing the first detection of a COM larger than \ce{CH3OH} in a protostar located outside the LMC bar region. 
  
  \item  A variety of molecules are detected in the hot cores, including \ce{CO}, \ce{HCO+}, \ce{H^{13}CO+}, \ce{HC^{15}N}, \ce{HC3N}, \ce{SiO}, \ce{SO}, \ce{SO+}, \ce{NS}, \ce{SO2}, \ce{^{34}SO2}, \ce{^{33}SO2}, \ce{CH3OH}, \ce{^{13}CH3OH}, \ce{HCOOH}, \ce{HCOOCH3}, \ce{CH3OCH3}, and \ce{C2H5OH}. 
  \ce{H2CCO} is tentatively detected in one source, and hydrogen radio recombination lines are detected in two sources. 

  \item  \ce{^{13}CH3OH} is detected in the most COM-rich hot core (Lh07, or N113 A1). 
  The \ce{^{12}CH3OH} emission in this source is estimated to be only moderately optically thick ($\tau$ $\sim$0.7). 
  In contrast, \ce{SO2} emission is optically thick in several hot cores ($\tau$ up to 2.2), and thus \ce{^{34}SO2} is used for column density estimates. 

  \item  A tight positive correlation is observed between the HCO$^+$(4-3) line widths and 850 $\mu$m continuum brightnesses. 
  They follow the same correlation irrespective of the presence or absence of hot cores, suggesting that more massive cores induce stronger feedback, resulting in more pronounced non-thermal gas motions.

  \item  All hot-core sources show stronger emission in the high-excitation SO line compared to non-hot-core sources, suggesting that its strong detection will be useful for identifying hot-core candidates in the LMC. 

  \item  The abundance of \ce{CH3OH} shows large source-to-source variations among LMC hot cores, with some sources deficient in COMs despite being associated with high-temperature gas. 
  In contrast, such a large chemical diversity is not seen in \ce{SO2}. 

  \item  A good positive correlation is found between \ce{SO2} abundances and rotational temperatures, suggesting that the gas temperature contributes to its chemical enrichment. 
  On the other hand, such correlation is not seen in \ce{CH3OH}. 
  No clear correlation is found between molecular abundances and infrared spectral characteristics. 
  
  \item  The present LMC hot cores without \ce{CH3OH} detections are all located outside the LMC bar region and are characterized by either high luminosity or active star formation in their surroundings. 
  We speculate that a combination of multiple factors, including locally low metallicity, active star formation in the vicinity, and high protostellar luminosity, may jointly trigger the unique COM-poor hot core chemistry observed in some LMC sources. 

\end{itemize}

\begin{acknowledgments}
This paper makes use of the following ALMA data: ADS/JAO.ALMA$\#$2019.1.01770.S. 
ALMA is a partnership of ESO (representing its member states), NSF (USA) and NINS (Japan), together with NRC (Canada), NSTC and ASIAA (Taiwan), and KASI (Republic of Korea), in cooperation with the Republic of Chile. 
The Joint ALMA Observatory is operated by ESO, AUI/NRAO and NAOJ.
This work has made extensive use of the Cologne Database for Molecular Spectroscopy and the molecular database of the Jet Propulsion Laboratory. 
This work was supported by JSPS KAKENHI grant Nos. 20H05845, 20H05847, 21H01145, 25K07364, and 25K07365.
This work was supported by NAOJ ALMA Scientific Research Grant Code 2025-29B.
T. S. gratefully acknowledges support from the Uchida Energy Science Promotion Foundation. 
K.E.I.T. gratefully acknowledges support from Center for Fundamental Research, Institute of Science Tokyo, through a Grant-in-Aid for Challenging Research and a Grant-in-Aid for Pioneering Research.
A. S. acknowledges support by the grant CNS2023-145240 (Consolida), PID2020-117710GB-I00, and PID2023-146675NB-I00 (MCI-AEI-FEDER, UE).
Finally, we would like to thank an anonymous referee for insightful comments, which substantially improved this paper. 
\end{acknowledgments}

\software{CASA \citep{McM07})}




\appendix

\restartappendixnumbering
\onecolumngrid
\section{Continuum and SO emission for all sources} \label{sec_app_image_all} 
\begin{figure*}[tp!]
\begin{center}
\includegraphics[width=17.8cm]{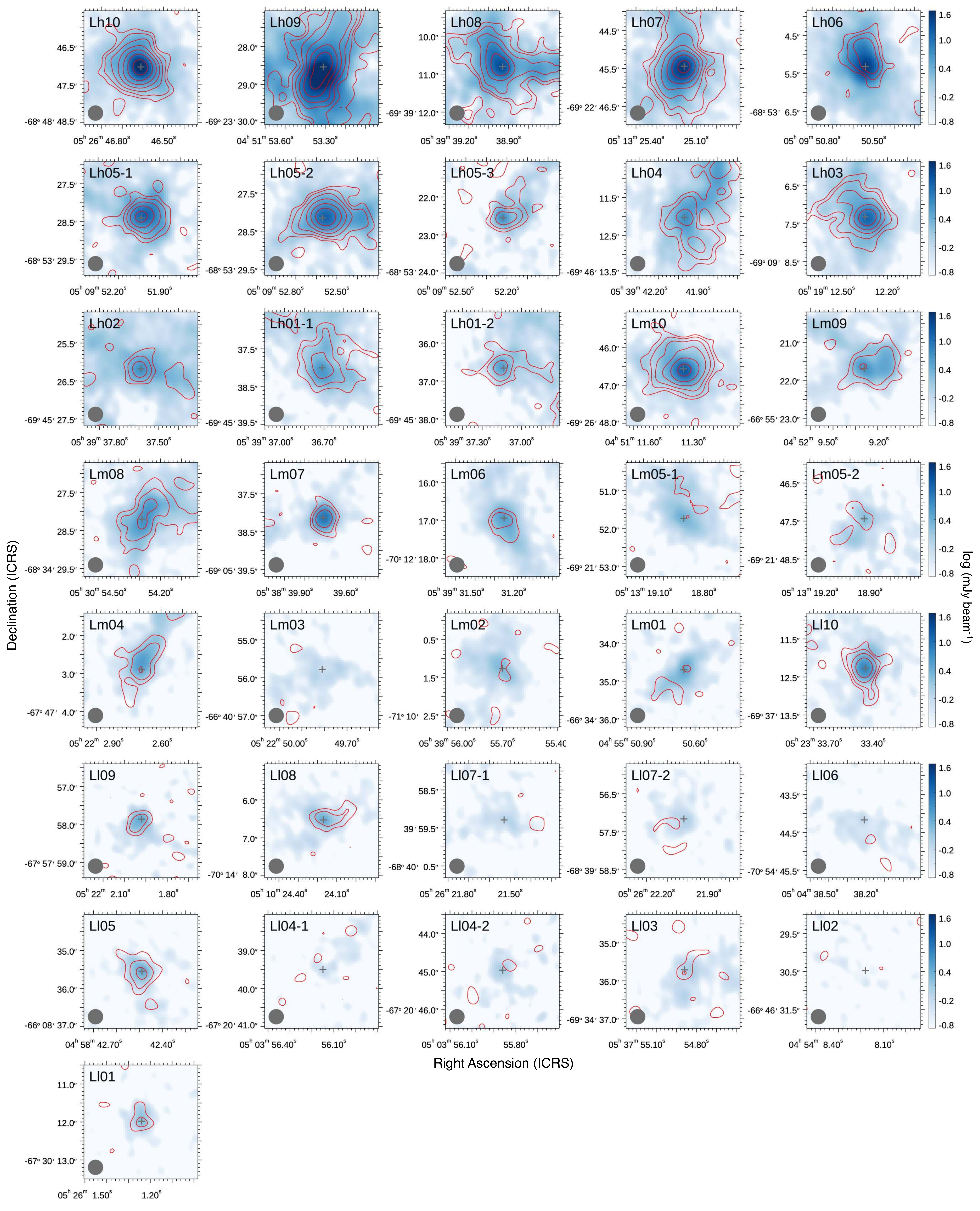}
\caption{
Intensity distributions of the SO(8${_8}$--7${_7}$) line (red contours) and the 850~$\mu$m continuum (blue background) for all sources. 
The contour levels correspond to 10$^{0.5}$, 10$^{0.8}$, 10$^{1.1}$, 10$^{1.4}$, 10$^{1.7}$, 10$^{2.0}$, and 10$^{2.3}$ K km s$^{-1}$. 
The color bar for the continuum is a logarithmic scale and in units of mJy beam$^{-1}$. 
The gray cross indicates the peak position of the molecular or continuum emission. 
The synthesized beam is shown as a gray filled circle. 
}
\label{image_all}
\end{center}
\end{figure*}

Figure~\ref{image_all} shows the intensity distributions of the 850~$\mu$m continuum and SO(8$_{8}$--7$_{7}$) line for all sources. 
The SO images are obtained by integrating the emission over a velocity range of $V_{\rm sys}$ $\pm$7 km s$^{-1}$, where $V_{\rm sys}$ of each source is listed in Table~\ref{sourcelist}. 

\restartappendixnumbering
\onecolumngrid
\section{Spectra of hot cores} \label{sec_app_spec} 
\begin{figure*}[tp!]
\begin{center}
\includegraphics[width=17.5cm]{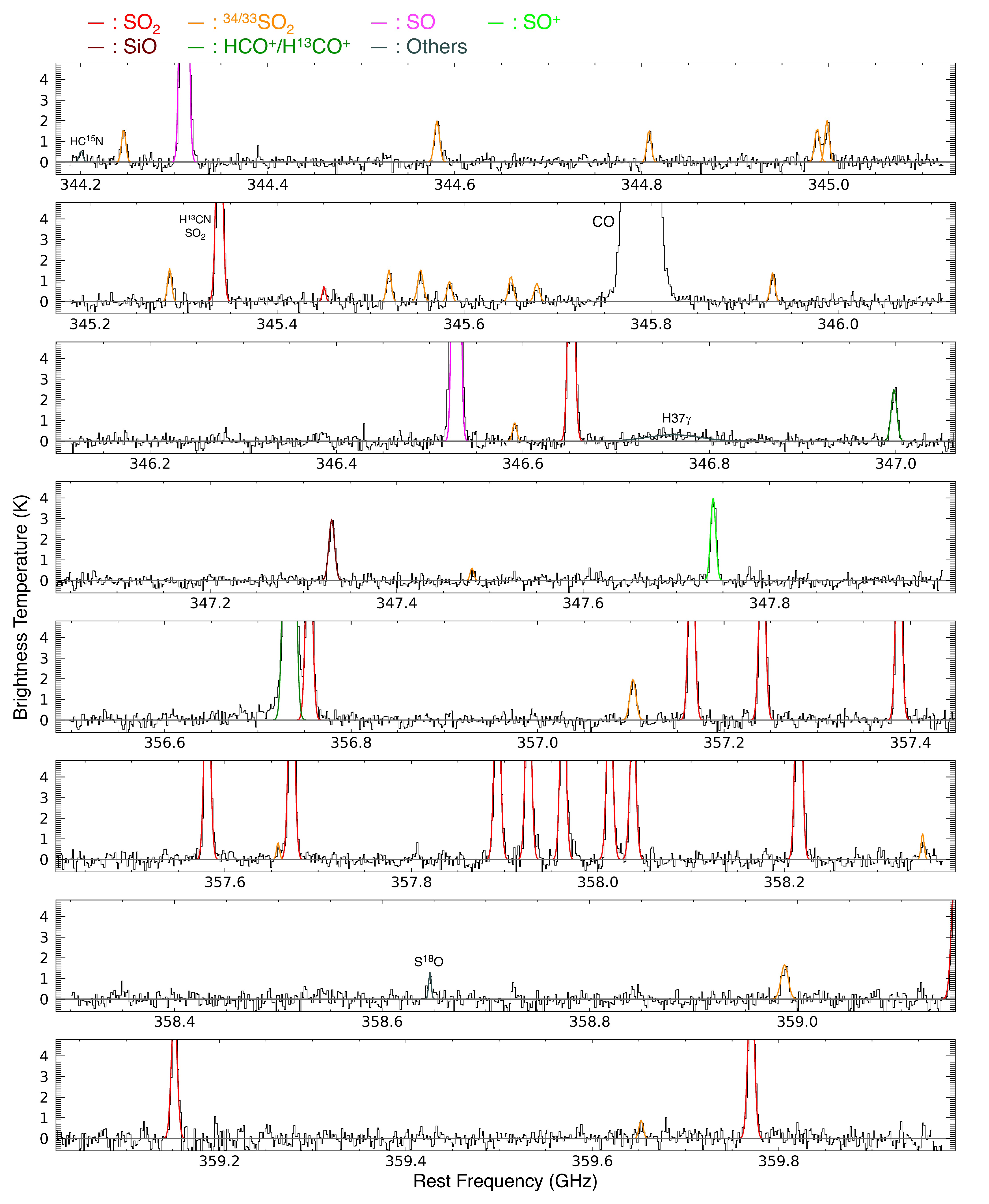}
\caption{
ALMA Band 7 spectra of Lh10. 
The black line represents the observed spectra, while the colored lines indicate the line fitting results. 
Tentative detections are indicated by ``?". 
}
\label{spec1}
\end{center}
\end{figure*}

\begin{figure*}[tp!]
\begin{center}
\includegraphics[width=17.5cm]{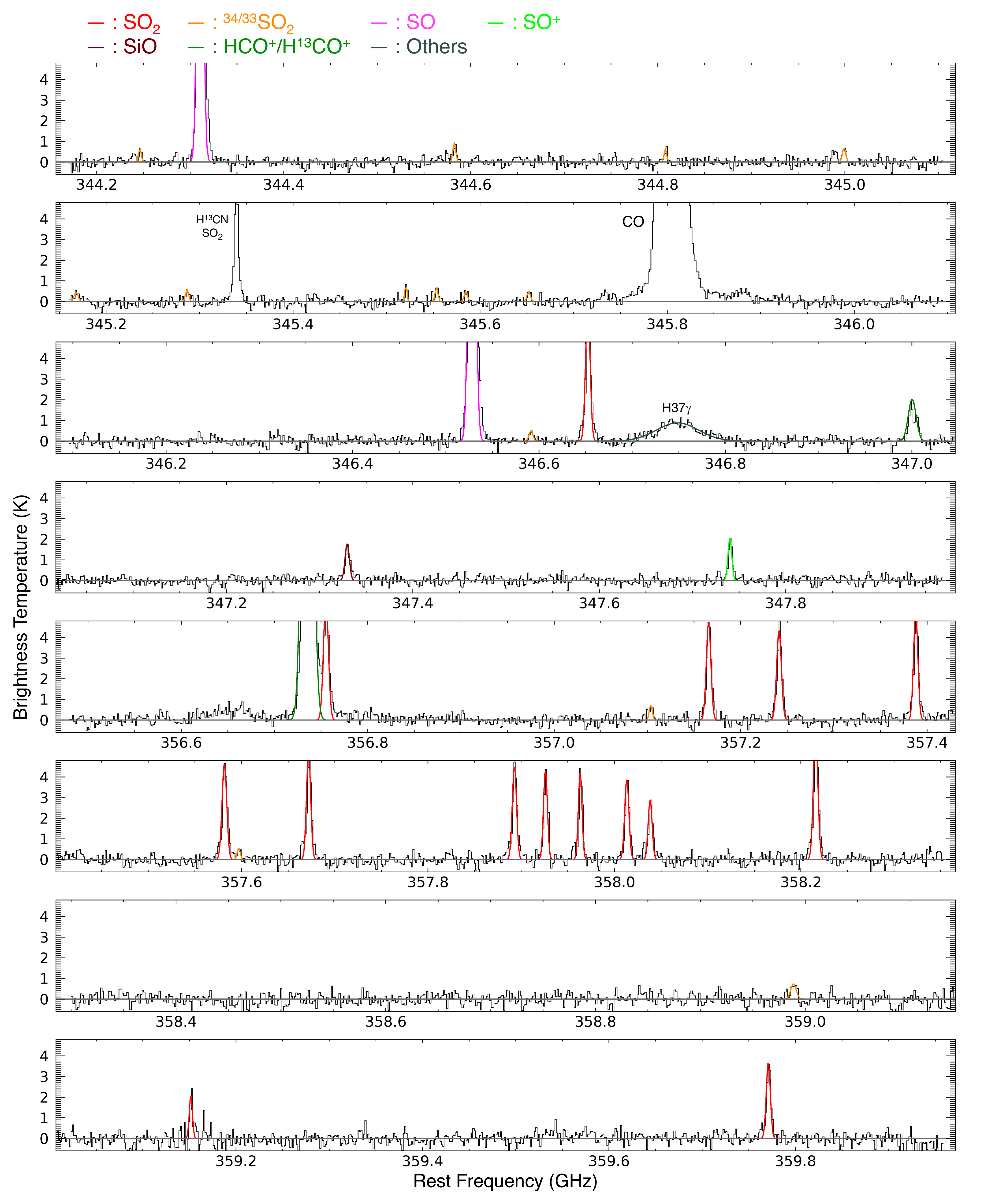}
\caption{Same as in Figure \ref{spec1}, but for Lh09. 
}
\label{spec2}
\end{center}
\end{figure*}

\begin{figure*}[tp!]
\begin{center}
\includegraphics[width=17.5cm]{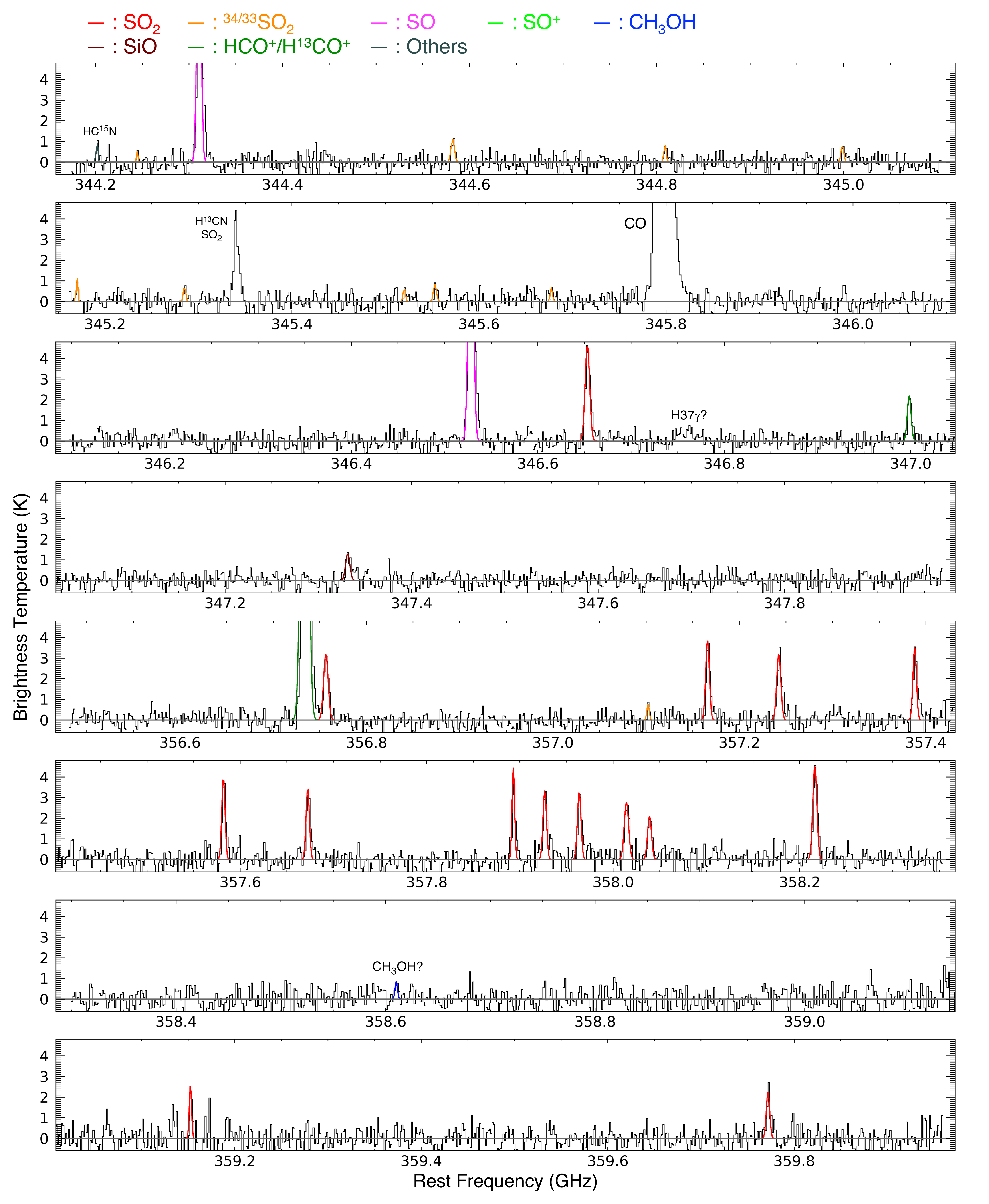}
\caption{Same as in Figure \ref{spec1}, but for Lh08. 
}
\label{spec3}
\end{center}
\end{figure*}

\begin{figure*}[tp!]
\begin{center}
\includegraphics[width=17.5cm]{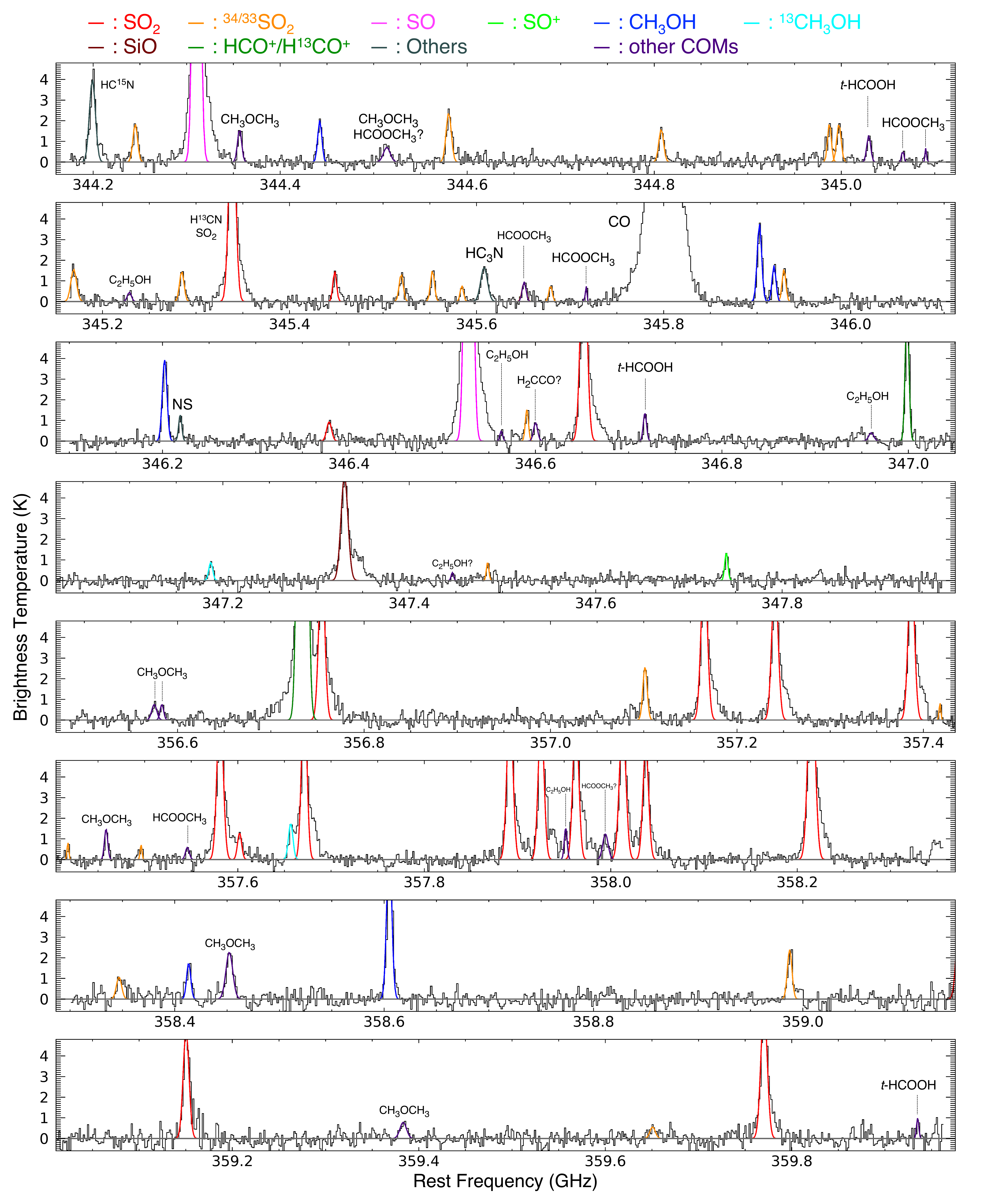}
\caption{Same as in Figure \ref{spec1}, but for Lh07.  
}
\label{spec4}
\end{center}
\end{figure*}

\begin{figure*}[tp!]
\begin{center}
\includegraphics[width=17.5cm]{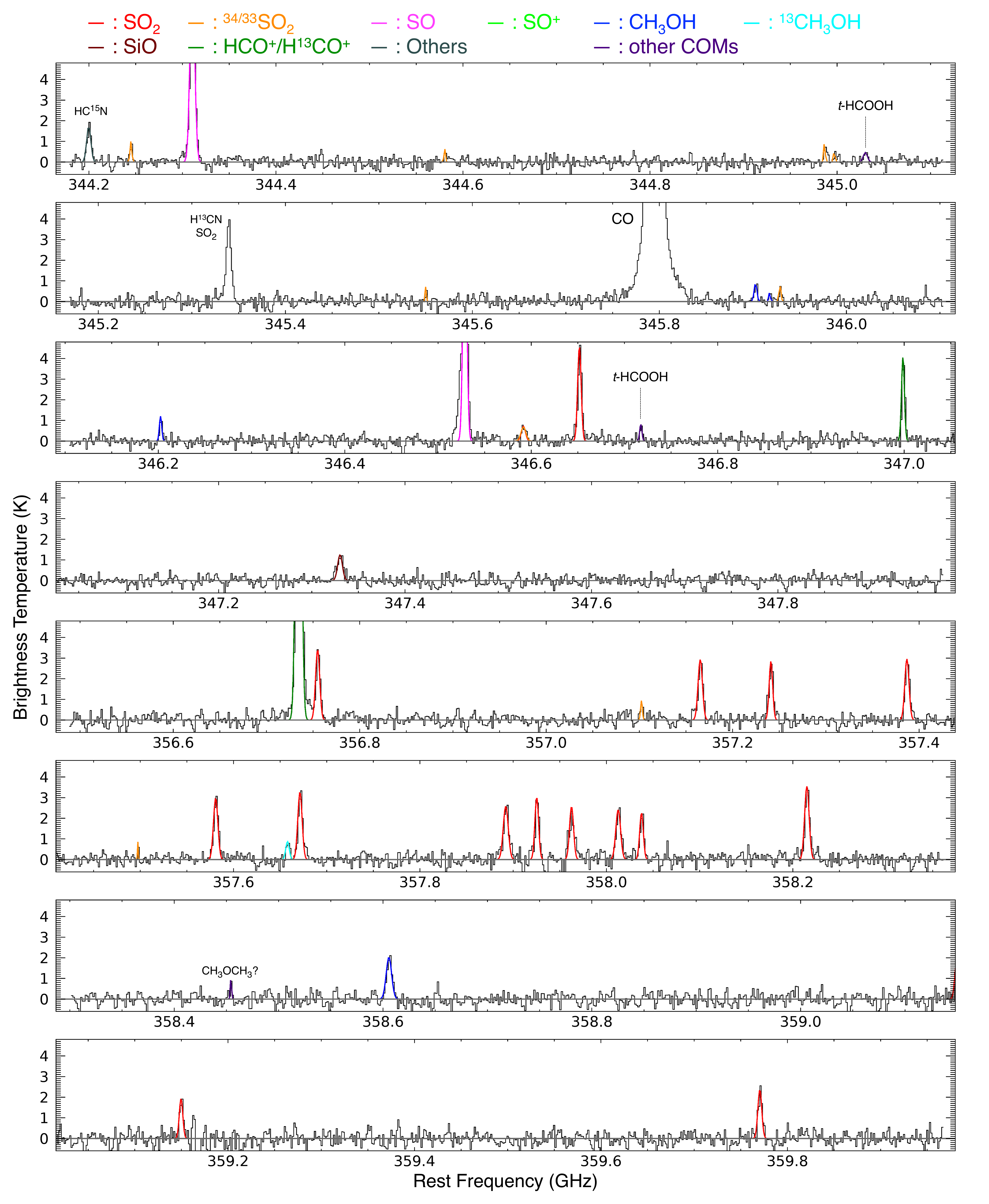}
\caption{Same as in Figure \ref{spec1}, but for Lh05-1.  
}
\label{spec5}
\end{center}
\end{figure*}

\begin{figure*}[tp!]
\begin{center}
\includegraphics[width=17.5cm]{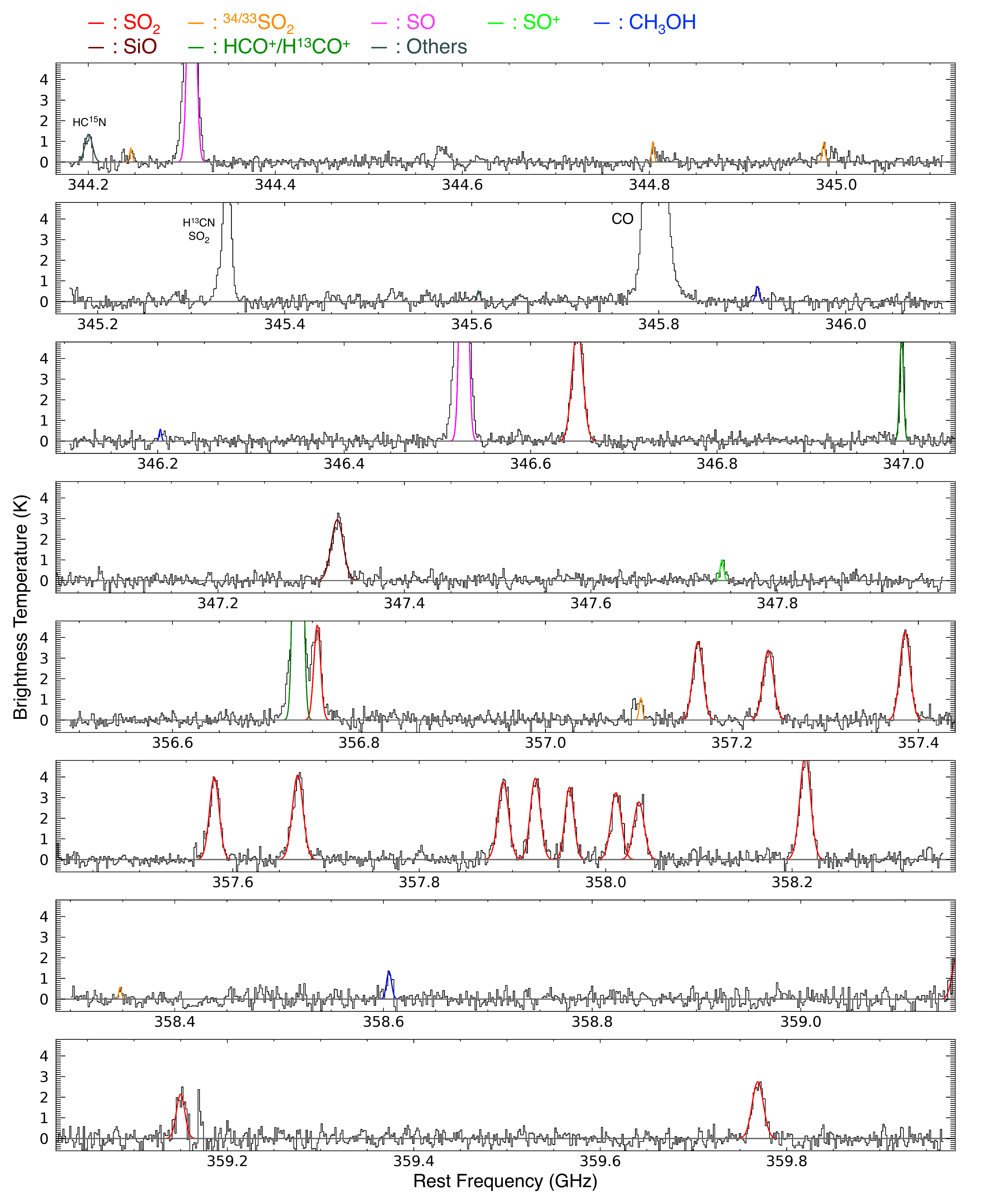}
\caption{Same as in Figure \ref{spec1}, but for Lh05-2.  
}
\label{spec6}
\end{center}
\end{figure*}

\begin{figure*}[tp!]
\begin{center}
\includegraphics[width=17.5cm]{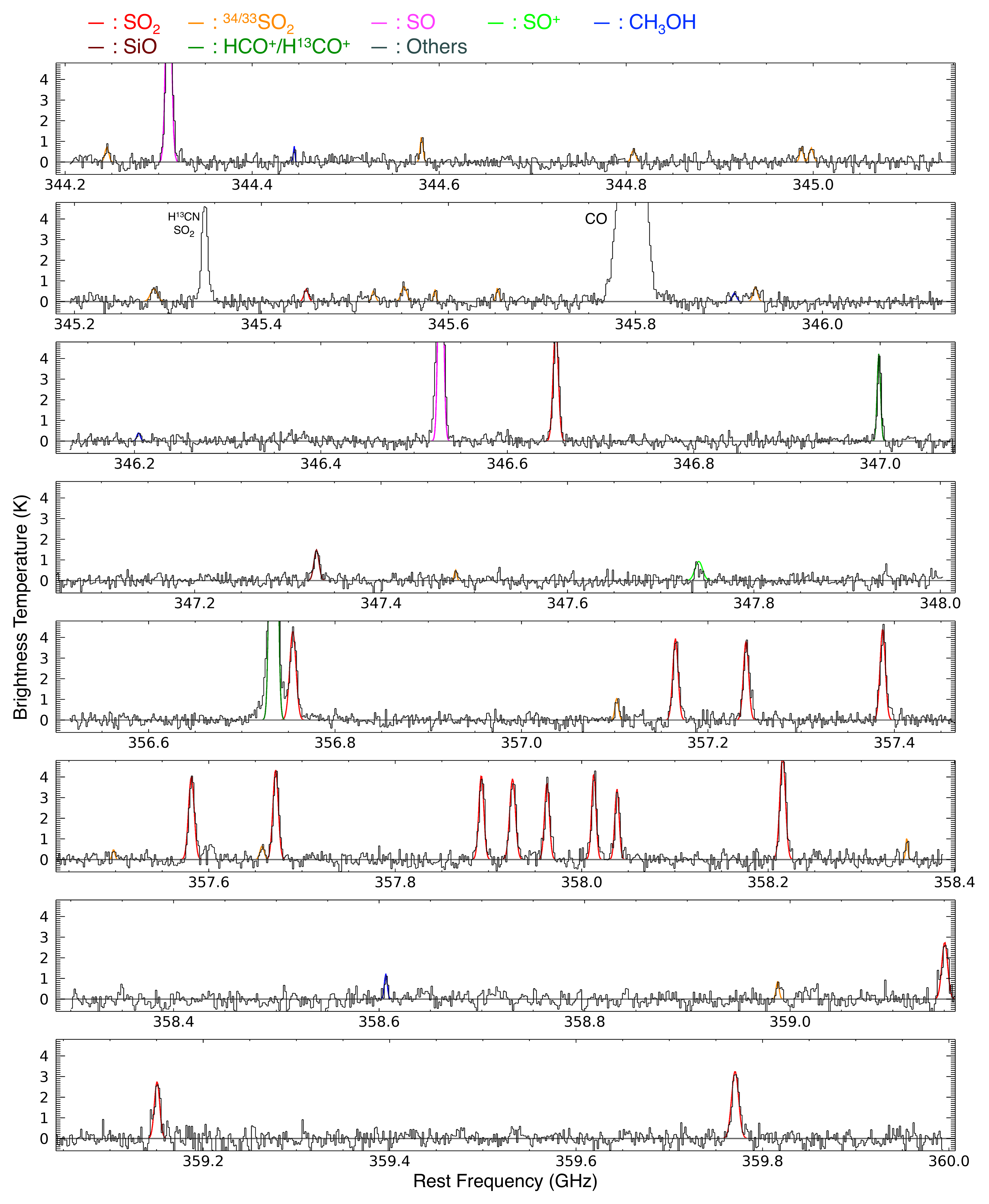}
\caption{Same as in Figure \ref{spec1}, but for Lh03.  
}
\label{spec7}
\end{center}
\end{figure*}

\begin{figure*}[tp!]
\begin{center}
\includegraphics[width=17.5cm]{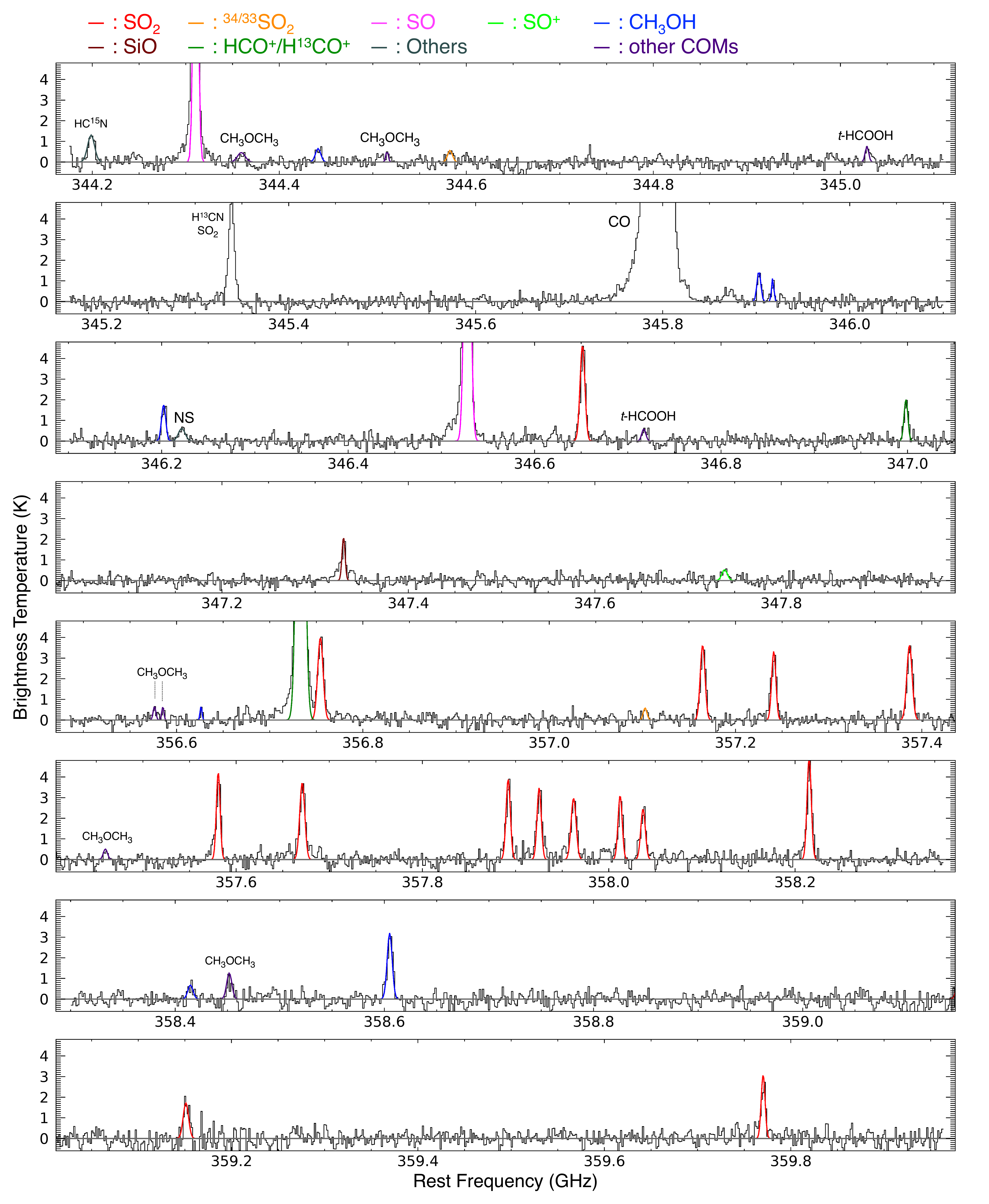}
\caption{Same as in Figure \ref{spec1}, but for Lm10.  
}
\label{spec8}
\end{center}
\end{figure*}

\begin{figure*}[tp!]
\begin{center}
\includegraphics[width=17.5cm]{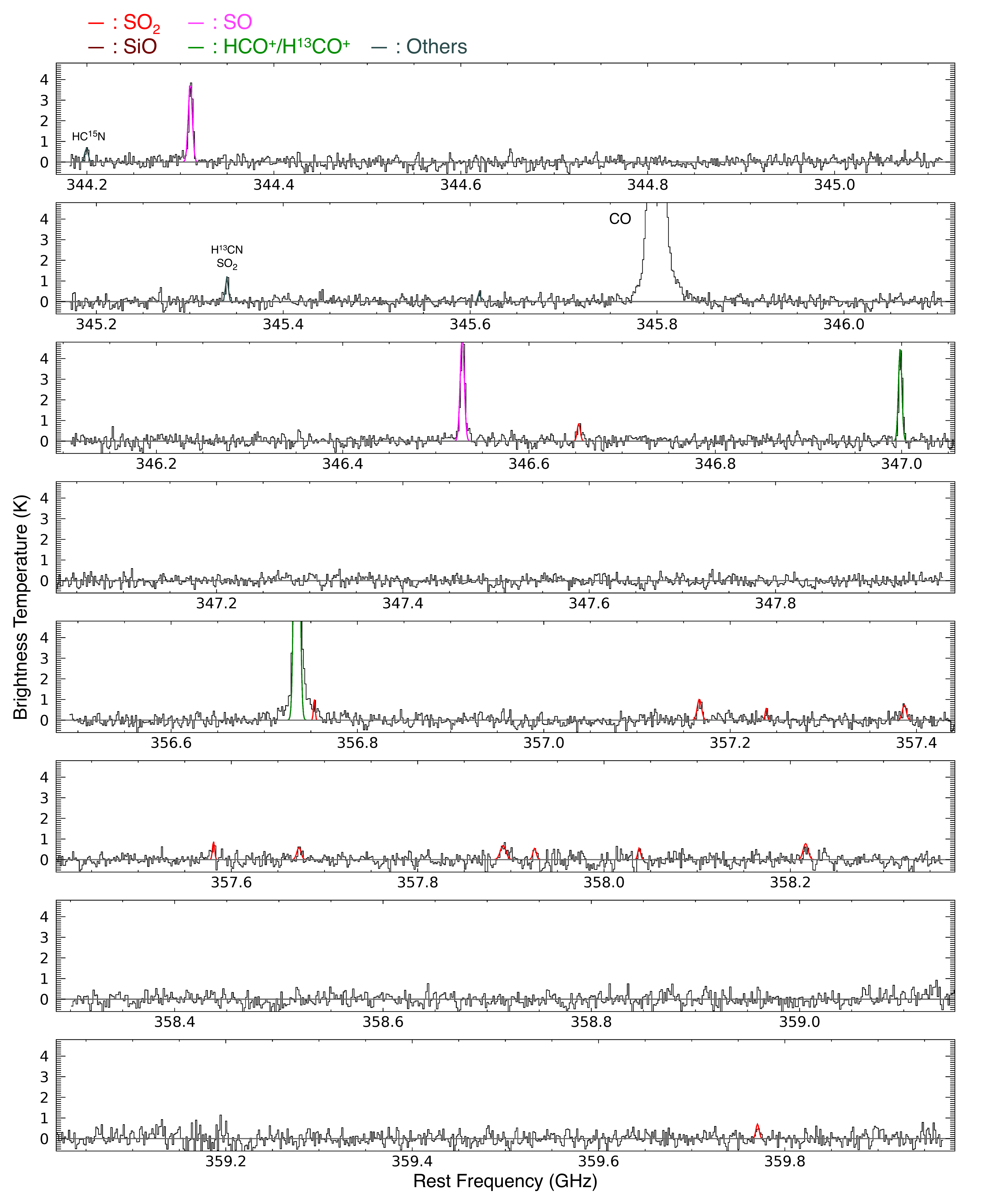}
\caption{Same as in Figure \ref{spec1}, but for Lm07. 
}
\label{spec9}
\end{center}
\end{figure*}

\begin{figure*}[tp!]
\begin{center}
\includegraphics[width=17.5cm]{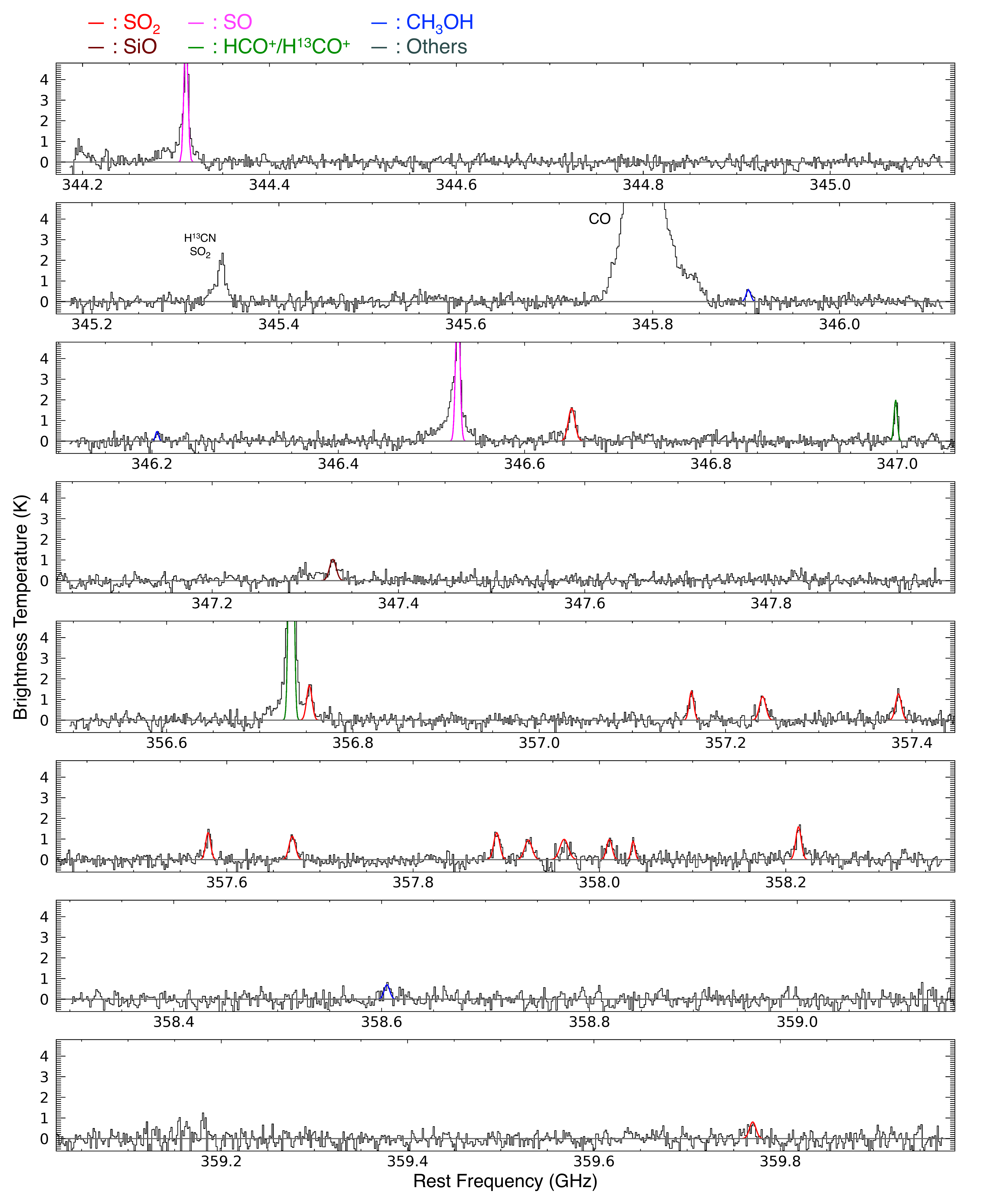}
\caption{Same as in Figure \ref{spec1}, but for Ll10.  
}
\label{spec10}
\end{center}
\end{figure*}

Figures~\ref{spec1}–\ref{spec10} show the ALMA Band 7 spectra of the hot cores and the hot core candidate observed in this survey. 
Table~\ref{tab_intensity} summarizes the measured line parameters of the selected key transitions. 
The detection criterion adopted here is a 3$\sigma$ significance level combined with velocity consistency with the systemic velocities. 
For weak lines, two or three channels are binned to improve sensitivity.

\begin{deluxetable*}{@{\extracolsep{4pt}} lcccccccccc}
\tablecaption{Continuum and Line intensities \label{tab_intensity}}
\tabletypesize{\footnotesize} 
\tablewidth{0pt}
\tablehead{
\colhead{Source} & \colhead{}  & \multicolumn{7}{c}{Integrated Intensity} & \multicolumn{2}{c}{Peak Intensity} \\
\cline{3-9} \cline{10-11}
\colhead{} & \colhead{Continuum}  & 
\colhead{SO}                            & \colhead{SO$_2$}                    & \colhead{SO$_2$}                          & \colhead{CH$_3$OH}                 & \colhead{CH$_3$OH}                                   & \colhead{SiO} & \colhead{H$^{13}$CO$^+$} & \colhead{CO} & \colhead{HCO$^+$} \\
\colhead{} & \colhead{850 $\mu$m} & 
\colhead{(8$_{8}$--7$_{7}$)} & \colhead{(4$_{4,0}$--4$_{3,1}$)} & \colhead{(19$_{1,19}$--18$_{0,18}$)} & \colhead{(4$_{1}$ E--3$_{0}$ E)}  & \colhead{(16$_{1}$ A$^-$--15$_{2}$ A$^-$)} & \colhead{(8--7)} & \colhead{(4--3)}                 & \colhead{(3--2)} & \colhead{(4--3)} \\
\colhead{} & \colhead{(mJy/beam)} & \colhead{(K km/s)} & \colhead{(K km/s)} & \colhead{(K km/s)} & \colhead{(K km/s)} & \colhead{(K km/s)}  & \colhead{(K)} & \colhead{(K)} 
}
\startdata
      Lh10 &   35.1 &  241.2 &   45.8 &   93.3 &  $<$   0.8 &  $<$   0.7 &   19.6 &   16.4 &   77.4 &   41.3 \\
      Lh09 &   67.5 &  138.8 &   12.9 &   34.4 &  $<$   0.7 &  $<$   0.5 &    9.0 &   13.9 &   86.9 &   35.1 \\
      Lh08 &   15.7 &   62.1 &    9.3 &   24.5 &    2.1? &  $<$   1.0 &    6.4 &    9.6 &   72.9 &   31.0 \\
      Lh07 &   25.5 &  204.2 &   34.8 &   76.4 &   38.6 &   19.6 &   40.7 &   28.1 &   69.2 &   50.6 \\
      Lh06 &   24.2 &   16.9 &  $<$   0.7 &  $<$   0.6 &  $<$   0.8 &  $<$   1.1 &  $<$   1.1 &    7.1 &   42.1 &   18.9 \\
    Lh05-1 &   10.4 &   50.2 &    9.3 &   18.9 &   11.0 &    2.9 &    6.8 &   13.8 &   47.9 &   26.4 \\
    Lh05-2 &   10.9 &  143.6 &   33.1 &   68.5 &    6.7 &    2.9 &   40.8 &   22.7 &   56.2 &   30.7 \\
    Lh05-3 &   3.4 &  13.4 &   $<$ 0.8 &   $<$ 1.0 &    $<$ 1.0 &    $<$ 0.8 &   $<$ 1.0 &   2.7 &   25.6 &   11.2 \\
      Lh04 &    4.3 &   15.3 &  $<$   0.8 &  $<$   0.7 &  $<$   0.9 &  $<$   0.7 &  $<$   0.7 &    5.5 &   53.7 &   16.5 \\
      Lh03 &    9.9 &   73.3 &   16.5 &   33.1 &    3.6 &    2.0 &    8.7 &   16.6 &   55.8 &   37.2 \\
      Lh02 &    4.0 &   16.0 &  $<$   1.2 &    2.6 &  $<$   0.8 &  $<$   0.6 &  $<$1.9 &    9.4 &   29.8 &   20.0 \\
    Lh01-1 &    3.4 &   17.2 &  $<$   0.7 &  $<$   0.7 &    5.3 &  $<$   0.7 &  $<$   0.7 &    3.6 &   21.0 &   14.4 \\
    Lh01-2 &    2.6 &   12.9 &    1.1 &  $<$   0.7 &    1.1 &  $<$   0.7 &    0.7? &    8.7 &   45.4 &   21.8 \\
      Lm10 &   12.0 &  90.3 &   12.7 &   26.1 &   16.1 &    6.5 &    7.5 &    9.0 &   51.0 &   26.8 \\
      Lm09 &    3.7 &   11.5 &  $<$   0.7 &  $<$   0.6 &  $<$   0.8 &  $<$   0.6 &  $<$   0.6 &    6.1 &   40.5 &   15.0 \\
      Lm08 &    4.1 &   12.7 &    1.2 &  $<$   0.6 &  $<$   0.8 &  $<$   0.6 &  $<$   0.6 &    5.4 &   48.7 &   16.5 \\
      Lm07 &    6.7 &   18.0 &    1.8 &    3.0 &  $<$   0.8 &  $<$   0.6 &  $<$   1.0 &   17.9 &   52.9 &   24.4 \\
      Lm06 &    2.9 &    9.3 &  $<$   1.2 &  $<$   0.6 &  $<$   0.8 &  $<$   0.6 &  $<$   0.6 &    3.6 &   21.1 &   12.4 \\
    Lm05-1 &    2.0 &    1.5 &  $<$   0.7 &  $<$   1.1 &  $<$   0.8 &  $<$   0.6 &  $<$   0.6 &  $<$   0.6 &   30.0 &    5.9 \\
    Lm05-2 &    1.4 &    4.4 &  $<$   0.3 &  $<$   1.1 &  $<$   1.3 &  $<$  1.6 &  $<$   0.6 &  $<$   1.1 &   46.3 &    8.4 \\
      Lm04 &    2.6 &   10.8 &  $<$   0.7 &  $<$   0.7 &    4.8 &  $<$   0.7 &  $<$   0.7 &    2.6 &   31.9 &    8.9 \\
      Lm03 &    0.6 &  $<$   0.6 &  $<$   0.8 &  $<$   0.7 &  $<$   0.9 &  $<$   0.7 &  $<$   0.7 &    1.5 &   42.3 &    5.2 \\
      Lm02 &    1.6 &    3.5 &  $<$   0.7 &  $<$   0.6 &    2.2 &  $<$   0.6 &  $<$   0.6 &    2.7 &   20.0 &    9.8 \\
      Lm01 &    2.2 &    2.3 &  $<$   0.5 &  $<$   0.6 &  $<$   0.8 &  $<$   0.6 &  $<$   1.4 &  $<$   0.6 &   34.1 &    6.0 \\
      Ll10 &    4.7 &   31.1 &    3.8 &   12.5 &    4.0 &    3.1 &    8.0 &    7.7 &   42.3 &   22.2 \\
      Ll09 &    2.0 &    7.6 &  $<$   2.0 &  $<$   0.6 &  $<$   0.8 &  $<$   0.6 &  $<$   0.6 &  $<$   1.1 &   22.3 &   10.1 \\
      Ll08 &    1.8 &    8.6 &  $<$   0.7 &    2.8 &  $<$   0.8 &  $<$   0.6 &  $<$   0.6 &    2.8 &   23.4 &    9.5 \\
    Ll07-1 &    0.7 &  $<$   0.6 &  $<$   0.7 &  $<$   0.6 &  $<$   0.8 &  $<$   0.6 &  $<$   1.4 &  $<$   0.6 &   14.4 &    5.5 \\
    Ll07-2 &    0.8 &  $<$   0.7 &  $<$   0.7 &  $<$   0.6 &  $<$   1.3 &  $<$   0.6 &  $<$   0.6 &  $<$   0.6 &   17.8 &    5.4 \\
      Ll06 &    0.6 &  $<$   0.6 &  $<$   0.7 &  $<$   0.6 &  $<$   0.8 &  $<$   0.6 &  $<$   0.3 &  $<$   0.2 &   32.8 &    3.3 \\
      Ll05 &    2.0 &   12.2 &  $<$   0.7 &    2.6 &  $<$   0.8 &  $<$   0.6 &  $<$   0.6 &    4.5 &   26.6 &   10.6 \\
    Ll04-1 &    0.4 &  $<$   0.6 &  $<$   0.7 &  $<$   0.6 &  $<$   0.8 &  $<$   0.7 &  $<$   0.6 &  $<$   0.6 &   39.5 &    4.0 \\
    Ll04-2 &    0.9 &  $<$   1.4 &  $<$   0.9 &  $<$   0.6 &  $<$   0.8 &  $<$   0.6 &  $<$   1.4 &  $<$   0.6 &   32.0 &    6.8 \\
      Ll03 &    0.8 &    4.6 &  $<$   0.6 &  $<$   0.6 &  $<$   0.8 &    1.1? &  $<$   0.6 &  $<$   0.6 &   18.7 &    5.7 \\
      Ll02 &   0.2? &  $<$   0.6 &  $<$   0.6 &  $<$   0.6 &  $<$   0.8 &  $<$   0.6 &  $<$   0.6 &  $<$   0.6 &   29.0 &    3.0 \\
      Ll01 &    1.2 &    5.9 &    $<$1.8 &  $<$   0.6 &  $<$   0.8 &  $<$   0.6 &  $<$   0.6 &    3.4 &   23.7 &    9.2 \\
\hline
Error &    \nodata &  $\pm$1.4 & $\pm$1.2 & $\pm$1.1 & $\pm$1.1 & $\pm$0.9 & $\pm$1.3 & $\pm$1.1 & $\pm$0.3 & $\pm$0.4 \\
\enddata
\tablecomments{The last row indicates the typical error. ``?'' indicates tentative detection. 
}
\end{deluxetable*}
%
\clearpage

\clearpage
\restartappendixnumbering
\onecolumngrid
\section{Derived column densities and fractional abundances} \label{sec_app_n} 
Table~\ref{tab_n} summarizes derived column densities and rotational temperatures for the hot cores and the hot core candidate observed in this survey. 

\startlongtable
\begin{deluxetable}{l l c c c c }
\tablecaption{Estimated rotation temperatures, column densities, and fractional abundances \label{tab_n}}
\tabletypesize{\small} 
\tablehead{
\colhead{Object}   & \colhead{Molecule}        & \colhead{$T_{\mathrm{rot}}$}                  &       \colhead{$N$(X)}            &  \colhead{$N$(X)/$N_{\mathrm{H_2}}$}   & Note  \\
\colhead{}                 &  \colhead{}                  &\colhead{(K)}                    &        \colhead{(cm$^{-2}$)}     & \colhead{}                                                  & \colhead{}  
}
\startdata 
Lh10 & H$_2$                          & 172             & (7.9   $\pm$ 2.4  ) $\times$ 10$^{23}$  & \nodata & (1) \\
  & CH$_3$OH                       & 172                       & $<$7     $\times$ 10$^{14}$ & $<$9     $\times$ 10$^{-10}$ & (2) \\
  & SO$_2$ (high-$T$)              & 172$^{+2}_{-2}$           & (7.8   $\pm$ 0.1  ) $\times$ 10$^{15}$  &  \nodata  & (3) \\
&  SO$_2$ (low-$T$)                      &   22$^{+1}_{-1}$            &     (4.9 $\pm$ 0.9) $\times$ 10$^{15}$   & \nodata                              & (3) \\
&  SO$_2$ ($^{34}$S, high-$T$)    &   \nodata                       & (1.3   $\pm$ 0.3  ) $\times$ 10$^{16}$  & (1.6   $\pm$ 0.6  ) $\times$ 10$^{-8 }$  & (4) \\
  & $^{34}$SO$_2$ (high-$T$)       & 138$^{+30}_{-21}$         & (8.6   $\pm$ 2.0  ) $\times$ 10$^{14}$  & (1.1   $\pm$ 0.4  ) $\times$ 10$^{-9 }$  & (3) \\
&  $^{34}$SO$_2$ (low-$T$)          &   $\sim$40                    &     $\sim$6 $\times$ 10$^{14}$         &  \nodata                             & (3) \\
  & SO                             & 172                       & (5.0   $\pm$ 0.1  ) $\times$ 10$^{15}$  & (6.3   $\pm$ 1.9  ) $\times$ 10$^{-9 }$  & (2) \\
  & SO$^+$                         & 172                       & (7.7   $\pm$ 0.4  ) $\times$ 10$^{14}$  & (9.7   $\pm$ 3.0  ) $\times$ 10$^{-10}$  & (2) \\
  & SiO                            & 172                       & (3.1   $\pm$ 1.0  ) $\times$ 10$^{13}$  & (3.9   $\pm$ 1.7  ) $\times$ 10$^{-11}$  & (5) \\
  & HCO$^+$                 & 30--60                    & (3.5   $\pm$ 0.4  ) $\times$ 10$^{14}$  & (4.4   $\pm$ 1.4  ) $\times$ 10$^{-10}$  & (6) \\
& & & & & \\
Lh09 & H$_2$                          & \nodata             & (8.0   $\pm$ 4.0  ) $\times$ 10$^{23}$  & \nodata & (7) \\
  & CH$_3$OH                       & 113                       & $<$3     $\times$ 10$^{14}$ & $<$4     $\times$ 10$^{-10}$ & (2) \\
  & SO$_2$ (high-$T$)              & 113$^{+3}_{-3}$           & (2.4   $\pm$ 0.1  ) $\times$ 10$^{15}$  & \nodata  & (3) \\
&  SO$_2$ (low-$T$)                      &   18$^{+3}_{-2}$            &   $\sim$1 $\times$ 10$^{15}$          & \nodata        & (3) \\
&  SO$_2$ ($^{34}$S, high-$T$)    &   \nodata                       & (3.2   $\pm$ 1.5  ) $\times$ 10$^{15}$  & (4.0   $\pm$ 2.7  ) $\times$ 10$^{-9 }$  & (4) \\
  & $^{34}$SO$_2$ (high-$T$)       & 86$^{+37}_{-20}$          & (2.1   $\pm$ 1.0  ) $\times$ 10$^{14}$  & (2.6   $\pm$ 1.8  ) $\times$ 10$^{-10}$  & (3) \\
&  $^{34}$SO$_2$ (low-$T$)          &   $\sim$30                    &     $\sim$2 $\times$ 10$^{14}$         &  \nodata                                   & (3) \\
  & SO                             & 113                       & (2.4   $\pm$ 0.1  ) $\times$ 10$^{15}$  & (3.0   $\pm$ 1.5  ) $\times$ 10$^{-9 }$       & (2) \\
  & SO$^+$                         & 113                       & (2.3   $\pm$ 0.2  ) $\times$ 10$^{14}$  & (2.9   $\pm$ 1.5  ) $\times$ 10$^{-10}$  & (2) \\
  & SiO                            & 113                       & (1.2   $\pm$ 0.3  ) $\times$ 10$^{13}$  & (1.5   $\pm$ 0.8  ) $\times$ 10$^{-11}$  & (5) \\
  & HCO$^+$                 & 30--60                    & (3.0   $\pm$ 0.3  ) $\times$ 10$^{14}$  & (3.8   $\pm$ 1.9  ) $\times$ 10$^{-10}$  & (6) \\
& & & & & \\
Lh08 & H$_2$                          & 123             & (5.0   $\pm$ 1.5  ) $\times$ 10$^{23}$  & \nodata & (1) \\
  & CH$_3$OH                       & 123                       & $<$2     $\times$ 10$^{15}$ & $<$4     $\times$ 10$^{-9 }$ & (2) \\
  & SO$_2$ (high-$T$)              & 123$^{+18}_{-14}$         & (1.6   $\pm$ 0.3  ) $\times$ 10$^{15}$  & \nodata  & (3) \\
&  SO$_2$ (low-$T$)                      &   18$^{+4}_{-3}$            &    $\sim$8 $\times$ 10$^{14}$   & \nodata                              & (3) \\
&  SO$_2$ ($^{34}$S, high-$T$)    &   \nodata                       & (3.8   $\pm$ 2.3  ) $\times$ 10$^{15}$  & (7.6   $\pm$ 5.1  ) $\times$ 10$^{-9 }$  & (4) \\
  & $^{34}$SO$_2$ (high-$T$)       & 108$^{+81}_{-32}$         & (2.5   $\pm$ 1.5  ) $\times$ 10$^{14}$  & (5.0   $\pm$ 3.4  ) $\times$ 10$^{-10}$  & (3) \\
&  $^{34}$SO$_2$ (low-$T$)          &   $\sim$20                    &     $\sim$1 $\times$ 10$^{14}$         &  \nodata                             & (3) \\
  & SO                             & 123                       & (1.1   $\pm$ 0.1  ) $\times$ 10$^{15}$  & (2.2   $\pm$ 0.7  ) $\times$ 10$^{-9 }$  & (2) \\
  & SO$^+$                         & 123                       & $<$3     $\times$ 10$^{13}$ & $<$5     $\times$ 10$^{-11}$ & (2) \\
  & SiO                            & 123                       & (8.7   $\pm$ 3.4  ) $\times$ 10$^{12}$  & (1.7   $\pm$ 0.9  ) $\times$ 10$^{-11}$  & (5) \\
  & HCO$^+$                 & 30--60                    & (2.1   $\pm$ 0.4  ) $\times$ 10$^{14}$  & (4.2   $\pm$ 1.5  ) $\times$ 10$^{-10}$  & (6) \\
& & & & & \\
& & & & & \\
& & & & & \\
Lh07 & H$_2$                          & 259             & (3.7   $\pm$ 1.1  ) $\times$ 10$^{23}$  & \nodata & (1) \\
  & CH$_3$OH                        & 155$^{+4}_{-4}$           & (3.4   $\pm$ 0.1  ) $\times$ 10$^{16}$  & \nodata                                                  & (3) \\
  & CH$_3$OH ($^{13}$C)      & \nodata                         & (4.9   $\pm$ 1.8  ) $\times$ 10$^{16}$  & (1.3   $\pm$ 0.6  ) $\times$ 10$^{-7 }$  & (8) \\
  & $^{13}$CH$_3$OH                & 155           & (1.0   $\pm$ 0.2  ) $\times$ 10$^{15}$  & (2.7   $\pm$ 1.0  ) $\times$ 10$^{-9 }$  & (9) \\
 & SO$_2$ (high-$T$)              & 363$^{+10}_{-10}$         & (1.2   $\pm$ 0.1  ) $\times$ 10$^{16}$  & \nodata  & (3) \\
&  SO$_2$ (low-$T$)                      &   27$^{+1}_{-1}$            &     (3.9 $\pm$ 0.5) $\times$ 10$^{15}$   & \nodata                              & (3) \\
&  SO$_2$ ($^{34}$S, high-$T$)    &   \nodata                       & (1.7   $\pm$ 0.2  ) $\times$ 10$^{16}$  & (4.6   $\pm$ 1.5  ) $\times$ 10$^{-8 }$  & (4) \\
  & $^{34}$SO$_2$ (high-$T$)       & 145$^{+6}_{-5}$           & (1.1   $\pm$ 0.1  ) $\times$ 10$^{15}$  & (3.0   $\pm$ 0.9  ) $\times$ 10$^{-9 }$  & (3) \\
  & SO                             & 363                       & (7.0   $\pm$ 0.2  ) $\times$ 10$^{15}$  & (1.9   $\pm$ 0.6  ) $\times$ 10$^{-8 }$  & (2) \\
  & SO$^+$                         & 363                       & (3.1   $\pm$ 0.7  ) $\times$ 10$^{14}$  & (8.4   $\pm$ 3.1  ) $\times$ 10$^{-10}$  & (2) \\
  & SiO                            & 259                       & (8.5   $\pm$ 3.3  ) $\times$ 10$^{13}$  & (2.3   $\pm$ 1.1  ) $\times$ 10$^{-10}$  & (5) \\
  & HCO$^+$                 & 30--60                    & (6.0   $\pm$ 0.4  ) $\times$ 10$^{14}$  & (1.6   $\pm$ 0.5  ) $\times$ 10$^{-9 }$  & (6) \\
  & CH$_3$OCH$_3$                  & 97$^{+13}_{-10}$ & (2.6   $\pm$ 0.3  ) $\times$ 10$^{15}$  & (7.0   $\pm$ 2.2  ) $\times$ 10$^{-9 }$  & (3) \\
  & HCOOCH$_3$                     & 114$^{+30}_{-20}$ & (2.9   $\pm$ 1.6  ) $\times$ 10$^{15}$  & (7.8   $\pm$ 4.9  ) $\times$ 10$^{-9 }$  & (3) \\
  & C$_2$H$_5$OH                   & 83$^{+25}_{-16}$ & (2.0   $\pm$ 1.3  ) $\times$ 10$^{15}$  & (5.4   $\pm$ 3.9  ) $\times$ 10$^{-9 }$  & (3)  \\
  & \textit{trans}-HCOOH            & 141$^{+41}_{-26}$ & (6.9   $\pm$ 2.0  ) $\times$ 10$^{14}$  & (1.9   $\pm$ 0.8  ) $\times$ 10$^{-9 }$  &  (3)  \\
  & HC$_3$N                           & 155                         & (6.8   $\pm$ 0.6  ) $\times$ 10$^{13}$  & (1.8   $\pm$ 0.6  ) $\times$ 10$^{-10}$  & (9) \\
  & H$_2$CCO                         & 155                         & (2.2   $\pm$ 0.5  ) $\times$ 10$^{14}$  & (5.9   $\pm$ 2.2  ) $\times$ 10$^{-10}$  & (9) \\
  & CH$_3$CHO                       & 155                         & $<$2     $\times$ 10$^{14}$ & $<$5     $\times$ 10$^{-10 }$                                   &      (9) \\
& & & & & \\
Lh05-1 & H$_2$                          & 146             & (2.7   $\pm$ 0.8  ) $\times$ 10$^{23}$  & \nodata & (1) \\
  & CH$_3$OH                       & 121$^{+15}_{-13}$         & (6.1   $\pm$ 1.7  ) $\times$ 10$^{15}$  & (2.3   $\pm$ 0.9  ) $\times$ 10$^{-8 }$  & (3) \\
  & SO$_2$ (high-$T$)              & 170$^{+15}_{-13}$         & (1.6   $\pm$ 0.2  ) $\times$ 10$^{15}$  & \nodata     & (3) \\
&  SO$_2$ (low-$T$)                      &   23$^{+5}_{-3}$            &   $\sim$1 $\times$ 10$^{15}$           & \nodata                              & (3) \\
&  SO$_2$ ($^{34}$S, high-$T$)    &   \nodata                       & (3.5   $\pm$ 1.1  ) $\times$ 10$^{15}$  & (1.3   $\pm$ 0.6  ) $\times$ 10$^{-8 }$  & (4) \\
  & $^{34}$SO$_2$ (high-$T$)       & 174$^{+70}_{-39}$         & (2.3   $\pm$ 0.7  ) $\times$ 10$^{14}$  & (8.5   $\pm$ 3.6  ) $\times$ 10$^{-10}$  & (3) \\
  & SO                             & 170                       & (1.0   $\pm$ 0.1  ) $\times$ 10$^{15}$  & (3.7   $\pm$ 1.2  ) $\times$ 10$^{-9 }$  & (2) \\
  & SO$^+$                         & 170                       & $<$2     $\times$ 10$^{13}$ & $<$7     $\times$ 10$^{-11}$ & (2) \\
  & SiO                            & 146                       & (3.0   $\pm$ 0.3  ) $\times$ 10$^{13}$  & (1.1   $\pm$ 0.3  ) $\times$ 10$^{-10}$  & (5) \\
  & HCO$^+$                 & 30--60                    & (3.0   $\pm$ 0.3  ) $\times$ 10$^{14}$  & (1.1   $\pm$ 0.3  ) $\times$ 10$^{-9 }$  & (6) \\
  & \textit{trans}-HCOOH                    & 121               & (2.5   $\pm$ 1.1  ) $\times$ 10$^{14}$  & (9.3   $\pm$ 4.9  ) $\times$ 10$^{-10}$  &  (9)  \\
& & & & & \\
Lh05-2 & H$_2$                          & 148             & (2.8   $\pm$ 0.8  ) $\times$ 10$^{23}$  & \nodata & (1) \\
  & CH$_3$OH                       & 149$^{+36}_{-24}$         & (4.3   $\pm$ 1.2  ) $\times$ 10$^{15}$  & (1.5   $\pm$ 0.6  ) $\times$ 10$^{-8 }$  & (3) \\
  & SO$_2$ (high-$T$)              & 148$^{+5}_{-5}$           & (4.9   $\pm$ 0.3  ) $\times$ 10$^{15}$  & \nodata & (3) \\
&  SO$_2$ (low-$T$)                      &   21$^{+2}_{-1}$            &     (3.4 $\pm$ 1.2) $\times$ 10$^{15}$   & \nodata                              & (3) \\
&  SO$_2$ ($^{34}$S, high-$T$)    &   \nodata                       & (4.9   $\pm$ 0.3  ) $\times$ 10$^{15}$  & (1.7   $\pm$ 0.5  ) $\times$ 10$^{-8 }$  & (4) \\
  & $^{34}$SO$_2$ (high-$T$)       & 134$^{+78}_{-36}$         & (3.2   $\pm$ 1.5  ) $\times$ 10$^{14}$  & (1.1   $\pm$ 0.6  ) $\times$ 10$^{-9 }$  & (3) \\
  & SO                             & 148                       & (2.8   $\pm$ 0.1  ) $\times$ 10$^{15}$  & (1.0   $\pm$ 0.3  ) $\times$ 10$^{-9 }$  & (2) \\
  & SO$^+$                         & 148                       & (1.2   $\pm$ 0.3  ) $\times$ 10$^{14}$  & (4.3   $\pm$ 1.6  ) $\times$ 10$^{-10}$  & (2) \\
  & SiO                            & 148                       & (6.0   $\pm$ 1.6  ) $\times$ 10$^{13}$  & (2.1   $\pm$ 0.8  ) $\times$ 10$^{-10}$  & (5) \\
  & HCO$^+$                 & 30--60                    & (4.9   $\pm$ 0.4  ) $\times$ 10$^{14}$  & (1.8   $\pm$ 0.5  ) $\times$ 10$^{-9 }$  & (6) \\
  & \textit{trans}-HCOOH                    & 149                 & $<$8     $\times$ 10$^{13}$               & $<$3     $\times$ 10$^{-10}$ &  (9)  \\
& & & & & \\
Lh03 & H$_2$                          & 246                            & (1.5   $\pm$ 0.5  ) $\times$ 10$^{23}$  & \nodata & (1) \\
  & CH$_3$OH                       & 185$^{+61}_{-37}$         & (4.0   $\pm$ 1.6  ) $\times$ 10$^{15}$  & (2.7   $\pm$ 1.4  ) $\times$ 10$^{-8 }$  & (3) \\
  & SO$_2$ (high-$T$)              & 306$^{+12}_{-1}$          & (4.7   $\pm$ 0.2  ) $\times$ 10$^{15}$  & \nodata & (3) \\
 &  SO$_2$ (low-$T$)                      &   31$^{+2}_{-2}$            &     (1.6 $\pm$ 0.3) $\times$ 10$^{15}$   & \nodata                              & (3) \\
 &  SO$_2$ ($^{34}$S, high-$T$)    &   \nodata                       & (6.5   $\pm$ 3.1  ) $\times$ 10$^{15}$  & (4.3   $\pm$ 2.5  ) $\times$ 10$^{-8 }$  &  (4) \\
  & $^{34}$SO$_2$ (high-$T$)       & 245$^{+90}_{-52}$         & (4.3   $\pm$ 1.2  ) $\times$ 10$^{14}$  & (2.9   $\pm$ 1.2  ) $\times$ 10$^{-9 }$  & (3) \\
 &  $^{34}$SO$_2$ (low-$T$)          &   $\sim$45                    &     $\sim$2 $\times$ 10$^{14}$         &  \nodata                             & (3) \\
  & SO                             & 306                       & (2.1   $\pm$ 0.1  ) $\times$ 10$^{15}$  & (1.4   $\pm$ 0.5  ) $\times$ 10$^{-8 }$  & (2) \\
  & SO$^+$                         & 306                       & (2.6   $\pm$ 0.6  ) $\times$ 10$^{14}$  & (1.7   $\pm$ 0.7  ) $\times$ 10$^{-9 }$  & (2) \\
  & SiO                            & 246                       & (1.8   $\pm$ 0.8  ) $\times$ 10$^{13}$  & (1.2   $\pm$ 0.7  ) $\times$ 10$^{-10}$  & (5) \\
  & HCO$^+$                 & 30--60                    & (3.6   $\pm$ 0.3  ) $\times$ 10$^{14}$  & (2.4   $\pm$ 0.8  ) $\times$ 10$^{-9 }$  & (6) \\
& & & & & \\
Lm10 & H$_2$                          & 182             & (2.4   $\pm$ 0.7  ) $\times$ 10$^{23}$  & \nodata & (1) \\
  & CH$_3$OH                       & 169$^{+15}_{-13}$         & (1.6   $\pm$ 0.3  ) $\times$ 10$^{16}$  & (6.7   $\pm$ 2.3  ) $\times$ 10$^{-8 }$  & (3) \\
  & SO$_2$ (high-$T$)              & 192$^{+16}_{-14}$         & (2.2   $\pm$ 0.2  ) $\times$ 10$^{15}$  & (9.2   $\pm$ 2.8  ) $\times$ 10$^{-9 }$  & (3) \\
  &  SO$_2$ (low-$T$)                      &   26$^{+2}_{-2}$            &     (1.2 $\pm$ 0.4) $\times$ 10$^{15}$   & \nodata                              & (3) \\
  & SO                             & 192                       & (2.0   $\pm$ 0.1  ) $\times$ 10$^{15}$  & (8.3   $\pm$ 2.5  ) $\times$ 10$^{-9 }$  & (2) \\
  & SO$^+$                         & 192                       & (1.2   $\pm$ 0.4  ) $\times$ 10$^{14}$  & (5.0   $\pm$ 2.2  ) $\times$ 10$^{-10}$  & (2) \\
  & SiO                            & 182                       & (1.3   $\pm$ 0.5  ) $\times$ 10$^{13}$  & (5.4   $\pm$ 2.6  ) $\times$ 10$^{-11}$  & (5) \\
  & HCO$^+$                 & 30--60                    & (1.9   $\pm$ 0.3  ) $\times$ 10$^{14}$  & (7.9   $\pm$ 2.6  ) $\times$ 10$^{-10}$  & (6) \\
  & CH$_3$OCH$_3$                  & 83$^{+30}_{-17}$ & (8.0   $\pm$ 2.4  ) $\times$ 10$^{14}$  & (3.3   $\pm$ 1.4  ) $\times$ 10$^{-9 }$  & (3) \\
  & \textit{trans}-HCOOH              & 169 & (4.1   $\pm$ 1.3  ) $\times$ 10$^{14}$  & (1.7   $\pm$ 0.7  ) $\times$ 10$^{-9 }$  & (9)  \\
  & HCOOCH$_3$                     & 169 & $<$2     $\times$ 10$^{15}$ & $<$8     $\times$ 10$^{-9 }$ & (3) \\
  & C$_2$H$_5$OH                   & 169 & $<$7     $\times$ 10$^{14}$ & $<$3     $\times$ 10$^{-9 }$ &  (9) \\
  & CH$_3$CHO                      & 169 & $<$2     $\times$ 10$^{14}$ & $<$8     $\times$ 10$^{-10}$ & (9)  \\
& & & & & \\
Lm07 & H$_2$                          & 78              & (3.5   $\pm$ 1.1  ) $\times$ 10$^{23}$  & \nodata & (1) \\
  & CH$_3$OH                       & 78                        & $<$2     $\times$ 10$^{14}$ & $<$7     $\times$ 10$^{-10}$ & (2) \\
  & SO$_2$                  & 78$^{+13}_{-10}$          & (3.1   $\pm$ 0.1  ) $\times$ 10$^{14}$  & (8.9   $\pm$ 2.8  ) $\times$ 10$^{-10}$  & (3) \\
  & SO                             & 78                        & (3.0   $\pm$ 0.2  ) $\times$ 10$^{14}$  & (8.6   $\pm$ 2.8  ) $\times$ 10$^{-10}$  & (2) \\
  & SO$^+$                         & 78                        & $<$2     $\times$ 10$^{14}$ & $<$4     $\times$ 10$^{-10}$ & (2) \\
  & SiO                            & 78                        & $<$2     $\times$ 10$^{12}$ & $<$5     $\times$ 10$^{-12}$ & (5) \\
  & HCO$^+$                 & 30--60                    & (3.8   $\pm$ 0.3  ) $\times$ 10$^{14}$  & (1.1   $\pm$ 0.4  ) $\times$ 10$^{-9}$  & (6) \\
& & & & & \\
Ll10 & H$_2$                          & 143             & (1.2   $\pm$ 0.4  ) $\times$ 10$^{23}$  & \nodata & (1) \\
  & CH$_3$OH                       & 200$^{+130}_{-57}$        & (4.7   $\pm$ 2.0  ) $\times$ 10$^{15}$  & (3.6   $\pm$ 1.9  ) $\times$ 10$^{-8 }$  & (3) \\
  & SO$_2$ (high-$T$)              & 87$^{+8}_{-7}$            & (7.5   $\pm$ 1.1  ) $\times$ 10$^{14}$  & (5.8   $\pm$ 2.0  ) $\times$ 10$^{-9 }$  & (3) \\
 &  SO$_2$ (low-$T$)                      &   36$^{+15}_{-8}$            &    $\sim$2 $\times$ 10$^{14}$    & \nodata                              & (3) \\
  & SO                             & 87                        & (5.2   $\pm$ 0.3  ) $\times$ 10$^{14}$  & (4.0   $\pm$ 1.3  ) $\times$ 10$^{-9 }$  & (2) \\
  & SO$^+$                         & 87                        & $<$2     $\times$ 10$^{13}$ & $<$2     $\times$ 10$^{-10}$ & (2) \\
  & SiO                            & 143                       & (1.2   $\pm$ 0.5  ) $\times$ 10$^{13}$  & (9.2   $\pm$ 4.8  ) $\times$ 10$^{-11}$  & (5) \\
  & HCO$^+$                 & 30--60                    & (1.6   $\pm$ 0.3  ) $\times$ 10$^{14}$  & (1.2   $\pm$ 0.4  ) $\times$ 10$^{-9 }$  & (6) \\
\enddata
\tablecomments{
Uncertainties and upper limits are of 2$\sigma$ level and do not include systematic errors due to adopted spectroscopic constants. 
See Section \ref{sec_ana} for more details. 
(1) $N_{\mathrm{H_2}}$ is derived from dust continuum. The dust temperature is assumed to be equal to the rotational temperature of SO$_2$ or CH$_3$OH, or the average of the two if both are available. A 30$\%$ uncertainty is assumed for $N_{\mathrm{H_2}}$. 
(2) $T_{\mathrm{rot}}$ is assumed to be the same as that of the high-temperature component of SO$_2$. 
(3) Based on the rotation diagram. 
(4) Estimated from $^{34}$SO$_2$ assuming $^{32}$S/$^{34}$S = 15. 
(5) A 50$\%$ uncertainty is assumed for the adopted $T_{\mathrm{rot}}$. 
(6) Estimated from H$^{13}$CO$^+$ assuming the adopted $T_{\mathrm{rot}}$. 
(7) See Section \ref{sec_ana} for the derivation of $N_{\mathrm{H_2}}$ for Lh09. 
(8) Estimated from  $^{13}$CH$_3$OH assuming $^{12}$C/$^{13}$C = 49. 
(9) $T_{\mathrm{rot}}$ is assumed to be the same as that of CH$_3$OH. 
}
\end{deluxetable}

\end{document}